\documentclass[pra,letterpaper,onecolumn,nofootinbib,10pt]{revtex4}
	\usepackage{graphicx}
	\usepackage{epstopdf}
\usepackage{amsmath} 		% better math formatting
\usepackage{amssymb}		% additional math symbols	
\usepackage{amsthm}   		% theorem-formatting

\usepackage[mathscr]{eucal}	% extra Euler Script font
\usepackage{mathrsfs}		% Ralph Smith's Formal Script : defines mathscr{}
\DeclareMathAlphabet{\mathpzc}{OT1}{pzc}{m}{it}  % postscript Zapf Chancery math alphabet

\usepackage{bm}			% more robust and more general bold-face math: use \boldsymbol{}
\usepackage{verbatim}     % allows long comment blocks: use "\begin{comment}" and "\end{comment}"
\usepackage{xspace}	% better handling of spaces in macros

\usepackage{epigraph}    		% allows epigraphs

\usepackage{xfrac}

\providecommand{\textquote}[1]{\textsl{#1}}      % quotation
\providecommand{\textforeign}[1]{\textit{#1}}     % foreign language quotation
\providecommand{\textperson}[1]{\textsc{#1}}   % author
\providecommand{\textsource}[1]{\textit{#1}}     % work
\def\be{\begin{equation}}      % with numbering
\def\ee{\end{equation}}

\def\beu{\begin{equation*}}   % without numbering
\def\eeu{\end{equation*}}

\def\bsub{\begin{subequations}}  % AMS subequations
\def\esub{\end{subequations}}

\def\ie{i.e.}  
\def\eg{e.g.}
\providecommand{\stext}[1]{\text{\tiny{#1}}}  % with AMS-LaTeX v. 2, can use \Tiny for even smaller text....
\def\iff{\Leftrightarrow}  % this constitutes a re-defintiion; remove for a longer arrow with more space

\def\realsymbol{\mathbb{R}}

\DeclareMathOperator{\realpart}{Re} 
\DeclareMathOperator{\impart}{Im} 
\providecommand{\abs}[1]{\left\lvert#1\right\rvert}   % absolute value
\DeclareMathOperator*{\argmin}{arg\,min}	% arguments minimizing
\DeclareMathOperator*{\argmax}{arg\,max}	% arguments maximizing
\providecommand{\bv}[1]{\boldsymbol{#1}}        % alternate shorthand for same
\providecommand{\vv}[1]{\bv{#1}}  % choose either standard, modified-arrow,  or bold-faced format....
\providecommand{\unitvec}[1]{\hat{\boldsymbol{#1}}} % unit vector format
\providecommand{\opdet}[1]{\abs}  %  bar form of determinant
\def\transsymbol{\text{\tiny T}}  		% symbol for matrix transpose
\providecommand{\cc}{^{\ast}}		% complex conjugate
\providecommand{\hc}{^{\dag}}		% hermitian conjugate
\providecommand{\trans}{^{\transsymbol}}  % transpose
\providecommand{\inv}{^{-1}}			% inverse
\providecommand{\innerp}[2]{\left\langle#1\left\lvert\vphantom{#1#2}\right.\!#2\right\rangle}  %  Dirac form for inner product

\providecommand{\innerpd}[2]{\left\langle#1\left\lvert\vphantom{#1#2}\right.\!#2\right\rangle}     % Dirac (angle bracket and bar) form
\providecommand{\innerpr}[2]{\left(#1 ,  #2\right)}  % standard math rounded-parenthesis-and-comma  form 
\providecommand{\grad}{\vv{\nabla}}
\providecommand{\divergence}{\grad\cdot}
\providecommand{\curl}{\grad\times}
\providecommand{\laplacian}{\nabla^{2}}

\providecommand{\del}{\partial}

\DeclareMathSymbol{\bigtimes}{\mathop}{symbols}{"02}
\begin{document}

% title page

%\preprint{working draft}

\title{Technical Report:\\
A Variational Principle for Spontaneous Wiggler and Synchrotron Radiation}
\author{A.E.~Charman}
\affiliation{Department of Physics, U.C.\ Berkeley, Berkeley, CA 94720}%
\email{acharman@physics.berkeley.edu}
\author{G.~Penn}
\affiliation{Center for Beam Physics, Lawrence Berkeley National Laboratory, Berkeley, CA 94720}
\author{J.S.~Wurtele}
\affiliation{Department of Physics, U.C.\ Berkeley, Berkeley, CA 94720}
%
% \date{5/01/2020}
% \date{05/20/2020}
\date{06/14/2020}

% abstract

\begin{abstract}
Within the framework of a Hilbert space theory, we develop a maximum-``power'' variational principle (MPVP) applicable to classical spontaneous electromagnetic radiation from relativistic electron beams or other prescribed classical current sources.  A simple proof is summarized for the case of three-dimensional fields propagating in vacuum, and specialization to the important case of paraxial optics is also discussed.  The techniques have been developed to model undulator radiation from relativistic electron beams, but are more broadly applicable to synchrotron or other radiation problems, and may generalize to certain structured media.  We illustrate applications with a simple, mostly analytic example involving spontaneous undulator radiation (requiring a few additional approximations), as well as a mostly numerical example involving x-ray generation via high harmonic generation in sequenced undulators.
\end{abstract}

\maketitle

\begin{epigraphs}
\qitem{ \textforeign{Mehr Licht!} \\ \textquote{(More light!)} }{ \textperson{Johann Wolfgang Von Goethe}\\\textsource{(attributed last words)}}
\vspace{1.4 ex}
\end{epigraphs}

\section{introduction}

Although perhaps more familiar in classical and quantum mechanics, variational principles are also ubiquitous in electromagnetism \cite{harrington:61, mikhlin:64, jackson:75, kong:86, davies:90, wang:91, zhang:91, vago_gyimesi:98, hanson_yakovlev:02, schwinger:06}, and variational techniques or approaches enjoy several advantages: they can provide unified theoretical treatments and compact mathematical descriptions of many physical phenomena;  they often suggest appealing physical interpretations of the physical behaviors governed by them; they allow flexibility and freedom, and facilitate changes of coordinates and imposition of constraints or incorporation of conservation laws; they can reveal substantial connections between classical and quantum mechanical descriptions;  and from a practical standpoint, they offer starting points for efficient approximations or economical numerical computations, replacing systems of complicated PDEs or integro-differential equations with more tractable quadratures, ODEs, algebraic or even linear equations, and/or ordinary function minimization.  Although approximate, parameterized variational solutions may be more easily interpreted or offer more insights than even exact forms.

Here we review another variational principle, which we call, in a slight abuse of terminology, the Maximum-Power Variational Principle (MPVP), along with some surrounding mathematical formalism.  Although relatively simple in its statement and scope, somewhere between intuitively plausible and obvious depending upon one's point of view, the MPVP may be of some use in the classical theory of radiation, in particular in the analysis of light sources relying on radiation from relativistic electron beams in undulators, or more generally for the approximation of features of various forms of synchrotron or ``magnetic Bremsstrahlung'' emission, such in problems involving coherent synchrotron radiation (CSR) effects for short electron bunches in storage rings.  After some suitable further generalization, we anticipate that these ideas could also be of use in contexts of antenna, \v{C}erenkov, transition, wave-guide, Smith-Purcell, photonic crystal, or other types of radiation emitted by electric currents in the vicinity of conductors, or by charged particle beams traveling through certain media or structures.

Motivated by well-known parallels between the Schr{\"o}dinger equation in non-relativistic quantum mechanics and the paraxial wave equation of classical physical optics, we originally introduced a Hilbert space formalism for wiggler fields and derived the MPVP in the paraxial limit.  Guided by these results, we generalized these results to the case of non-paraxial fields in free space\cite{charman:05}.  Here, after discussion of our main assumptions and basic governing equations, we will present a simplified proof for the general free-space geometry, discuss its specialization to the important limit of paraxial optics, translate between time-domain and frequency-domain versions, and briefly speculate on possible further generalizations.
% to media containing conducting or dielectric structures.
We offer some physical interpretations of the MPVP, compare and contrast it with better known variational principles in electromagnetism, and briefly summarize its application to treatments of radiation from relativistic electron beams in magnetic undulators.

The variational principle explicated here may be summarized by saying that, in effect, 
\textit{classical charges radiate spontaneously ``as much as possible,'' consistent with energy conservation.}  Given prescribed sources in the form of an electric current density in either the time domain or frequency domain, the MPVP can supply approximations to the spatial profile and polarization of the radiation fields extrapolated over all space, as well as lower bounds on actual radiated energy flux.  As such, the MPVP may provide an alternative to techniques that involve numerical solutions for fields via Li{\'e}nard-Wiechart, Panofsky, Jefimenko, Feynman-Heaviside, or similar integral expressions (perhaps with additional approximations, like that popularized by Wang \cite{wang:93}), as well as to asymptotic series expansions for radiative fields, such as that of Wilcox.\cite{wilcox:56}  Although quite simple to state and interpret, the MPVP can serve as a practical approximation technique---at least in the important special case of paraxial radiation fields, where, for example, it has been successfully applied to an analysis of coherent x-ray generation via harmonic cascade by a radiating electron beam traveling through a sequence of undulators.

We hope that this variational principle and the associated formalism may find further application in the analysis of few-electron or even single-electron bunches in Fermilab's IOTA storage ring, in particular regarding questions of classical versus quantum behavior in emission of radiation, and of optical stochastic cooling.  We review and publicize our previous results with these larger goals in mind.

\subsection{Assumptions and Applicability}

The maximum-power variational principle developed here is applicable to classical, spontaneous electromagnetic radiation emitted from prescribed, localized current sources.  By classical, we mean that any quantum dynamical, quantum optical, or quantum statistical effects may be ignored.  In this context, by spontaneous emission from prescribed sources, we mean that the spacetime trajectories of the charged particles constituting the sources for the radiation are, in principle, to be considered prescribed functions of time, determined by initial conditions, external guiding fields (wigglers, bending magnets, quadrupoles, cavities, etc.) and possibly space-charge self-fields (either exact Coulomb fields or self-consistent mean-fields), but remain independent of the actual radiation fields emitted.  This means that the self-consistent effects of any radiation reaction or recoil, multiple scattering, energy gain or loss, or any other feedback of the radiation itself on the space-time trajectories of the source charges will be neglected.  Obviously, radiating particles must lose energy and possibly linear or angular momentum, but nevertheless, neglecting such back-action remains a good approximation in many situations, especially for relativistic particle beams, where kinematic effects tend to suppress observable changes in a particle's velocity, even if more substantial changes may occur to its energy or momentum.   Throughout, we also assume that the charge density $\rho(\bv{x},t)$ and current density $\bv{J}(\bv{x},t)$ are not only explicitly prescribed, but remain sufficiently localized in space so that the far-field can be meaningfully defined, and also remain at least weakly localized in time, so that Fourier transforms to or from the frequency domain can be performed and remain well-behaved. (See Appendix \ref{fourier} for  sign and scaling conventions).

For simplicity, we have thus far assumed that, apart from its generation by the prescribed sources in a bounded region, the emitted radiation otherwise propagates in vacuum.  Further generalizations to allow for non-uniform dielectric or permeability tensors, or perfectly conducting boundaries, representing wave-guides, lenses, windows, or other optical devices, might also be possible, but have not yet been studied in detail.  Possibilities for including effects of active, lossy, or non-reciprocal optical media, or of some dynamical recoil or bunching or other feedback effects in the emitting beams, are less clear.

\def\epsz{\epsilon_{0}}
\def\muz{\mu_{0}}
\def \greenop{\mathcal{G}}
\def\gadv{G_{\stext{adv}}}
\def\gret{G_{\stext{ret}}}
\def\gbar{\bar{G}}
\def\dop{\mathcal{D}}

\section{Mathematical and Physical Preliminaries}

The framework described here rests on the ability to decompose (at least mentally or mathematically) the electromagnetic fields from specified sources into irrotational (curl-free) field and solenoidal (divergence-free) fields, and then further decompose the latter into 
so-called reactive fields and radiation fields.  Only the radiation fields will contribute to any outgoing electromagnetic energy flux
detectable in the far field, and these are the fields which may be approximated by our variational principle.  We begin with an overview of formal solutions to Maxwell's equations with these distinctions in mind.

\subsection{Maxwell and Helmholtz Equations}

It will be convenient to work via Fourier transform within the (positive) frequency domain, and make use of Helmholtz-Hodge theorem and formally decompose vector fields and sources into their solenoidal (\ie, divergence-free, or functionally transverse) and irrotational (\ie, curl-free, or functionally longitudinal) contributions.  The microscopic Maxwell's equations can then be written (in \textit{SI} units) as
\bsub
\label{eqn:maxwell_1}
\begin{align}
\bv{E}_{\|} &= -\grad \phi =   -\tfrac{1}{\epsz} \tfrac{i}{\omega} \bv{J}_{\|},        \label{eqn:e_long}\\
\bv{B}_{\|} &= \phantom{+}\bv{0},\label{eqn:b_long}\\
\curl\bv{E}_{\perp} &= \phantom{+}i\omega \curl \bv{A}_{\perp} = i\omega \bv{B}_{\perp} = i\omega \bv{B} ,\label{eqn:e_perp}\\
\curl\bv{B}_{\perp} &= -\laplacian \bv{A}_{\perp} = \muz \bv{J}_{\perp} - i \muz \epsz \omega\bv{E}_{\perp} = \muz \bv{J}_{\perp} +  k^2 \bv{A}_{\perp}   ;\label{eqn:b_perp}
\end{align}
\esub
where $c = \tfrac{1}{\sqrt{\muz \epsz}}$ is the speed of light \textit{in vacuo}, the frequency $\omega$ and wavenumber are related by the free-space dispersion relation $\omega^2 = c^2 k^2$, and we have also introduced the usual \textit{Coulomb-gauge} scalar potential $\phi(\bv{x}, \omega)$ and transverse vector potential $\bv{A}_{\perp}(\bv{x}, \omega)$, from which the (frequency-domain) electromagnetic fields may be derived in the usual fashion:
\bsub
\begin{align}
\bv{B}(\bv{x}, \omega) &= \bv{B}_{\perp}(\bv{x}, \omega)  = \curl \bv{A}_{\perp}(\bv{x}, \omega),\\
\bv{E}(\bv{x}, \omega) &= \bv{E}_{\perp}(\bv{x}, \omega) + \bv{E}_{\|}(\bv{x}, \omega) = 
i\omega \bv{A}_{\perp}(\bv{x}, \omega) - \grad\phi(\bv{x}, \omega).
\end{align}
\esub
For convenience we choose the spatial origin to lie somewhere in the vicinity of the support of what are assumed to be spatially localized sources (at any non-zero frequency),
\be
\bv{J}(\bv{x}, \omega) = \bv{J}_{\perp}(\bv{x}, \omega) + \bv{J}_{\|}(\bv{x}, \omega).
\ee
Suitable limits to handle the possibility of infinitely extended sources can be considered at the end of our calculations.

It will be convenient to express the three-dimensional position $\bv{x} = r \unitvec{r}$ in either Cartesian coordinates $(x, y, z)$ or spherical coordinates $(r, \theta, \phi)$, with respect to the chosen origin and some suitable orientation of the axes.  Unless otherwise noted, we will assume that $\omega = c k > 0$.  Because Cartesian components of physical fields in the time domain are real-valued, negative-frequency components can always be inferred by relationships such as $\bv{E}_{\perp}(\bv{x}, -\omega) = \bv{E}_{\perp}(\bv{x}, \omega)\cc$ and $\bv{B}_{\perp}(\bv{x}, -\omega) = \bv{B}_{\perp}(\bv{x}, \omega)\cc$.  Zero-frequency (i.e., static) field components are not associated with radiation, and are not of direct interest here.

The irrotational electric field $\bv{E}_{\|}(\bv{x}, \omega)$ consists of the unretarded Coulomb fields, associated (back in the time domain) with the instantaneous positions of the charges, and as such contains no information about actual radiation.  Of primary interest is the transverse vector potential $\bv{A}_{\perp}(\bv{x}, \omega)$, which contains all the outgoing radiation fields (but also in general  some non-radiative fields, before the far-field is reached). 

In the frequency-domain, and in otherwise free space (apart from the prescribed sources), the Coulomb-gauge vector potential $\bv{A} _{\perp}(\bv{x}, \omega)$ will satisfy the inhomogeneous (i.e., sourced, or driven) vector Helmholtz equation:
\be\label{h1}
\left( \laplacian + k^2 \right) \bv{A} _{\perp}(\bv{x}; \omega) = -\muz \bv{J}_{\perp}(\bv{x}; \omega),
\ee
together with the transverse gauge constraint 
\be\label{gauge1}
\grad\!\cdot\! \bv{A}_{\perp}(\bv{x}; \omega) = 0.
\ee
In the absence of sources, the transverse vector potential would satisfy a homogeneous (i.e., source-free) Helmholtz equation,
\be\label{h0}
\left( \laplacian + k^2 \right) \bv{A} _{\perp}(\bv{x}; \omega) = \bv{0},
\ee
as well as the gauge constraint \eqref{gauge1}.  For any $k \neq 0$, these governing equations can be combined to assert that any transverse, source-free solution $\bv{A}_{\perp}(\bv{x}; \omega)$ must be an eigenfunction of the double-curl operator,
\be
\grad\!\times\! \grad \!\times\! \bv{A}_{\perp}(\bv{x}; \omega) = k^2\, \bv{A}_{\perp}(\bv{x}; \omega).
\ee

For any well-behaved current sources, there will be a unique solution to \eqref{h1} satisfying \textit{outgoing Sommerfeld boundary conditions}, such that 
\be
\lim\limits_{r \to \infty} r \left( \tfrac{\del}{\del r} - ik\right)\bv{A}_{\perp}(r,\theta, \phi; \omega) = \bv{0}
\ee
uniformly in $\theta$ and $\phi$ (still assuming that $k > 0$), ensuring that the currents act as sources for outgoing radiation rather than as sinks for ingoing radiation.

\subsection{Green Functions, Sources, and Fields}

Formally, solutions to the inhomogeneous Helmholtz equation can be expressed in terms of a scalar Green function, representing the response to an impulsive source.  In free space, the causal, or retarded, Green function, given by
\be\label{green_function_1}
\gret(\bv{x}, \bv{x}'; \omega) = \frac{ e^{+ik \abs{ \bv{x} - \bv{x'}}}  } { 4\pi \abs{ \bv{x} - \bv{x'} } }
\ee 
(still assuming $\omega = c k > 0$), will satisfy the impulsively sourced Helmholtz equation
\be\label{impulsiveHelmholtz}
\left( \laplacian + k^2 \right) G(\bv{x}; \bv{x}'; \omega)  = - \delta(\bv{x} - \bv{x}'),
\ee
along with an outgoing Sommerfeld radiative boundary condition
\be
\lim\limits_{r \to \infty} r \left( \tfrac{\del}{\del r} - ik \right)\gret(\bv{x}, \bv{x}'; \omega) = \bv{0},
\ee
with respect to observation position $\bv{x}$, for any fixed source position $\bv{x}'$, (continuing to assume that $k > 0$). Notice that we are here using a sign convention of Jackson, for which the Green function corresponds to the response to a negative impulse.

As a function of the distance $\abs{ \bv{x} - \bv{x'} }$, this Green function is symmetric under interchange of $\bv{x}$ and $\bv{x}'$, and is also translationally and rotationally invariant under common transformations to both $\bv{x}$ and $\bv{x}'$.

As a consequence of linearity, the full Coulomb-gauge vector potential can then be expressed in terms of a convolution integral over a solenoidal source:
\be\label{green_sol_ret}
\begin{split}
\bv{A}_{\perp}(\bv{x}; \omega) &= \bv{A}_{\stext{in}}(\bv{x}; \omega) + \bv{A}_{\stext{ret}}(\bv{x}; \omega) \\
&=  \bv{A}_{\stext{in}}(\bv{x}; \omega) + \muz \! \int \!d^3\bv{x}'\, \gret(\bv{x}; \bv{x}'; \omega)\, \bv{J}_{\perp}(\bv{x}'; \omega).
\end{split}
\ee
Back in the time domain, use of $\gret(\bv{x}; \bv{x}'; \omega)$ ensures that the response at a given spacetime point will depend on the sources only on the past light-cone.  In addition, we can add in an ``incident'' or ``incipient'' or ``input'' contribution $\bv{A}_{\stext{in}}(\bv{x}; \omega)$ which is a solenoidal solution to the homogeneous Helmholtz equation, and incorporates initial conditions, effectively determining the fields in the remote past before the sources in question turned on, then evolved forward in time via the \textit{source-free} Maxwell equations.  (Often, these are referred to as ``ingoing'' fields, but that can be misleading, as they may not be exclusively directed toward the sources).

Using the symmetry of the Green function under interchange of $\bv{x}$ and $\bv{x}'$, and some integration by parts, it is straightforward to verify that $\bv{A}_{\perp}(\bv{x}; \omega)$ will indeed be divergence-free whenever $\bv{J}_{\perp}(\bv{x}; \omega)$ is.  Explicit use of the Coulomb gauge constraint and the solenoidal current density as source thereby offers the convenience of working with a scalar rather than dyadic Green function, and with just a vector potential rather than both vector and scalar potentials.  Manifest Lorentz covariance is lost, but seems a worthwhile price to pay here.

As a consequence of this transversality as well as the boundary conditions satisfied by the causal Green function, the vector potential $\bv{A}_{\stext{ret}}(\bv{x}; \omega)$ itself will asymptotically satisfy outgoing Sommerfeld boundary conditions, 
\be
\lim\limits_{r \to \infty} r \left( \tfrac{\del}{\del r} - ik \right)\bv{A}_{\stext{ret}}({r} \unitvec{r}; \omega)  = \bv{0},
\ee
as well as the slightly stronger \textit{Silver-M{\"u}ller} boundary conditions,
\be
\lim\limits_{r \to \infty} r \left( -\unitvec{r} \!\times\! \grad \!\times \,-\; ik \right) \bv{A}_{\stext{ret}}({r} \unitvec{r}; \omega)  = \bv{0},
\ee
uniformly in angular direction.

Of course, the Green function \eqref{green_function_1} is not the only solution to the impulsively-sourced Helmholtz equation \eqref{impulsiveHelmholtz}.  We can add any scalar solutions to the homogeneous (source-free) Helmholtz equation to obtain other Green functions with different intrinsic boundary conditions.  In particular, the time-reversed, or advanced Green function,
\be
\gadv(\bv{x}, \bv{x}'; \omega) = \gret(\bv{x}, \bv{x}'; \omega)\cc = \frac{e^{-ik \abs{\bv{x} - \bv{x'} }} } {4\pi  \abs{\bv{x} - \bv{x'}}},
\ee 
also satisfies equation \eqref{impulsiveHelmholtz}, but with \textit{ingoing} Sommerfeld radiation boundary conditions, such that,
\be
\lim\limits_{r \to \infty} r \left( \tfrac{\del}{\del r} + ik \right)\gadv(\bv{x}, \bv{x}'; \omega) = \bv{0}
\ee
(still assuming $k > 0$).  In effect, it  generates radiation fields converging from infinity, which are absorbed by the currents now acting as sinks.  

In terms of this advanced Green function, the same Coulomb-gauge vector potential $\bv{A}_{\perp}(\bv{x}; \omega)$ can also be written as
\be\label{green_sol_adv}
\begin{split}
\bv{A}_{\perp}(\bv{x}; \omega) &= \bv{A}_{\stext{out}}(\bv{x}; \omega) + \bv{A}_{\stext{adv}}(\bv{x}; \omega) \\
&= \bv{A}_{\stext{out}}(\bv{x}; \omega) + \muz \! \int \!d^3\bv{x}'\, \gadv(\bv{x}; \bv{x}'; \omega)\, \bv{J}_{\perp}(\bv{x}'; \omega),
\end{split}
\ee
where $\bv{A}_{\stext{out}}(\bv{x}; \omega)$ is a source-free solution intended to capture the ``output'' or ``outcome'' or ``outstanding'' fields at times far in the future, after the sources have turned off, then propagated back in time via the source-free Maxwell equations.  (Often, these are called ``outgoing'' fields, but that is a bit of a misnomer, as they may not be directed exclusively away from our sources).  Back in the time domain, use of $\gadv$ means that $\bv{A}_{\stext{adv}}$ will depend on sources behavior on the future light cone, but this is so it can properly subtract out the correct contributions from $\bv{A}_{\stext{out}}$, which after all, represented a final condition extrapolated backward in time.  That is, in the time domain, convolution of the retarded Green function with the current sources would tell us what is to be added to any source-free fields initially present (as $t \to -\infty$), whereas convolution with the advanced Green function would tell us what to remove from the free fields present far in the future (as $t \to +\infty$).

Notice that the causal Green function can also be re-written as
\be
\gret(\bv{x}, \bv{x}'; \omega) = \dop(\bv{x}, \bv{x}'; \omega) + \gbar(\bv{x}, \bv{x}'; \omega),
\ee
where
\be
\gbar(\bv{x}, \bv{x}'; \omega) = \tfrac{1}{2} \bigl[ \, \gret(\bv{x}, \bv{x}'; \omega)  + \gadv(\bv{x}, \bv{x}'; \omega) \, \bigr] =  \frac{ \cos k\abs{\bv{x} - \bv{x'} }} {4\pi  \abs{\bv{x} - \bv{x'}}}
\ee
is the time-symmetric, principal-value, half-advanced/half-retarded, or Wheeler-Feynman Green function, which like the other Green functions  satisfies equation \eqref{impulsiveHelmholtz}, and  
\be
\dop(\bv{x}, \bv{x}'; \omega) = \tfrac{1}{2} \bigl[ \, \gret(\bv{x}, \bv{x}'; \omega)  - \gadv(\bv{x}, \bv{x}'; \omega) \, \bigr] =  \frac{i \,\sin k\abs{\bv{x} - \bv{x'}}} {4\pi  \abs{\bv{x} - \bv{x'}}}
\ee
is half the difference between the retarded and advanced Green functions, thereby canceling their singularities, so as to  satisfy the source-free Helmholtz equation
\be\label{HelmholtzD}
\left( \laplacian + k^2 \right) \dop(\bv{x}; \bv{x}'; \omega) = 0
\ee
everywhere, including precisely at $\bv{x} = \bv{x}'$.

The function $\dop(\bv{x}, \bv{x}'; \omega)$  can generate source-free solutions in the same way---namely, via convolution over some real or effective current density---as the Green functions determine advanced or retarded solutions from actual sources.  For lack of any better terminology, we will call $\dop$ the \textit{radiation kernel}, for reasons that will hopefully become clear momentarily.

Notice that, like any of the other Green functions, the time-symmetric Green function $\gbar(\bv{x}, \bv{x}'; \omega)$ diverges as $\bv{x} \to \bv{x}'$, in order to account for the Dirac delta function source term, whereas the radiation kernel $\dop(\bv{x}, \bv{x}; \omega)$ is bounded everywhere, and in particular, $\abs{ \dop(\bv{x}, \bv{x}'; \omega) } \le \abs{ \dop(\bv{x}, \bv{x}; \omega) } =  \tfrac{k}{4\pi}$.  As functions of the distance $\abs{\bv{x} - \bv{x}'}$ between source and observation position, both $\gbar$ and $\dop$ are symmetric under interchange of $\bv{x}$ and $\bv{x}'$, and also invariant under their common rotation or translation, but $\gbar$ is symmetric under $k \to -k$, while $\dop$ is antisymmetric under $k \to -k$. Similarly, $\gbar$ is real-valued (symmetric under complex conjugation), while $\dop$ is imaginary (antisymmetric under complex conjugation).

\subsection{Radiation}

If we ask ourselves what exactly characterizes radiation fields, the following familiar properties come to mind:
\begin{enumerate}
\item radiation consists of fields that have been ?shaken loose? from the emitting charges and take on  an independent dynamical existence, so 
should solve the source-free Maxwell equations everywhere, including on the actual worldlines of sources;
\item radiation fields can (irreversibly) transport energy, linear and angular momentum, and information ``to infinity;''
\item they depend on the acceleration of source charges, not just velocities and positions;
\item they can be expressed as superpositions of \textit{null} fields, meaning superpositions of fields for which the Lorentz invariants $\bv{E}(\bv{x},t) \!\cdot\! \bv{B}(\bv{x},t)$ and $\abs{\bv{E}(\bv{x},t)}^2 - c^2 \abs{\bv{B}(\bv{x},t)}^2$ vanish everywhere;
\item the radiation fields account for all radiated power as revealed by Larmor-Li{\'e}nard formula,
\item and account for finite radiation reaction forces;
\item in the asymptotic far field:  the amplitude of radiation fields  emitted from one source charge exhibits $O(\sfrac{1}{r})$ fall-off in distance between observation and emission points,
\item and electric and magnetic fields will be perpendicular to each other and to line of sight between the point of emission  and observation,
\item and satisfy outgoing Sommerfeld or Silver-M{\"u}ller radiation conditions.
\end{enumerate}

Following Dirac and other authors, we will define the \textit{radiation fields} associated with the source $\bv{J}(\bv{x}; \omega)$ in terms of the difference between the outgoing and ingoing fields at each frequency, or equivalently, in terms of the fields associated with the difference between the retarded and advanced Coulomb-gauge vector potentials:
\be
\begin{split}
\bv{A}_{\stext{rad}}(\bv{x}; \omega) &= \bv{A}_{\stext{out}}(\bv{x}; \omega) - \bv{A}_{\stext{in}}(\bv{x}; \omega) \\
&=  \bv{A}_{\stext{ret}}(\bv{x}; \omega) - \bv{A}_{\stext{adv}}(\bv{x}; \omega) \\
&= 2 \,\muz \! \int \!d^3\bv{x}'\, \dop(\bv{x}; \bv{x}'; \omega)\, \bv{J}_{\perp}(\bv{x}'; \omega).
\end{split}
\ee

Radiation fields in the sense of Dirac will satisfy all of the characteristic properties mentioned above except for the very last, as clearly they include both ingoing and outgoing (or retarded and advanced) field components, about which we will have more to say.
Although generated via convolution between the radiation kernel and the actual sources, $\bv{A}_{\stext{rad}}(\bv{x}; \omega)$ is a solenoidal solution to the homogeneous Helmholtz equation everywhere in space, and the corresponding electromagnetic fields will be solutions to the source-free Maxwell equations everywhere---not just in the far-field, and not just at spacetime points away from the actual sources. 
% It makes sense to think of these fields as radiation, since electromagnetic radiation should consist of fields which are ``shaken loose'' from accelerating charges, and take on an independent dynamical existence.
Furthermore, if non-vanishing, $\bv{E}_{\stext{rad}}(\bv{x}; \omega)$ and  $\bv{B}_{\stext{rad}}(\bv{x}; \omega)$ will have components which fall off in amplitude inversely with distance from the actual physical sources, which is another characteristic of radiative fields distinguishing them from the near-zone or intermediate-zone fields associated with $\gbar$, which fall off inversely with the square of the distance, or faster.  As we will see, the corresponding fields can always be written as superpositions of transverse plane waves satisfying the vacuum dispersion relation, which are null fields, and also account properly for all of the Poyning flux in the far field.

But while the causal, or retarded, part of $\bv{A}_{\stext{rad}}$ coincides exactly with the vector potential $\bv{A}_{ret}$ produced by the sources, $\bv{A}_{\stext{rad}}$ also contains an equal amount of incoming power in an advanced component $\bv{A}_{\stext{adv}}$, which is needed to cancel any singularities and ensure that $\bv{A}_{\stext{ret}}$ satisfies the homogeneous Helmholtz equation everywhere, including right on top of source charges, where the near-fields produced by the other part of $\gret$, namely $\gbar$, would diverge.  So $\bv{A}_{\stext{rad}}(\bv{x},\omega)$ is in effect a homogeneous or source-free extrapolant of the actual outgoing solution $\bv{A}_{\perp}(\bv{x}; \omega)$, propagated from the far field back throughout the rest of space and time, including into the near-zones of the actual sources.

For example, If the retarded fields look like plane waves propagating downstream from the sources, then the radiation fields will also look like plane waves upstream from the sources, as if we ignored the sources and just extrapolated these plane eaves everywhere in space and time.  If the retarded fields from localized sources look like expanding spherical waves at late times, then the radiation fields will also include converging spherical waves at earlier times.

%%%%%

The non-radiative fields, generated by convolution of the solenoidal source $\bv{J}_{\perp}$ with the time-symmetric Green function $\gbar$, might be called \textit{bound} fields, as they remain partially bound to the sources and cannot radiate away, or as near or intermediate-zone fields, because they fall off inversely with the square of the distance from the sources, or even more rapidly.  Or perhaps better, they might be described as \textit{reactive} fields, since they can temporarily and reversibly exchange energy with nearby sources or other non-radiative fields but cannot transport energy to infinity.

This decomposition of the solenoidal fields into radiative and reactive contributions is employed in the Dirac-Lorentz approach to electrodynamics. The radiation fields are regular everywhere, including on the wordline of any source charge, and are responsible for a finite radiation reaction force, while the reactive fields diverge on the worldline of any source, which can be interpreted as an infinite mass renormalization.

Their very different nature and different roles in energy conservation arise from different behavior of the time-symmetric Green function $\gbar$ and the radiation kernel $\dop$ under time reversal.  As a consequence, it turns out that overall (that is, integrated over all space and either time or frequency), the transverse reactive fields cannot irreversibly exchange any net energy with the sources (in the absence of other, resistive media), but are associated with reversible borrowing or recovery of  energy from sources or other non-radiative fields in their vicinity. In contrast, the radiation fields do involve an irreversible energy transport.

%%%%

Not all current sources will necessarily emit electromagnetic radiation, and because of the possibility of such non-radiating sources and non-radiative fields, the ingoing and outgoing contributions to $\bv{A}_{\stext{rad}}(\bv{x}; \omega)$ cannot necessarily be unambiguously separated everywhere in space,  just from knowing $\bv{A}_{\stext{rad}}(\bv{x}; \omega)$ itself.  But we can project out the purely outgoing part of $\bv{A}_{\stext{rad}}$ in the asymptotic far-field, by using the Sommerfeld radiation condition, that is:
\be
\bv{E}_{\stext{ret}} \sim \tfrac{c}{2}(ik + \tfrac{\del}{\del r}) \bv{A}_{\stext{rad}} \text{ asymptotically as } kr \to \infty.
\ee
On the other hand, knowledge of just  $\bv{A}_{\stext{ret}}(\bv{x}; \omega)$  in just the far-field does in principle determine $\bv{A}_{\stext{rad}}(\bv{x}; \omega)$ uniquely.  We will derive an explicit construction below, after introducing the useful notion of the angular radiation pattern.

However, it is important to remember that, if examined over all space and time, the radiative extrapolation $\bv{E}_{\stext{rad}}(\bv{x}; \omega)$ will contain incoming fields never actually present in the physical system, as it replaces the physical fields in the near-zone associated with the sources by fields that in effect were emitted by arbitrarily distant sources at arbitrarily remote times in the past.  

In some configurations, we may be able to separate out the retarded and advanced contributions easily, by looking in different parts of spacetime.   For instance, in a well-collimated beam of radiation, the advanced fields from $\bv{A}_{\stext{adv}}(\bv{x}; \omega)$ may be predominately found upstream from the sources, while the fields from $\bv{A}_{\stext{ret}}(\bv{x}; \omega)$ may be predominately downstream.  Or in multipolar radiation from a well-localized source, back in the time domain, $\bv{A}_{\stext{adv}}(\bv{x}; \omega)$ may be converging towards the sources at early times, while $\bv{A}_{\stext{ret}}(\bv{x}; \omega)$ may be propagating out from the sources at later times.  Or we may be interested only in the asymptotic far-field, where we can extract the outgoing fields in the far-field region using $\bv{A}_{\stext{ret}}(\bv{x}; \omega)$ and what we know about the Sommerfeld boundary conditions.  Otherwise, perhaps just having an approximation to $\bv{A}_{\stext{rad}}(\bv{x}; \omega)$ itself may be sufficient, or even desirable, for instance because it can be directly imaged through lenses or other optical systems.

Also note that, while outgoing radiation-zone fields produced by prescribed sources will in principle be uniquely determined by those current sources, the far-fields cannot uniquely determine the sources, because of the possibility of non-radiating sources (at some or all frequencies), which only produce fields with $O(\sfrac{1}{r^2})$ or faster fall-off in amplitude.  Such non-radiating sources constitute the linear-algebraic nullspace of $\dop(\bv{x}, \bv{x}'; \omega)$, regarded as a linear operator mapping sources to radiative vector potentials.  Because the Sommerfeld boundary conditions differentiate fields only in the asymptotic radiation zone, both $\gret$ and $\gadv$ can produce the same non-radiative (i.e., near-zone and intermediate-zone) fields from a given source $\bv{J}_{\perp}$, contributions which will then cancel from their difference.

So as to focus on the fields actually radiated outward to infinity (in the absence of further absorbers) by the given sources $\bv{J}_{\perp}(\bv{x}; \omega)$, hereafter we shall assume that $\bv{A}_{\stext{in}}(\bv{x}; \omega) = \bv{0}$, so that $\bv{A}_{\perp}(\bv{x}; \omega) = \bv{A}_{\stext{ret}}(\bv{x}; \omega)$, and $\bv{A}_{\stext{rad}}(\bv{x}; \omega)  = \bv{A}_{\stext{ret}}(\bv{x}; \omega)  - \bv{A}_{\stext{adv}}(\bv{x}; \omega) = \bv{A}_{\stext{out}}(\bv{x}; \omega)$.  The corresponding fields associated with this vector potential then include only transverse fields generated directly by the source $\bv{J}_{\perp}(\bv{x}; \omega)$, and exclude any external or background fields, including any fields needed to induce or maintain the currents, or any incident radiation from remote sources.

Although the vector field $\bv{A}_{\stext{rad}}(\bv{x}; \omega)$ satisfies a source-free Helmholtz equation, it is still associated with the source $\bv{J}_{\perp}(\bv{x}; \omega)$, and is generated by the latter via convolution with the radiation kernel.  So we often speak of $\bv{J}_{\perp}(\bv{x}; \omega)$ as the generator of the $\bv{A}_{\stext{rad}}(\bv{x}; \omega)$, or even as the ``source'' of what are technically source-free fields.

In passing, we also note that if determining the radiation fields from the sources, it may actually be more efficient to start with the radiative magnetic field rather than the radiative vector potential or radiative electric field, because this avoids having to first calculate the solenoidal part of the current density.   With some suitable integration by parts, we find
\be
\begin{split}
\bv{B}_{\stext{rad}}(\bv{x}; \omega) &=  \grad \!\times\!  2 \,\muz \! \int \!d^3\bv{x}'\, \dop(\bv{x}; \bv{x}'; \omega)\, \bv{J}_{\perp}(\bv{x}'; \omega) \\
&=  2 \,\muz \! \int \!d^3\bv{x}'\,  \grad \!\times\! \bigl[ \dop(\bv{x}; \bv{x}'; \omega)\, \bv{J}_{\perp}(\bv{x}'; \omega) ] =  -2 \,\muz \! \int \!d^3\bv{x}'\,  \grad' \dop(\bv{x}; \bv{x}'; \omega)  \times  \bv{J}_{\perp}(\bv{x}'; \omega) \\
&= -2 \,\muz \! \int \!d^3\bv{x}'\,  \grad' \!\times\! \bigl[ \dop(\bv{x}; \bv{x}'; \omega) \, \bv{J}_{\perp}(\bv{x}'; \omega) \bigr] + 2 \,\muz \! \int \!d^3\bv{x}'\,  \dop(\bv{x}; \bv{x}'; \omega) \, \grad' \!\times\! \bv{J}_{\perp}(\bv{x}'; \omega) \\
&= + 2 \,\muz \! \int \!d^3\bv{x}'\,  \dop(\bv{x}; \bv{x}'; \omega) \, \grad' \!\times\! \bv{J}_{}(\bv{x}'; \omega),
\end{split}
\ee
where the boundary terms resulting from the integral of the total curl will vanish, because the radiation kernel falls of inversely with distance, and the solenoidal current density falls off at least with with the inverse square of distance (and ypically with the inverse cube of distance for $\omega \neq 0$, particularly for the relevant non-radial field components).

\subsection{Surjectivity of the Radiation Kernel}

We have mentioned that not every current density leads to non-zero radiation fields.  It follows that the radiation fields cannot \textit{uniquely} determine a current density that will reproduce them, because any current densities which differ by some non-radiating source would generate the same radiation fields.

But apart from this non-uniqueness, an essential existence question remains, as to whether we can always generate any radiation fields via a convolution between the radiation kernel and \textit{some} effective source.  That is to say, we know that for any divergence-free current density $\bv{J}_{\perp}(\bv{x}; \omega)$ for which the required integrals exist, the convolution $\bv{A}_{\stext{rad}}(\bv{x}; \omega) = 2 \,\muz \! \int \!d^3\bv{x}'\, \dop(\bv{x}; \bv{x}'; \omega)\, \bv{J}_{\perp}(\bv{x}'; \omega)$ represents a solution to the source-free Maxwell equations.  But given any source-free solution $\bv{A}_{0}(\bv{x}; \omega)$, is it possible to find some (solenoidal) vector field $\bv{s}_{\perp}(\bv{x}; \omega)$ such that $\bv{A}_{0}(\bv{x}; \omega) = 2 \,\muz \! \int \!d^3\bv{x}'\, \dop(\bv{x}; \bv{x}'; \omega)\, \bv{s}_{\perp}(\bv{x}'; \omega)$?

Intuition suggests that every possible radiation field should be associated with some possible source.  It turns out that the radiation kernel is surjective with respect to almost the entire space of % essentially any
source-free solutions to Maxwell's equations, provided $\omega \neq 0$ (so as to exclude static fields, which are non-radiative).  We already know that any such free-space solutions can be formally written as superpositions of transverse, harmonic plane waves, so if we can generate arbitrary plane waves via convolution of $\dop(\bv{x},\bv{x}; \omega)$ with a suitable effective source $\bv{s}_{\perp}(\bv{x}'; \omega)$, then we can reproduce any free-space solution by suitable Fourier superposition.  (We make no claims about the physical realization of the source, in terms of moving point charges).

Suppose we want to generate a radiation vector potential of the form 
\be
\bv{A}_0(\bv{x}; \omega) = \bv{A}_0(\bv{x}; -\omega)\cc  = \int \! d^3\bv{\kappa} \; a(\bv{\kappa}) \, \delta(\omega - c\abs{\bv{\kappa}}) \, \unitvec{\epsilon}(\bv{\kappa}) \, e^{i \bv{\kappa} \cdot \bv{x} },
\ee
valid for all $\omega > 0$, where we assume the polarization satisfies $\unitvec{\epsilon}(\bv{\kappa})\cc \!\cdot \unitvec{\epsilon}(\bv{\kappa}) = 1$, $\unitvec{\kappa} \!\cdot  \unitvec{\epsilon}(\bv{\kappa}) = 0$, and $\unitvec{\epsilon}(-\bv{\kappa}) = \unitvec{\epsilon}(\bv{\kappa})\cc$, while the amplitude satisfies $\lim\limits_{\bv{\kappa} \to \bv{0}} a(\bv{\kappa}) = {0}$.  Consider a solenoidal vector field of similar form, namely
\be
\bv{s}_{\perp}(\bv{x}; \omega) =  \int \! d^3\bv{\kappa} \;   \sigma(\bv{\kappa}; \omega)  \, \unitvec{\epsilon}(\bv{\kappa}) \, e^{i \bv{\kappa} \cdot \bv{x} }.
\ee
Then the corresponding radiation associated with this source current density is
\be
\begin{split}
\bv{A}_{\stext{rad}}(\bv{x}; \omega) &= 2 \muz \! \int \!d^3\bv{x}'\, \dop(\bv{x}; \bv{x}'; \omega)\, \bv{s}_{\perp}(\bv{x}'; \omega)  \\
&= 2 \muz \! \int \!d^3\bv{x}'\, \tfrac{i \,\sin  \frac{\omega}{c} \abs{\bv{x} - \bv{x'}}} {4\pi \abs{\bv{x} - \bv{x'}}} \, \int \! d^3\bv{\kappa} \;   \sigma(\bv{\kappa}; \omega) \, \unitvec{\epsilon}(\bv{\kappa}) \, e^{i \bv{\kappa} \cdot \bv{x}' } \\
&= \int \! d^3\bv{\kappa} \,  \tfrac{i\muz}{2\pi} \, k\, \sigma(\bv{\kappa}; \omega)  \, \unitvec{\epsilon}(\bv{\kappa}) \, e^{i \bv{\kappa} \cdot \bv{x} }   \!  \int \!d^3\bv{x}' \,  \tfrac{\sin k \abs{\bv{x}' - \bv{x}}} {k \abs{\bv{x}' - \bv{x}}} e^{i \bv{\kappa} \cdot (\bv{x}' - \bv{x}) },
\end{split}
\ee
where we have made the substitution $\omega = c k$.  After defining $\bv{\xi} = (\bv{x}'-\bv{x})$, and introducing $\eta$ as the angle between $\bv{\kappa}$ and $\bv{\xi}$, we can re-write the radiation as
\be
\begin{split}
\bv{A}_{\stext{rad}}(\bv{x}; \omega)
&= \int \! d^3\bv{\kappa} \,  \tfrac{i\muz}{2\pi} \, k\, \sigma(\bv{\kappa}; \omega)  \, \unitvec{\epsilon}(\bv{\kappa}) \, e^{i \bv{\kappa} \cdot \bv{x} }   \!  \int \!d^3\bv{\xi} \,  \tfrac{\sin k \abs{\bv{\xi} }} {k \abs{\bv{\xi}'}} e^{i \bv{\kappa} \cdot \bv{\xi} } \\
&= \int \! d^3\bv{\kappa} \,  \tfrac{i\muz}{2\pi} \, k\, \sigma(\bv{\kappa}; \omega)  \, \unitvec{\epsilon}(\bv{\kappa}) \, e^{i \bv{\kappa} \cdot \bv{x} } \, 2\pi \! \int\limits_{0}^{\infty} \! \xi^2 \, d\xi \, \tfrac{\sin (k \xi)} { k\xi } \! \int\limits_{-1}^{+1} \! d(\cos\eta) \; e^{i \kappa \xi \,\cos \eta} \\
&= \int \! d^3\bv{\kappa} \,  \tfrac{i\muz}{2\pi} \, k\, \sigma(\bv{\kappa}; \omega)  \, \unitvec{\epsilon}(\bv{\kappa}) \, e^{i \bv{\kappa} \cdot \bv{x} } \, 4\pi \! \int\limits_{0}^{\infty} \! \xi^2 \, d\xi \, \tfrac{\sin (k \xi)} { k\xi } \, \tfrac{\sin (\kappa \xi)} { \kappa \xi } \\
&= \int \! d^3\bv{\kappa} \,  2 i \muz  \, k\, \sigma(\bv{\kappa}; \omega)  \, \unitvec{\epsilon}(\bv{\kappa}) \, e^{i \bv{\kappa} \cdot \bv{x} }  \! \int\limits_{0}^{\infty} \! \xi^2 \, d\xi \, j_0(k \xi)\, j_0(\kappa \xi),
\end{split}
\ee
where $j_0(x) = \frac{\sin(x)}{x}$ is the $0$th-order spherical Bessel function (or ``sinc'' function).  Using the closure relation
\be
\int\limits_{0}^{\infty} \!  d\xi \, \xi^2  \, j_{\ell} (k \xi) \, j_{\ell}(\kappa \xi) =  \tfrac{\pi \kappa^{\ell}}{2 k^{\ell+2}}  \, \delta( k - \kappa) = \tfrac{\pi}{2 k^{2}}  \, \delta( k - \kappa) 
\ee
for spherical Bessel functions, we find that
\be
\bv{A}_{\stext{rad}}(\bv{x}; \omega) = 
\int \! d^3\bv{\kappa} \,  2 i \muz  \, k\, \sigma(\bv{\kappa}; \omega)  \, \unitvec{\epsilon}(\bv{\kappa}) \, e^{i \bv{\kappa} \cdot \bv{x} } \, \tfrac{\pi}{2 k^{2}}  \, \delta( k - \kappa),
\ee
which, remembering that $\omega = ck$, can be further re-arranged to obtain
\be
\bv{A}_{\stext{rad}}(\bv{x}; \omega) = 
\int \! d^3\bv{\kappa} \;   i \pi \muz \, c^2 \, \tfrac{1}{\omega} \sigma(\bv{\kappa}; \omega) \,  \delta( \omega - c\abs{\bv{\kappa}} )  \, \unitvec{\epsilon}(\bv{\kappa}) \, e^{i \bv{\kappa} \cdot \bv{x} } .
\ee
So if we choose
\be 
\sigma(\bv{k}; \omega) = \tfrac{-1}{\pi \muz c^2 } \, i \omega\,  a(\bv{k}),
\ee
then this source will reproduce the desired radiation fields. Because of the delta function, we are actually free to add in any additional source contributions that lack ``on-shell'' content---that is, for which the frequencies and wavevectors in Fourier space do not satisfy the free-space electromagnetic dispersion relation $\omega = c \abs{\bv{k}}$.
 
\subsection{Poynting Theorems and Power Relations}

Next we will introduce two Hilbert-space inner products, the first a volumetric, ``Joule'' inner product, that can be used to determine the work done on or by the sources by electric fields, and the second a far-field surface or ``Poynting'' inner product, that can be used to determine the outgoing flux of electromagnetic energy in the fair field, and then relate these by energy conservation constraints that follow directly from Maxwell's equations and suitable boundary conditions.

\subsubsection{Inner and Other Products}

It will be convenient to introduce a ``quantum-mechanical''-like notation for two different sesquilinear products over pairs of vector fields, whereby
\be\label{ip1}
\innerpd{\bv{f}}{\bv{g}} \equiv \!\int\limits_{\realsymbol^3} \! d^{3}\bv{x}\, \bv{f}(\bv{x})\cc \cdot \bv{g}(\bv{x})
\ee
is a ``volumetric'' inner product for $3$-dimensional, complex vector fields (in which any parametric dependence on frequency or time is left implicit for the moment), and
\be\label{sp2}
\innerpr{ \bv{f} }{ \bv{g} } = \lim_{R \to \infty}  \!\! \int\limits_{r = R} \!\! r^2 \,d^2\Omega(\unitvec{r}) \,\, \unitvec{r} \cdot \left[\, \bv{f}( r \unitvec{r} )\cc \times \bv{g}( r \unitvec{r} ) \, \right]
\ee
is a far-field ``surface'' product, in which $d^2\Omega(\unitvec{r})$ is the differential element of solid angle.  The former integral \eqref{ip1} defines the standard $\mathcal{L}^2$ functional inner product on square-integrable vector fields, but the latter integral \eqref{sp2} is not complex-symmetric (but rather antisymmetric), and is therefore not positive-definite, since $\innerpr{ \bv{g} }{ \bv{f} } = -\innerpr{ \bv{f} }{ \bv{g} }\cc$, and $\innerpr{ \bv{f} }{ \bv{f} } = 0$ even when $\bv{f}(\bv{x}) \neq \bv{0}$.

However, as we will see, this surface integral will be nonnegative where it matters most, namely when applied to the Poynting flux associated with outgoing electric field and corresponding magnetic field from a radiating source.  As $kr \to \infty$, we may assume that the outgoing radiation fields, with amplitude falling off like $O(\sfrac{1}{r})$, increasingly dominate, such that the \textit{asymptotic} far-fields approach
\bsub
\begin{align}
\bv{E}_{\stext{ret}}(r \unitvec{r}; \omega) &= i\omega \bv{A}_{\stext{ret}}(r \unitvec{r}; \omega)  \;\;\; \sim  i \omega \frac{\muz}{4\pi}  \frac{e^{+ikr}}{r} \, \bv{N}_{\perp}(\unitvec{r}; \omega), \text{ and } \\
\bv{B}_{\stext{ret}}(r \unitvec{r}; \omega) &= \grad\! \times\!  \bv{A}_{\stext{ret}}(r \unitvec{r}; \omega) \sim +ik \,\unitvec{r}\! \times\!  \bv{A}_{\stext{ret}}(r \unitvec{r}; \omega)  \sim  \unitvec{r} \times \tfrac{1}{c} \bv{E}_{\stext{ret}}(r \unitvec{r}; \omega),
\end{align}
\esub
where $\bv{N}_{\perp}(\unitvec{r})$ is the transverse part of a so-called \textit{radiation vector}, or far-field angular \textit{radiation pattern}:
\be\label{eqn:rad_vector}
\bv{N}_{\perp}(\unitvec{r}; \omega ) =  \int\! d^3\bv{x}' \, \bv{J}_{\perp}(\bv{x}'; \omega) \,e^{-ik\unitvec{r}\cdot \bv{x}'} =
\left(1 - \unitvec{r}\unitvec{r}\trans \right)\!\int\! d^3\bv{x}' \, \bv{J}(\bv{x}'; \omega) \,e^{-ik\unitvec{r} \cdot \bv{x}'},
\ee
in which we have made use of the equivalence of functional transversality (divergencelessness) and geometric transversality in reciprocal (wavevector) space, as well as in the far-field limit---this avoids some pesky regularization issues in the integral over $\bv{J}_{\perp}(\bv{x}'; \omega)$, which can have slowly decaying ``tails''  (generically falling off like $O(\sfrac{1}{r^{3}})$ as $kr \to \infty$) even if $\bv{J}(\bv{x}; \omega)$ itself is strongly localized in space.

Using these far-field expressions, we find (at any specified frequency $\omega$) that
\be\label{eqn:power_r0}
\innerpr{\bv{E}_{\stext{ret}} }{\bv{B}_{\stext{ret}} } =  \tfrac{\muz^2}{16\pi^2  c} \,\omega^2 \! \int \! d^2\Omega(\unitvec{r}) \, \bv{N}_{\perp}(\unitvec{r}; \omega)\cc\!\cdot\!\bv{N}_{\perp}(\unitvec{r}; \omega) \; \ge \, 0,
\ee
which is nonnegative definite for electromagnetic far-fields satisfying outgoing Sommerfeld conditions, but is not strictly positive-definite with respect to physical fields, again because of the possibility of non-radiating sources which do not contribute to any far-field radiated power.

 In fact, the surface product \eqref{sp2} does give rise to a true inner product on a restricted subspace of vector fields.  If $\bv{a}_{\stext{rad}}(\bv{x}; \omega)$ and $\bv{a}'_{\stext{rad}}(\bv{x}; \omega)$ are both radiation vector fields of finite outgoing far-field flux spectral density at prescribed frequencies of interest, then $\bv{a}_{\stext{ret}}(\bv{x}; \omega)$ and $\bv{a}'_{\stext{ret}}(\bv{x}; \omega)$, their outgoing projections valid in the far field, may be defined by
\bsub
\begin{align}
\bv{a}_{\stext{ret}}(\bv{x}; \omega) &= \tfrac{1}{2ik}(ik + \tfrac{\del}{\del r}) \bv{a}_{\stext{rad}}(\bv{x}; \omega)  \\
\bv{a}'_{\stext{ret}}(\bv{x}; \omega) &= \tfrac{1}{2ik}(ik + \tfrac{\del}{\del r}) \bv{a}'_{\stext{rad}}(\bv{x}; \omega) 
\end{align}
\esub
asymptotically as $kr \to \infty$.  It then follows that 
\be\begin{split}
\tfrac{1}{\muz} \innerpr{ i\omega \,\bv{a}_{\stext{ret}}  }{ \grad\!\times\! \bv{a}'_{\stext{ret}}  } &= \tfrac{1}{\muz} \innerpr{ \bv{e}_{\stext{ret}}   }{  \bv{b}'_{\stext{ret}}  } \\
&= \tfrac{1}{\muz} \!\lim_{R \to \infty} \! r^2 \!\!\!\! \int\limits_{r = R} \!\!\!\!  \,d^2\Omega(\unitvec{r}) \,\, \unitvec{r} \cdot \left[\, \bigl\{ i\omega\, \bv{a}_{\stext{ret}}( r \unitvec{r}; 
\omega ) \bigr\} \cc \times \bigl\{ \grad\!\times\! \bv{a}'_{\stext{ret}} ( r \unitvec{r}; \omega ) \bigr\} \, \right] \\
&=  
\tfrac{\omega k}{\muz} \lim_{R \to \infty} \! r^2 \!\!\! \int\limits_{r = R} \!\!\!  \,d^2\Omega(\unitvec{r}) \,\, \unitvec{r} \cdot \left[\,  \bv{a}_{\stext{ret}}( r \unitvec{r}; \omega )\cc  \!\times\! \bigl\{ \unitvec{r}\!\times\! \bv{a}'_{\stext{ret}} ( r \unitvec{r} ; \omega) \bigr\} \, \right] \\
&=  \tfrac{\omega^2}{\muz c} \lim_{R \to \infty} \! r^2 \!\!\! \int\limits_{r = R} \!\!\!  \,d^2\Omega(\unitvec{r}) \; \bv{a}_{\stext{ret}}( r \unitvec{r}; 
\omega )\cc \cdot \bv{a}'_{\stext{ret}} ( r \unitvec{r}; \omega )
\end{split}
\ee
really does represent a well-defined inner product between the \textit{radiative} vector potentials $\bv{a}_{\stext{rad}}(\bv{x}; \omega)$ and $\bv{a}'_{\stext{rad}}(\bv{x}; \omega)$, in that it is obviously conjugate-linear in its first argument, linear in its second, conjugate-symmetric under interchange of arguments, and, less obviously, also positive definite, as can be inferred from the following argument.  If $\lim\limits_{R \to \infty}  r^2 \!\!\! \int\limits_{r = R} \!\!\!  \,d^2\Omega(\unitvec{r}) \, \bigl\lvert \bv{a}_{\stext{ret}}( r \unitvec{r}; \omega ) \bigr\rvert^2 = 0$, then asymptotically, it must be the case that $\bv{a}_{\stext{ret}}( r \unitvec{r} ; \omega) \sim O(\sfrac{1}{r^2})$ or smaller in almost all directions, so $\bv{N}_{\perp}(\unitvec{r}; \omega) = 0$ for almost all $\unitvec{r}$ at the given frequency or frequencies of interest, and hence there can be no outgoing radiation at these frequencies, which also implies that the total radiation fields must vanish at these same frequencies, as we will be able to verify explicitly in a moment. 

Crucially, when applied to electromagnetic fields and sources, these volumetric and surficial inner products can be related by energy conservation.  We do not want to evaluate the volumetric inner product for pairs of electric fields or magnetic fields, as such integrals would be related to the (spectral density of) electromagnetic field energy, which may diverge badly for harmonic fields extending over all space.  Instead, we will consider volumetric overlap integrals between electric fields and current densities, which yield the (spectral density) of mechanical or ``Joule work,'' associated with exchange of energy between moving source charges and electric fields.  The surface product will normally be evaluated for pairs of outgoing electric and magnetic fields, leading, as we have just seen, to expressions for the (spectral density) of electromagnetic energy flux in the far-field.

Application of the elementary vector identity $\divergence (\bv{f} \times \bv{g}) = \bv{g} \cdot \grad \!\times  \bv{f} - \bv{f} \cdot \grad \!\times \bv{g}$, followed by application of the frequency-domain Maxwell's equations, yields the relation
\be\label{poyntinga}
\divergence (\bv{E}_{\perp}\cc \times \bv{B}_{\perp}) = i\omega \bigl[ \tfrac{1}{c^2} \bv{E}_{\perp}\cc \!\cdot\! \bv{E}_{\perp} - \bv{B}_{\perp}\cc \!\cdot\! \bv{B}_{\perp} \bigr] - \mu_0 \bv{J}_{\perp} \!\cdot \! \bv{E}_{\perp}\cc
\ee
at each frequency $\omega$.  Integrating over a sphere of radius $R$, applying Gauss's law, taking real parts, and considering the limit as $R \to \infty$, we deduce a version of Poynting's theorem for solenoidal, frequency-domain fields, namely 
\be\label{eqn:power_1}
- \realpart \innerpd{\bv{E}_{\stext{ret}} }{\bv{J}_{\perp}} =  \tfrac{1}{\muz} \realpart \innerpr{\bv{E}_{\stext{ret}} }{\bv{B}_{\stext{ret}} } =  \tfrac{1}{\muz}  \innerpr{\bv{E}_{\stext{ret}} }{\bv{B}_{\stext{ret}} } \;\ge\, 0.
\ee
which relates outgoing (spectral density) of energy flux in the far field to (spectral density) of mechanical work exchanged between the sources and the fields in their vicinity.
 
\subsubsection{Hermiticity and Reciprocity}

\vspace{-10pt}
We may also observe that with respect to the volumetric inner product \eqref{ip1}, the Green functions $\gret$ and $\gadv$ (at any fixed $\omega > 0$) may be viewed as position-space representations of linear operators which are Hermitian adjoints of each other, and hence the radiation kernel $\dop = \tfrac{1}{2}(\gret - \gadv)$ will be an anti-Hermitian linear operator, and so $i\omega \, \dop$ will be Hermitian.

In particular, this implies that when $\bv{E}_{\stext{rad}}(\bv{x}; \omega)$ consists of radiation from the transverse source $\bv{J}_{\perp}(\bv{x}; \omega)$, and  $\bv{E}'_{\stext{rad}}(\bv{x}; \omega)$ is radiation from transverse source $\bv{J}'_{\perp}(\bv{x}; \omega)$, then (separately at each frequency $\omega$) the fields and sources must satisfy a conjugate-reciprocity relation of the form
\be\label{reciprocity2}
% \begin{split}
\innerp{ \bv{E}_{\stext{rad}}  }{  \bv{J}'_{\perp}  } = \innerp{i\omega\, \bv{A}_{\stext{rad}}  }{  \bv{J}'_{\perp}  }  = \innerp{  \bv{J}_{\perp}  }{ i\omega\, \bv{A}'_{\stext{rad}}  }
= \innerp{  \bv{J}_{\perp}  }{ \bv{E}'_{\stext{rad}}  }  = \innerp{ \bv{E}'_{\stext{rad}}  }{  \bv{J}_{\perp}  }\cc,
% \end{split}
\ee
which will have important implications for our results below.

\subsubsection{Radiation Fields and Radiation Patterns}
 
We can now also express the radiation vector potential $\bv{A}_{\stext{rad}}(\bv{x}; \omega)$ directly in terms of this same outgoing radiation pattern $\bv{N}_{\perp}(\unitvec{r}; \omega)$ introduced earlier.  Applying Green's second identity to the vector potential $\bv{A}_{\stext{ret}}(\bv{x}; \omega)$ and the \textit{advanced} Green function $\gadv(\bv{x}, \bv{x}'; \omega)$,  strategically adding and subtracting some terms, and integrating over a sphere of radius $R$, we arrive at a vector version of 
Kirchoff diffraction integral, which can be thought of as a manifestation of Huygens' principle, relating the vector potential in the interior of the sphere to components and derivatives on the boundary:
\be
\begin{split}
\oint\limits_{r=R} \!\! & r{}^2 \,d^2\Omega(\unitvec{r})  \, \bigl[ \, \gadv(r \unitvec{r}, \bv{x}',  \omega) \,\bigl(\tfrac{\del}{\del r} +ik \bigr) \bv{A}_{\stext{ret}}(r \unitvec{r}; \omega) - \bv{A}_{\stext{ret}}(r\ \unitvec{r}; \omega) \,\bigl(\tfrac{\del}{\del r} + ik \bigr) \gadv(r \unitvec{r}, \bv{x'}; \omega)  \,\bigr] \\
&= \int\limits_{r=R} \!\! d^3\bv{x} \, \bigl[ \, \gadv(\bv{x}, \bv{x'}; \omega) \,\bigl(\laplacian + k^2 \bigr) \bv{A}_{\stext{ret}}(\bv{x}; \omega) - \bv{A}_{\stext{ret}}(\bv{x}; \omega) \,\bigl(\laplacian + k^2 \bigr) \gadv(\bv{x}, \bv{x}'; \omega) \, \bigr]   \\
&=  \int\limits_{r=R} \!\! d^3\bv{x} \, \bv{A}_{\stext{ret}}(\bv{x}; \omega) \, \delta(\bv{x} - \bv{x'})   = \bv{A}_{\stext{ret}}(\bv{x}'; \omega).
\end{split}
\ee
 Taking the $kR \to \infty$ limit, and using the known asymptotic far-field forms approached by $\bv{A}_{\stext{rad}}(r \unitvec{r}; \omega)$ and $\gadv(r \unitvec{r}, \bv{x}'; \omega)$,  we can relate the full radiation fields to the outgoing far-fields according to:
\be\label{afromn}
\bv{A}_{\stext{rad}}(\bv{x}'; \omega) = \tfrac{\muz}{8\pi^2} \, i k \!\int \!\! d^2\Omega(\unitvec{r}) \,e^{ik\unitvec{r}\cdot \bv{x}' } \bv{N}_{\perp}(\unitvec{r}; \omega).
\ee
As a superposition of transverse, harmonic plane waves satisfying the dispersion relation $\omega = ck$, clearly this solves the source-free, harmonic Maxwell equations everywhere in space, but somewhat less obviously, it possesses just the right boundary conditions in the far-field to reproduce the radiation $\bv{A}_{\stext{rad}}(\bv{x}'; \omega)$ everywhere.  So the angular radiation pattern $N_{\perp}(\unitvec{r}; \omega)$ (if known along almost all directions, for frequencies of interest ) uniquely determines the radiation vector potential almost everywhere.  Conversely, we have already seen how the radiation vector potential $\bv{A}_{\stext{rad}}(\bv{x}'; \omega)$ known just in the far field will uniquely determine the radiation pattern $N_{\perp}(\unitvec{r}; \omega)$.  As a corollary, the representation \eqref{afromn} also verifies that, at any frequency $\omega$, the radiation vector potential $\bv{A}_{\stext{rad}}(\bv{x}'; \omega) $ will vanish almost everywhere (in spatial position), if and only if the angular radiation pattern $\bv{N}_{\perp}(\unitvec{r}; \omega) = 0$ in almost all directions.

In addition, by applying the source-free version of relation \eqref{poyntinga} to the radiation fields, we find that there is no \textit{net} power flow at any given frequency, in the sense that
\be
\tfrac{1}{\muz} \realpart  \innerpr{\bv{E}_{\stext{rad}} }{\bv{B}_{\stext{rad}} } = 0,
\ee
evidently because the inflowing and outflowing (or retarded and advanced) spectral flux densities must cancel (separately in every frequency band, as we will verify explicitly below.

\subsubsection{More on Energy Flow}

Now let us look more closely at the energy flow in the advanced electromagnetic fields, and compare this to the energy balance in the retarded fields.  As $kr \to \infty$, the advanced fields approach the asymptotic forms
\bsub
\begin{align}
\bv{E}_{\stext{adv}}(r \unitvec{r}; \omega) &= i\omega \bv{A}_{\stext{adv}}(r \unitvec{r}; \omega)  \;\;\;\sim \phantom{+} i \omega \frac{\muz}{4\pi}  \frac{e^{-ikr}}{r} \, \bv{N}_{\perp}(-\unitvec{r}; \omega), \text{ and } \\
\bv{B}_{\stext{adv}}(r \unitvec{r}; \omega) &= \grad\! \times\!  \bv{A}_{\stext{adv}}(r \unitvec{r}; \omega) \sim -ik\unitvec{r} \times \tfrac{1}{c} \bv{A}_{\stext{adv}}(r \unitvec{r}; \omega) \sim -\unitvec{r} \times \tfrac{1}{c} \bv{E}_{\stext{adv}}(r \unitvec{r}; \omega).
\end{align}
\esub
It follows that
\be\label{eqn:power_a0}
\innerpr{\bv{E}_{\stext{adv}} }{\bv{B}_{\stext{adv}} } =  -\tfrac{\muz^2}{16\pi^2  c} \,\omega^2 \! \int \! d^2\Omega(\unitvec{r}) \, \bv{N}_{\perp}(-\unitvec{r}; \omega)\cc\!\cdot\!\bv{N}_{\perp}(-\unitvec{r}; \omega), 
\ee
but a change of variables $\unitvec{r}' = -\unitvec{r}$ will verify that
\be
\int \! d^2\Omega(\unitvec{r}) \, \bv{N}_{\perp}(-\unitvec{r}; \omega)\cc\!\cdot\!\bv{N}_{\perp}(-\unitvec{r}; \omega) = \int \! d^2\Omega(\unitvec{r}') \, \bv{N}_{\perp}(\unitvec{r}'; \omega)\cc\!\cdot\!\bv{N}_{\perp}(\unitvec{r}'; \omega)
\ee
when integrated over all solid angles, so the advanced and retarded spectral densities of radiated power are indeed equal in magnitude but opposite in sign:
\be
\innerpr{\bv{E}_{\stext{adv}} }{\bv{B}_{\stext{adv}} } = - \innerpr{\bv{E}_{\stext{ret}} }{\bv{B}_{\stext{ret}} } \;\ge \, 0.
\ee
The advanced fields also satisfy the conservation law \eqref{poyntinga}, with the \textit{same} source $\bv{J}_{\perp}(\bv{x}; \omega)$ (not complex conjugated or negated or anything), so that
\be\label{eqn:power_2}
- \realpart \innerpd{\bv{E}_{\stext{adv}} }{\bv{J}_{\perp}} =  \tfrac{1}{\muz} \realpart \innerpr{\bv{E}_{\stext{adv}} }{\bv{B}_{\stext{adv}} } =  \tfrac{1}{\muz}  \innerpr{\bv{E}_{\stext{adv}} }{\bv{B}_{\stext{adv}} } = -\tfrac{1}{\muz}  \innerpr{\bv{E}_{\stext{ret}} }{\bv{B}_{\stext{ret}} }  \;\le\, 0.
\ee
Relating the radiated power to mechanical work, we find
\be\label{power3}
\begin{split}
- \realpart \innerpd{\bv{E}_{\stext{rad}} }{\bv{J}_{\perp}} &= - \realpart \innerpd{ \,(\bv{E}_{\stext{ret}} - \bv{E}_{\stext{adv}} )\,}{\bv{J}_{\perp}} \\
&=- \realpart \innerpd{\bv{E}_{\stext{ret}} }{\bv{J}_{\perp}} +  \realpart \innerpd{\bv{E}_{\stext{adv}} }{\bv{J}_{\perp}} \\
&= \tfrac{1}{\muz}  \innerpr{\bv{E}_{\stext{ret}} }{\bv{B}_{\stext{ret}} }  - \tfrac{1}{\muz}  \innerpr{\bv{E}_{\stext{adv}} }{\bv{B}_{\stext{adv}} }  \\
&= 2 \, \tfrac{1}{\muz}  \innerpr{\bv{E}_{\stext{ret}} }{\bv{B}_{\stext{ret}} }  \;\ge\, 0,  
\end{split}
\ee
or equivalently,
\be\label{power4}
-\tfrac{1}{2} \realpart \innerpd{\bv{E}_{\stext{rad}} }{\bv{J}_{\perp}} = \tfrac{1}{\muz}  \innerpr{\bv{E}_{\stext{ret}} }{\bv{B}_{\stext{ret}} } \ge 0,
\ee
where the factor of $\sfrac{1}{2}$ appearing in the last expression basically arises to avoid over-counting in the energetics: radiation fields satisfying the source-free Maxwell equations must contain equal amounts of outgoing and incoming energy (spectral density) in complementary angular patterns, so  the ``virtual'' energy exchange between the actual sources and radiation fields turns out to be exactly twice that between the same sources and the corresponding outgoing component alone.  This factor is essentially a radiative analog of the well-known factor of $\sfrac{1}{2}$ that appears in the expression for the potential self-energy of a charge distribution in electrostatics.

Because solenoidal and irrotational fields will be functionally transverse when integrated over all space, we know that $\innerpd{\bv{E}_{\stext{rad}} }{\bv{J}_{\|}} = 0$, so we can also write
\be\label{power4b}
- \tfrac{1}{2} \realpart \innerpd{\bv{E}_{\stext{rad}} }{\bv{J}_{\perp}}  = - \tfrac{1}{2} \realpart \innerpd{\bv{E}_{\stext{rad}} }{\bv{J}} = \tfrac{1}{\muz}  \innerpr{\bv{E}_{\stext{ret}} }{\bv{B}_{\stext{ret}} } \ge 0,
\ee
which is often convenient, as we do not have to explicitly determine the solenoidal current density $\bv{J}_{\perp}(\bv{x}; \omega)$ from the actual physical current density $\bv{J}(\bv{x}; \omega)$.  Poynting relation \eqref{power4b} will provide a key ingredient in the construction of our variational principle, by identifying the (spectral density of) radiant energy flux $\tfrac{1}{\muz}  \innerpr{\bv{E}_{\stext{ret}} }{\bv{B}_{\stext{ret}} }$ in the far-field, to the (spectral density) of mechanical work $-\tfrac{1}{2} \realpart \innerpd{\bv{E}_{\stext{rad}} }{\bv{J}}$ that would be exchanged between the actual sources and the radiative part (only) of the electric fields generated by those sources.  Because of this equality, we can also interpret the work $-\tfrac{1}{2} \realpart \innerpd{\bv{E}_{\stext{rad}} }{\bv{J}}$ as the amount of energy that would need to be supplied by external forces to keep the sources following the prescribed trajectories, due to the energy lost to radiation.

\section{Maximum ``Power'' Variational Principle}

With all this mathematical machinery in place, derivation of the actual variational principle becomes straightforward.  

We have established that any radiation $\bv{a}_{\stext{rad}}(\bv{x}; \omega)$ whatsoever can be generated by some solenoidal source, say $\bv{s}_{\perp}(\bv{x}; \omega)$.  We will call this the \textit{trial radiation profile}, generated by the \textit{trial source}, respectively, while the resulting electric field $\bv{e}_{\stext{rad}}(\bv{x}; \omega) = \bv{e}_{\stext{ret}}(\bv{x}; \omega) - \bv{e}_{\stext{adv}}(\bv{x}; \omega)$ and corresponding magnetic field $\bv{b}_{\stext{rad}}(\bv{x}; \omega) =  \bv{b}_{\stext{ret}}(\bv{x}; \omega) - \bv{b}_{\stext{adv}}(\bv{x}; \omega)$ will be referred to as the trial fields (of either radiative, retarded, or advanced character, depending on which components are retained).

By linearity,  we may also consider the radiation emitted by the difference $\bigl[ \bv{J}_{\perp}(\bv{x}; \omega) - \bv{s}_{\perp}(\bv{x}; \omega) \bigr]$ between the actual solenoidal current density source of interest and this trial source.  Were the resulting electromagnetic fields to be actually present, the outgoing power (spectral density) radiated by this difference source would, like any outgoing radiation from any source, necessarily be non-negative, and satisfy the Poynting relation
\be\label{powerv1}
 - \tfrac{1}{2} \realpart \innerpd{\bv{E}_{\stext{rad}} - \bv{e}_{\stext{rad}}}{\bv{J}_{\perp} - \bv{s}_{\perp}}  = \tfrac{1}{\muz}  \innerpr{\bv{E}_{\stext{ret}} - \bv{e}_{\stext{ret}} }{\bv{B}_{\stext{ret}} - \bv{b}_{\stext{ret}}} \;\ge 0,
\ee
where equality will hold (at any frequency or ranges of frequencies) if and only if the true transverse source $\bv{J}_{\perp}(\bv{x};\omega)$ radiating the actual fields and the trial transverse source $\bv{s}_{\perp}(\bv{x};\omega)$ generating the trial fields differ at most by some non-radiating source at the relevant frequencies, or equivalently if and only if the corresponding electric fields satisfy $\bv{e}_{\stext{ret}}(\bv{x};\omega) = \bv{E}_{\stext{ret}}(\bv{x};\omega)$ almost everywhere.

Expanding the expression on the left-hand side using linearity, we find 
\be
- \tfrac{1}{2} \realpart \innerpd{\bv{E}_{\stext{rad}} }{\bv{J}_{\perp}}  
+ \tfrac{1}{2} \realpart \innerpd{\bv{E}_{\stext{rad}}}{\bv{s}_{\perp}}  
+ \tfrac{1}{2} \realpart \innerpd{\bv{e}_{\stext{rad}}}{\bv{J}_{\perp}}  
 - \tfrac{1}{2} \realpart \innerpd{\bv{e}_{\stext{rad}}}{\bv{s}_{\perp}}  \ge 0,
\ee
but our Poynting relations can be applied to each source separately, to wit:
\bsub
\begin{align}
-\tfrac{1}{2} \realpart \innerpd{\bv{E}_{\stext{rad}} }{\bv{J}_{\perp}}  &= \tfrac{1}{\muz} \innerpr{\bv{E}_{\stext{ret}} }{\bv{B}_{\stext{ret}} }  \; \\  %     \;\ge 0, \\
 -\tfrac{1}{2} \realpart \innerpd{\bv{e}_{\stext{rad}} }{\bv{s}_{\perp}} &= \tfrac{1}{\muz} \innerpr{\bv{e}_{\stext{ret}} }{\bv{b}_{\stext{ret}} }  \;   % \;\ge 0,
\end{align}
\esub
so that
\be
\tfrac{1}{\muz} \innerpr{\bv{E}_{\stext{ret}} }{\bv{B}_{\stext{ret}} }  
+ \tfrac{1}{2} \realpart \innerpd{\bv{E}_{\stext{rad}}}{\bv{s}_{\perp}}  
+ \tfrac{1}{2} \realpart \innerpd{\bv{e}_{\stext{rad}}}{\bv{J}_{\perp}}  
+ \tfrac{1}{\muz} \innerpr{\bv{e}_{\stext{ret}} }{\bv{b}_{\stext{ret}} }  \ge 0.
\ee
Next, using the fact that the radiation kernel is anti-Hermitian with respect to the volumetric inner product, we can invoke the fundamental reciprocity relation \eqref{reciprocity2} alluded to above, by which
%\be
%\begin{split}
%\realpart \innerpd{\bv{E}_{\stext{rad}}}{\bv{s}_{\perp}} &= \innerpd{i\omega\, \bv{A}_{\stext{rad}}}{\bv{s}_{\perp}} 
%= -i\omega \innerpd{\bv{A}_{\stext{rad}}}{\bv{s}_{\perp}}
%= -i\omega \innerpd{\bv{J}_{\perp}}{-\bv{a}_{\stext{rad}} } \\
%&= \innerpd{\bv{J}_{\perp}}{i \omega\, \bv{a}_{\stext{rad}} }
%= \innerpd{\bv{J}_{\perp}}{\bv{e}_{\stext{rad}} }
%= \innerpd{\bv{e}_{\stext{rad}} }{\bv{J}_{\perp}}\cc,
%\end{split}\ee
%implying 
\be
\realpart \innerpd{\bv{E}_{\stext{rad}}}{\bv{s}_{\perp}} = \realpart  \innerpd{\bv{e}_{\stext{rad}} }{\bv{J}_{\perp}}\cc = \realpart \innerpd{\bv{e}_{\stext{rad}} }{\bv{J}_{\perp}},
\ee
so that we have in fact been able to deduce that
\be\label{vp1}
\tfrac{1}{\muz} \innerpr{\bv{E}_{\stext{ret}} }{\bv{B}_{\stext{ret}} }  \ge
- \tfrac{1}{\muz} \innerpr{\bv{e}_{\stext{ret}} }{\bv{b}_{\stext{ret}} } 
-  \realpart \innerpd{\bv{e}_{\stext{rad}} }{\bv{J}_{\perp}} ,
\ee
with equality if and only if the actual fields and corresponding trial fields agree.
The left-hand side of \eqref{vp1} represents the outgoing spectral density of far-field energy flux in the physical fields, and must be nonnegative, while the first term on the right-hand side is minus the corresponding energy spectral density in the trial fields, and is in fact non-positive, while the last term, representing the spectral density of ``virtual'' work that would be extracted from the actual sources by the source-free trial field, can actually be of either sign.  Because \eqref{vp1} no longer makes explicit reference to the trial source $\bv{s}_{\perp}(\bv{x}; \omega)$, one might worry worry as to how exactly to define $\bv{a}_{\stext{ret}}(\bv{x}; \omega)$ from $\bv{a}_{\stext{rad}}(\bv{x}; \omega)$.  But only the asymptotic far-field matters to the calculation of outgoing Poynting flux, and the outgoing far fields can be unambiguously extracted from the radiation fields.

Now suppose the trial vector potential and corresponding fields actually depend on some tuplet $\bv{\alpha}$ of adjustable parameters determining the field phase, amplitude, polarization, and mode shape at each frequency of interest.  We do not require that the parameterized family $\left\{\, \bv{a}_{\stext{rad}}(\bv{x};\omega; \bv{\alpha}) \, \right\}$ of trial  solutions constitutes a linear subspace of vector fields, but all members must consist of radiation, i.e., be divergence-free solutions to the homogeneous Helmholtz equation, or equivalently, be eigenfunctions of the double-curl operator.  Also, we will subsequently assume that the variational parameters are independent almost everywhere in the allowed parameter space, so the same radiation fields cannot be generated by two or more different parameter values (with negligibly few exceptions).  Finally, we will impose the essentially trivial requirement that the parameterization is sufficient to allow for arbitrary complex re-scalings, so the overall amplitude and overall phase offset of the trial radiation fields can always be arbitrarily varied.

% Since $k  \neq 0$, it is easy to verify that a trial solution $\bv{f}_{\perp}^{\stext{hom}}$ is feasible (satisfying the source-free Helmholtz equation and the solenoidal gauge constraint) if and only if $\curl\!\curl\!\bv{f}_{\perp}^{\stext{hom}} = k^2 \bv{f}_{\perp}^{\stext{hom}}.$ 

Because inequality \eqref{vp1} must hold true for all possible parameterized trial fields, it follows that
% then be expressed as  
\bsub\label{vp2}
\begin{align}
\tfrac{1}{\muz} \innerpr{\bv{E}_{\stext{ret}} }{\bv{B}_{\stext{ret}} }   &\ge \max_{\bv{\alpha}} \bigl[\,
- \tfrac{1}{\muz} \innerpr{\bv{e}_{\stext{ret}} }{\bv{b}_{\stext{ret}} } 
-  \realpart \innerpd{\bv{e}_{\stext{rad}} }{\bv{J}_{\perp}} \, \bigr]\\
\text{ such that}\!&: \text{ } \grad \! \times \!  \grad \! \times \!  \bv{a}_{\stext{rad}}(\bv{x}; \omega; \bv{\alpha}) = k^2 \, \bv{a}_{\stext{rad}}(\bv{x}; \omega; \bv{\alpha}),
\end{align}
\esub
where equality is achieved if and only if the outgoing part  $\bv{a}_{\stext{ret}}(\bv{x}; \omega; \bv{\alpha})$ of  $\bv{a}_{\stext{rad}}(\bv{x}; \omega; \bv{\alpha})$ coincides with the true emission pattern $\bv{A}_{\stext{ret}}(\bv{x}; \omega; \bv{\alpha})$ in the far field (at the frequencies of interest).   

Again, in principle, the trial radiation fields can be unambiguously decomposed into incoming and outgoing components in the far-field limit, which is where the Poynting flux needs to be calculated, so this is now a well-posed optimization problem.  Indeed, this is actually our desired result, although not in an especially transparent form.  But under our articulated assumptions, it turns out that \eqref{vp2} will be entirely equivalent to the following constrained maximization problem:
\bsub\label{vp3}
\begin{align}
\tfrac{1}{\muz} \innerpr{\bv{E}_{\stext{ret}} }{\bv{B}_{\stext{ret}} }  &\ge \max_{\bv{\alpha}} \bigl[\,
 \tfrac{1}{\muz} \innerpr{\bv{e}_{\stext{ret}} }{\bv{b}_{\stext{ret}} } \, \bigr] \\
\text{such that}\!&: \text{ } \tfrac{1}{\muz} \innerpr{\bv{e}_{\stext{ret}} }{\bv{b}_{\stext{ret}} } = -\tfrac{1}{2} \realpart \innerpd{\bv{e}_{\stext{rad}} }{\bv{J}_{}} , \\
\text{and}\!&: \text{ } \grad \! \times \!  \grad \! \times \! \bv{e}_{\stext{rad}}(\bv{x}; \omega; \bv{\alpha}) = k^2 \, \bv{e}_{\stext{rad}}(\bv{x}; \omega; \bv{\alpha}).
\end{align}
\esub
That is, the actual radiation fields radiate more outgoing energy flux, and therefore extract more energy from the actual sources, than could any trial radiation field, if it were present in the vicinity of the sources.  (While not immediately obvious,  the mathematical equivalence can be better understood by thinking in terms of a Lagrange multiplier associated with the energy conservation constraint).

Also, because $\bv{e}_{\stext{rad}}(\bv{x}; \omega)$ must be everywhere divergence-free by assumption, it will be functionally orthogonal to any irrotational vector field, so we have conveniently replaced the integral $\innerpd{ \bv{e}_{\stext{rad}} }{\bv{J}_{\perp} }$ with the integral $\innerpd{ \bv{e}_{\stext{rad}} }{\bv{J} }$.  Usually, it is the full current density $\bv{J}(\bv{x},\omega)$ rather than $\bv{J}_{\perp}(\bv{x},\omega)$ that is specified explicitly, so it comes as something of a relief that we can  avoid having to extract the solenoidal part explicilty, which would involve either solving a Poisson equation for each frequency $\omega$, or else a projection to and from $\bv{k}$-space, requiring three-dimensional Fourier transforms, which may be almost as difficult to evaluate as would be solving Maxwell's equations exactly.

To confirm the equivalence, under our assumptions, of these seemingly different optimization problems, we may verify that they are both equivalent to a third formulation.  We suppose the variational parameters are $\bv{\alpha} = (a, \theta, \bv{\zeta})$, and write $\bv{a}_{\stext{rad}}(\bv{x};\omega; \bv{\alpha}) = a \, i e^{i\theta} \, \bv{u}_{\stext{rad}}(\bv{x};\omega; \bv{\zeta}),$ where $a$ is an overall real, positive scaling, and $\theta$ is a real-valued phase, while $\bv{u}_{\stext{rad}}(\bv{x}; \omega; \bv{\zeta})$ is a free-space solution representing the relative spatial shape and polarization of the trial vector field, as determined by the remaining set of ``shape'' parameters $\bv{\zeta}$.  Then one will find that the solutions to the variational optimization problems defined either by \eqref{vp2} or by \eqref{vp3} are formally identical, and both are given by:
\bsub\label{eqn:optimization_2}
\begin{align}
\tilde{\bv{\zeta}}(\omega) &= \argmax_{ \bv{\zeta} }  \Bigl[ \frac{ \abs{  \innerpd{ \bv{u}_{\stext{rad}} }{\bv{J}_{} }  }^2 }{\abs{ \innerpr{ \bv{u}_{\stext{ret}}  }{\curl\! \bv{u}_{\stext{ret}} } } }  \Bigr],\\
\tilde{\theta}(\omega) &= \left.\arg\bigl[ \innerpd{ \bv{u}_{\stext{rad}} }{\bv{J}_{}} \bigr] \right._{\bv{\zeta}(\omega) = \tilde{\bv{\zeta}}(\omega) } ,\\
\tilde{a}(\omega) &=    \left. \frac{ \mu_0\, \lvert \innerpd{ \bv{u}_{\stext{rad}} }{\bv{J}_{} } \rvert }{ 2\lvert  \innerpr{ \bv{u}_{\stext{ret}} }{\curl\! \bv{u}_{\stext{ret}}  }  \rvert }   \right\rvert_{\bv{\zeta} = \tilde{\bv{\zeta}}(\omega)},
\end{align}
\esub
so in fact they must describe the same variational principle.  The optimization with respect to parameters $\bv{\zeta}(\omega)$, separately at each frequency $\omega$ of interest, determines the best \textit{relative} mode shape and polarization amongst the parameterized family of trial solutions.  Then the optimal choice of phase offset $\theta(\omega)$ for the trial fields ensures that the maximum energy transfer and hence radiated power is achieved, and finally the overall amplitude $a(\omega)$ is fixed so that the power balance constraint is exactly met.

One further formulation of the variational principle might be mentioned. From the underlying inequality \eqref{powerv1}, we can see that the variational principle must also be formally equivalent to
\bsub\label{vp5}
\begin{align}
\bv{\alpha}(\omega) = \argmin\limits_{\bv{\alpha}}& \, \tfrac{1}{\muz} \innerpr{\bv{E}_{\stext{ret}} - \bv{e}_{\stext{ret}} }{ \bv{B}_{\stext{ret}} - \bv{b}_{\stext{ret}}  }  , \\
\text{ such that}\!&: \text{ } \grad \! \times \!  \grad \! \times \!  \bv{a}_{\stext{rad}}(\bv{x}; \omega; \bv{\alpha}) = k^2 \, \bv{a}_{\stext{rad}}(\bv{x}; \omega; \bv{\alpha}),
\end{align}
\esub
which just says that we are seeking the trial radiation field which is closest to the actual radiation fields---in the sense of a Hilbert-space distance defined in terms of far-field spectral density of radiated energy flux.  While enjoying a very simple interpretation, this last version \eqref{vp5} of the variational principle would not be useful in practice, because the whole point in seeking a variational approximation is that we do not know the actual fields, so cannot directly calculate the inner product appearing in \eqref{vp5}.

%%%%%%

\def\j{\mathcal{J}}
\def\gparaxial{{G}}
\def\gplus{G_{d}}
\def\gminus{G_{u}}
\def\D{\mathcal{D}}
\def\xperp{\bv{x}_{\perp}}
\def\s{\bv{\mathcal{S}}}

\section{Paraxial Optics}

In the approximate but often applicable regime of paraxial optics, a version of the MPVP can be shown to hold \textit{exactly}, that is, without any further approximations beyond those of the paraxial expansion itself.  Working explicitly within a paraxial approximation when it is justified will prove quite useful, because finding free-space radiation solutions for paraxial beams is far simpler than in full 3D geometry.

When electromagnetic wave propagation remains collimated and beam-like, largely confined to wavevectors deviating in direction only slightly from a specified optic axis (say the $+\unitvec{z}$ direction), we may anticipate that a \textit{paraxial parameter} representing the characteristic diffraction angle will be small, meaning
\be
\Theta = \tfrac{1}{k \sigma} \ll 1,
\ee
in which $k = \tfrac{2\pi}{\lambda}$ is the wavenumber of interest, and $\sigma$ is a measure of the (focused) transverse spot size of the beam.  After Fourier transforming in time and factoring out the carrier oscillation, the waveform envelope will vary transversely with characteristic scale-length of about $\sigma \approx \Theta\inv \lambda$ near the focal plane (and progressively more slowly away from the focus), and will vary longitudinally with a still longer length-scale of about $z_{\stext{R}} = \tfrac{1}{2} k \sigma^2 \approx \Theta\inv \sigma \approx \Theta^{-2} \lambda$, the so-called Rayleigh range.  This assumed separation of scales between $\lambda$, $\sigma$, and $z_R$ is what allows for a self-consistent paraxial expansion,
which In Fourier space, amounts to supposing that $k_{\perp}^2 = (k_x^2 + k_y^2) \ll k_z^2$, so that
 dispersion relation can be approximated by a Taylor expansion of the form
\be
k_z = +\sqrt{\tfrac{\omega^2}{c^2} - k_x^2 - k_y^2}  =  +\sqrt{k^2 - k_x^2 - k_y^2} \; \approx +k \bigl[ 1 - \tfrac{1}{2 k^2}(k_x^2 + k_y^2)  \bigr] + \dotsb,
\ee
at real frequencies $\omega$ of interest.

\subsection{Governing Equations}

In the frequency domain, one can then develop an asymptotic expansion in powers of the characteristic diffraction angle $\Theta$,\cite{lax:75} resulting, at leading order, in an approximate vector potential of the form
\be
\bv{A}_{\perp}(\bv{x};\omega) \approx \bv{\psi}(\xperp, z; k)\, e^{+i kz} ,
\ee
where $\xperp = (x,y)$ are the transverse spatial Cartesian coordinates, $\omega$ and $k$ are related by the  free-space dispersion relation
\be
\omega = ck  = \tfrac{2\pi c}{\lambda},
\ee
while the slowly-varying wave envelope $\bv{\psi}(\bv{x}; k)$ satisfies an approximate Coulomb-gauge condition,
\be
\unitvec{z}\cdot \bv{\psi}(\xperp, z; k) = 0,
\ee
dictating that $\bv{\psi}(\xperp, z; k)$ is actually geometrically transverse, as well as the (right-moving) paraxial wave equation
\be
+i \tfrac{\del}{\del z} \bv{\psi}(\xperp, z; k) + \tfrac{1}{2k} \laplacian_{\perp} \bv{\psi}(\xperp, z; k) = \s(\xperp, z; k),
\ee
with a diffraction term involving the transverse Laplacian operator 
\be
\laplacian_{\perp} = (\laplacian - \tfrac{\del^2}{\del z^2}) =  (\tfrac{\del^2}{\del x^2} + \tfrac{\del^2}{\del y^2}),
\ee
and a source or driving term 
\be
\s(\xperp, z; k) = -\muz\, \tfrac{1}{2k}  (1 - \unitvec{z} \unitvec{z}\trans) \, \bv{J}_{\perp}(\xperp, z; ck) \,  e^{-ikz},
\ee
proportional to the geometrically transverse part of the functionally transverse current density.  The homogeneous part of this paraxial wave equation is reminiscent of the Schr{\"o}dinger equation for a non-relativistic, spin-$1$ particle moving in $2$ spatial dimensions, only where longitudinal position $z$ in the paraxial optical case plays the role of the temporal evolution variable $t$ in the quantum mechanical analog.

At the next order in the small parameter $\Theta$, the only modification would involve the addition of a term of the form
\be\label{gaugep0}
\psi_z(\xperp, z; k) = \tfrac{i}{k} \grad_{\perp} \! \cdot \bv{\psi}(\xperp,z;k)
\ee 
modifying the gauge constraint, in which  $\grad_{\perp} = \grad - \unitvec{z} \tfrac{\del}{\del z} = \unitvec{x} \tfrac{\del}{\del x} + \unitvec{y} \tfrac{\del}{\del y}$.   Conveniently, both versions of the gauge constraints can be written as
\be\label{gaugep2}
\unitvec{z} \cdot \bv{\psi}(\xperp, z; k) =  \epsilon \, \tfrac{i}{k} \grad_{\perp}\! \cdot \bv{\psi}(\xperp,z;k),
\ee
where the leading-order gauge constraint corresponds to the choice $\epsilon = 0$, while the next-order gauge constraint corresponds to $\epsilon = 1$.

Given the paraxial vector potential, the associated paraxial electric field is, to leading order,
\be
\bv{E}_{\perp}(\bv{x};\omega) = i\omega \, \bv{A}_{\perp}(\bv{x};\omega) \approx i\omega \, \bv{\psi}(\xperp, z; k)\, e^{i kz} ,
\ee
while the associated paraxial magnetic field becomes
\be\begin{split}
\bv{B}_{\perp}(\bv{x};\omega)  &= \grad\! \times\! \bv{A}_{\perp}(\bv{x};\omega) \\
&\approx  [\grad\!\times\! \bv{\psi}(\xperp, z; k) ]\, e^{+i kz} + [ ik \unitvec{z} \!\times \bv{\psi}(\xperp, z; k)]\, e^{ikz} \approx ik \unitvec{z} \!\times \bv{\psi}(\xperp, z; k)\, e^{ikz},
\end{split}
\ee
after dropping terms which are of higher order in the paraxial parameter $\Theta$.  The next-order corrections to the fields would involve a geometrically longitudinal term of the form $i\omega \,\unitvec{z} \psi_z \, e^{ikz}$ added to $\bv{E}(\xperp,z;k)$, and a longitudinal contribution of the form $\unitvec{z}\, \unitvec{z}\!\cdot\! (\grad\!\times\! \bv{\psi}) \,e^{ikz}$ added to $\bv{B}(\xperp,z; k)$.

\subsection{Green Functions}

In terms of its characteristics, the governing wave equation has now become parabolic rather than hyperbolic or elliptic, yet paraxial versions of most of the relations and constructions previously developed in the full three-dimensional geometry will still apply.

In particular, following the development of standard scattering theory in quantum mechanics, we can introduce scalar Green functions such that
\be
\bigl[ +i \tfrac{\del}{\del z} + \tfrac{1}{2k} \laplacian_{\perp} \bigr] \gparaxial(
\bv{x}, \bv{x}'; \omega) = i\, \delta(z - z')\, \delta(\xperp - \xperp') ,
\ee
where we are now switching to sign and phase conventions typical of non-relativistic quantum theory, to better leverage mathematical intuitions developed in that context.  Various choices for Green functions will involve different boundary conditions, and will differ by homogenous (source-free) solutions to the paraxial wave equation.

However, as we are confining attention only to right-moving paraxial waves, the natural pair of Green functions will not correspond to advanced and retarded solutions, as in the full three-dimensional case, but instead to upstream and downstream solutions.  (A similar parabolic partial differential equation, but with appropriate changes of sign, will govern left-moving paraxial waves.  But for typical light sources such as lasers or relativistic particle beams, emitted radiation will tend to be highly collimated in the forward, or rightward, or $+\unitvec{z}$ direction only.  Other light sources may be more dipolar or quadrupolar in their emission pattern, but relatively few sources of interest would be predominately bi-directional, so typically we would rely on either a rightward or a leftward paraxial wave equation, but not both, for any given source or emitted wave-packet---although some Raman or Brillouin scattering problems might be exceptions).

In a paraxial geometry, a retarded Green function would be associated with right-moving waves to the right of the impulsive source and with left-moving waves to the left of the source, while an advanced Green function would involve the reversed pattern, namely right-moving waves to the left of the impulse and left-moving waves to the right of the impulse.  Here we instead employ \textit{downstream} and \textit{upstream} Green functions, both of which only involve right-moving waves.  The former, downstream Green function describes the emission of paraxial waves moving rightward but only found downstream from the source, while the latter, upstream Green function would involve the absorption of right-moving paraxial waves arriving from the region upstream of a sink.

The outgoing, downstream, or causal paraxial Green function ($\gplus$) can be written as
 \be
\gplus(\xperp, z; \xperp', z'; \omega) = +\Theta( z-z' )\, \tfrac{k }{2\pi i [z-z']} \, e^{\frac{+ik\abs{ \xperp - \xperp'}^2}{ 2[z-z']}},
\ee
and satisfies the boundary conditions
\bsub
\begin{align}
\gplus(\xperp, z; \xperp', z'; \omega) &= \phantom{+}0 \;\; \text{ if } z < z' \\
\lim\limits_{z \to z'^{+}} \gplus(\xperp, z; \xperp', z'; \omega) &= +\delta(\xperp - \xperp'),
\end{align}
\esub
whereas the ingoing, upstream, or absorbing paraxial Green function ($\gminus$), is instead
\be
\gminus(\xperp, z; \xperp', z'; \omega) = -\Theta(z'-z)\, \tfrac{k }{2\pi i [z-z']} \, e^{\frac{+ik\abs{ \xperp - \xperp'}^2}{ 2[z-z']}},
\ee
and satisfies the boundary conditions
\bsub
\begin{align}
\gminus(\xperp, z; \xperp', z'; \omega) &= \phantom{-}0 \;\; \text{ if } z > z' \\
\lim\limits_{z \to z'^{-}} \gminus(\xperp, z; \xperp', z'; \omega) &= -\delta(\xperp - \xperp').
\end{align}
\esub

Formally, the causal solution to the driven paraxial equation can be written in terms of the downstream Green function as
\be
\begin{split}
\bv{\psi}_{}(\xperp, z; k) &=  \bv{\psi}_{\stext{in}}(\xperp, z; k) + \bv{\psi}_{d}(\xperp, z; k)\\
&= \bv{\psi}_{\stext{in}}(\xperp, z; k)  +  \tfrac{1}{i}\!\! \int\limits_{-\infty}^{z} \!\!dz' \! \int \! d^2\xperp' \, \gplus(\xperp, z; \xperp', z'; \omega) \, \s(\xperp', z'; k),
\end{split}
\ee
where $\bv{\psi}_{\stext{in}}(\xperp, z; k)$ includes any free-space contributions that were already present upstream of the sources, and may be subsequently dropped for our purposes, as we are focusing attention on the radiation emitted by the prescribed sources.

We can also use the very same Green function as a \textit{propagator} for the source-free (radiation) fields, in the following sense.  Suppose we know $\bv{\psi}(\xperp, z'; k)$ in some transverse plane specified by longitudinal position $z'$.  Then in the absence of intervening sources, $\bv{\psi}(\xperp, z'; k)$ in any transverse plane further downstream, and specified by some longitudinal position $z$, where $z > z'$, would be
\be
\bv{\psi}_{}(\xperp, z; k) =  +\! \int \! d^2  \xperp' \, \gplus(\xperp, z; \xperp', z'; \omega) \, \bv{\psi}(\xperp', z'; k)  \;\; \text{ when } z > z',
\ee
which can be though of as a paraxial version of Huygens' principle, for downstream wave propagation.  In fact this is equivalent to the usual Fresnel diffraction integral, and reveals that a paraxial radiation field is in principle determined everywhere by knowledge of just the transverse components of the envelope of the vector potential in just one transverse plane.

Formally, the same vector potential envelope $\bv{\psi}_{}(\xperp, z; k)$ can also be written in terms of the upstream Green function, according to
\be
\begin{split}
\bv{\psi}_{}(\xperp, z; k) &=  \bv{\psi}_{\stext{out}}(\xperp, z; k) + \bv{\psi}_{u}(\xperp, z; k)\\
&= \bv{\psi}_{\stext{out}}(\xperp, z; k) + \tfrac{1}{i} \!\! \int\limits_{z}^{+\infty} \!\! dz' \! \int \! d^2\xperp' \, \gminus(\xperp, z; \xperp', z'; \omega) \, \s(\xperp', z'; k),
\end{split}
\ee
where $\bv{\psi}_{\stext{out}}(\xperp, z; k)$ represents free-space fields present downstream of the sources, then propagated everywhere else according to the free-space wave equation.  As it depends on downstream ``final'' boundary conditions rather than upstream ``initial'' boundary conditions, this construction is less useful in practice, but will be important to our formalism.  In the absence of intervening sources, $\bv{\psi}(\xperp, z; k)$ in some upstream transverse plane can be specified in terms of information on a downstream plane $z'$, where $z' > z$, by using $\gminus$ as a propagator, resulting in 
\be
\bv{\psi}_{}(\xperp, z; k) = -\!\int d^2 \! \xperp' \, \gminus(\xperp, z; \xperp', z'; \omega) \, \bv{\psi}(\xperp', z'; k)   \;\; \text{ when } z < z',
\ee
another paraxial version of Huygens' principle, but for backwards, or upstream inference of right-moving waves.

Right-moving paraxial radiation fields, satisfying the source-free paraxial wave equation everywhere, can be defined in terms of the difference
\be
\begin{split}
\bv{\psi}_{\stext{rad}}(\xperp, z; k) &= \bv{\psi}_{\stext{out}}(\xperp, z; k)  - \bv{\psi}_{\stext{in}}(\xperp, z; k) \\
&=  \bv{\psi}_{d}(\xperp, z; k) -  \bv{\psi}_{d}(\xperp, z; k) \\
&=  \tfrac{2}{i} \!\! \int\limits_{-\infty}^{+\infty} \!\!dz' \! \int \! d^2\xperp' \, \D(\xperp, z; \xperp', z'; \omega) \, \s(\xperp', z'; k),
\end{split}
\ee
where
\be
\D(\xperp, z; \xperp', z'; \omega)  = \tfrac{1}{2} 
\bigl[ \gplus(\xperp, z; \xperp', z'; \omega) - \gminus(\xperp, z; \xperp', z'; \omega)\bigr]
= \tfrac{k }{2\pi i [z-z']} \, e^{\frac{+ik\abs{ \xperp - \xperp'}^2}{ 2[z-z']}}
\ee
is the (right-moving) paraxial radiation kernel, which satisfies the source-free, right-moving,  paraxial wave equation
\be
\bigl[ +i \tfrac{\del}{\del z} + \tfrac{1}{2k} \laplacian_{\perp} \bigr] \D(
\bv{x}, \bv{x}'; \omega) = 0
\ee
everywhere in space, including as $z \to z'$, although $\D(\bv{x}, \bv{x}'; \omega)$ will approach a Dirac delta function $\delta(\xperp - \xperp')$ in the resulting singular limit---in the weak sense of distributions, not pointwise convergence.

\subsection{Power Balance and Other Relations}

In thinking about energy balance in this right-traveling paraxial geometry, it will be natural to integrate over a cylinder centered on the optic axis, and then consider the limit as both the length and radius of this cylinder increase without bound.  Because paraxial fields of finite power will decay rapidly in transverse distance away from the optic axis, only the ``end caps'' and not the sides of the cylinder will contribute non-vanishing flux in the infinite limit.  The end result is that we should continue to integrate our volumetric inner product over all space, but integrate the surface products only over transverse planes infinitely far upstream and/or downstream.

Specifically, the leading-order, right-moving, paraxial Poynting vector can be defined as
\be
\tfrac{1}{\muz} \bv{E}\cc \!\times\! \bv{B} = \tfrac{1}{\muz} \bigl[ i \omega\, \bv{\psi} e^{i kz } \bigr]\cc \!\times\!   \bigl[ ik 
\unitvec{z} \!\times\! \bv{\psi} e^{i kz } \bigr] =   +\tfrac{\omega^2}{\muz c} \abs{\bv{\psi} }^2 \unitvec{z},
\ee
which satsifies
\be\label{pdiv2}
\begin{split}
\divergence \tfrac{1}{\muz} (\bv{E}\cc \!\times\! \bv{B} ) &=  \tfrac{\omega^2}{\muz c}  \tfrac{\del}{\del z}  \abs{\bv{\psi} }^2 =  \tfrac{\omega^2}{\muz c}\bigr[ \bv{\psi}\cc \!\cdot\! \tfrac{\del \bv{\psi}}{\del z} + \bv{\psi} \!\cdot\! \tfrac{\del \bv{\psi}\cc}{\del z} \bigr] \\
&=  \tfrac{\omega^2}{\muz c} \tfrac{i}{2k} \bigl[ \bv{\psi}\cc \! \cdot \laplacian_{\perp} \bv{\psi} -  \bv{\psi} \! \cdot \laplacian_{\perp} \bv{\psi}\cc \bigr]
+  \tfrac{\omega^2}{\muz c}  i \bigl[ \bv{\psi} \! \cdot \s\cc -  \bv{\psi}\cc \!\cdot \s  \bigr].
\end{split}
\ee
But  elementary vector identities confirm that
\be
\bigl[ \bv{\psi}\cc \! \cdot \laplacian_{\perp} \bv{\psi} -  \bv{\psi} \! \cdot \laplacian_{\perp} \bv{\psi}\cc \bigr] = \grad_{\perp} \!\cdot \bigl[ (\bv{\psi}\cc \! \cdot \grad_{\perp} )\,\bv{\psi}
- (\bv{\psi} \! \cdot \grad_{\perp}) \,\bv{\psi}\cc  \bigr]
\ee
is a pure transverse divergence.  Assuming the Poynting flux (spectral density) in the beam is finite in any transverse plane, in the sense that
\be
\int \! d^2\xperp \, \bv{\psi}(\xperp,z; k )\cc \! \cdot \bv{\psi}(\xperp, z; k) < \infty
\ee
for any fixed $z$ (and fixed $k$), we may be assured that $\bv{\psi}(\xperp,z; k)$ will decay sufficiently rapidly at large transverse positions so that
\be
\lim\limits_{r_{\perp} \to \infty} \int \! d\theta \, \xperp \! \cdot \bigl[ \bv{\psi}\cc \! \cdot \grad_{\perp} \bv{\psi} -  \bv{\psi} \! \cdot \grad_{\perp} \bv{\psi}\cc \bigr] = 0,
\ee
where $r_{\perp} = \abs{\xperp} = \sqrt{x^2 + y^2}$ is the transverse distance from the optic axis, and $\theta = \tan\inv\bigl( \tfrac{y}{x} \bigr)$ is the azimuthal angle around the optic axis.  Therefore, upon integrating \eqref{pdiv2} over a right circular cylinder centered on the $\unitvec{z}$ axis, applying Gauss's law, and  taking the limit as both the length and radius of this cylinder go to infinity, contributions to energy flux through the curved sides of the cylinder vanish, while contributions from the transverse end-caps need not.

As a consequence, a frequency-domain, ``paraxial Poynting theorem'' says
\be
-\realpart \innerpd{ \bv{E}_{\perp} }{ \bv{J}_{\perp} }_{\Theta} =  \tfrac{1}{\muz} \realpart \innerpr{ \bv{E}_{\perp} }{ \bv{B}_{\perp} }_{\Theta} = \tfrac{1}{\muz} \realpart \innerpr{ \bv{E}_{\perp} }{ \bv{B}_{\perp} }_{\Theta},
\ee
in which
\be
\begin{split}
\innerpd{ \bv{E}_{\perp} }{ \bv{J}_{\perp} }_{\Theta} &= \lim\limits_{Z \to \infty} \int\limits_{-Z}^{+Z} \!\! dz  \! \int \! d^2\xperp\, \bv{E}_{\perp}(\xperp,z; \omega)\cc  \cdot \bv{J}_{\perp} (\xperp, z; \omega) \\
&= -2i \tfrac{\omega^2}{\muz c} \lim\limits_{Z \to \infty} \int\limits_{-Z}^{+Z} \!\! dz  \! \int \! d^2\xperp\, \bv{\psi}(\xperp,z; k)\cc \cdot \s(\xperp, z; k),
\end{split}
\ee
and
\be
\begin{split}
\tfrac{1}{\muz}\innerpr{\bv{E}_{\perp}}{\bv{B}_{\perp}}_{\Theta} = 
\lim\limits_{Z \to \infty} \Bigr[\, &+ \!\!\!\! \int\limits_{z = +Z} \!\!\! d^2 \xperp \; \unitvec{z} \!\cdot\! \bigl( \bigl[ i\omega \, \bv{\psi}(\xperp, z; k)  \bigr]\cc \times \bigl[ ik \unitvec{z} \! \times \!  \bv{\psi}(\xperp, z; k) \bigr]   \bigr)    \\
&- \!\!\! \int\limits_{z = -Z} \!\!\! d^2 \xperp \; \unitvec{z} \!\cdot\! \bigl( \bigl[ i\omega \, \bv{\psi}(\xperp, z; k)  \bigr]\cc \times \bigl[ ik \unitvec{z} \! \times \!  \bv{\psi}(\xperp, z; k) \bigr]   \bigr)  \,\Bigr] \\
= \tfrac{\omega^2}{c \muz}  \lim_{Z \to \infty} \Bigr[\,  &+ \!\!\! \int\limits_{z = +Z} \!\!\!  d^2 \xperp \, \abs{ \bv{\psi}(\xperp, z; k)  }^2 \,- \!\! \int\limits_{z = -Z} \!\!\!  d^2 \xperp \, \abs{ \bv{\psi}(\xperp, z; k)  }^2 \, \Bigr] .
\end{split}
\ee
Within the paraxial approximation, for the downstream fields (i.e., those fields generated from the actual transverse sources using the downstream Green function), no electromagnetic power enters or leaves the transverse plane infinitely far upstream ($z \to -\infty$) from the sources, so 
\be
-\realpart \innerpd{ \bv{E}_{d} }{ \bv{J}_{\perp} }_{\Theta} = \tfrac{1}{\muz}\innerpr{\bv{E}_{d}}{\bv{B}_{d}}_{\Theta} = \tfrac{\omega^2}{c \muz}  \lim_{Z \to +\infty} \! \int\limits_{z = Z} \!\!  d^2 \xperp \, \abs{ \bv{\psi}_d(\xperp, z; k)  }^2  \ge 0;
\ee
but for the upstream fields (i.e., those generated from the actual transverse sources by the upstream Green function), no electromagnetic power would enter or leave the transverse plane infinitely far downstream from the sources ($z \to +\infty$), so that
\be
-\realpart \innerpd{ \bv{E}_{u} }{ \bv{J}_{\perp} }_{\Theta} = \tfrac{1}{\muz}\innerpr{\bv{E}_{u}}{\bv{B}_{u}}_{\Theta} = -\tfrac{\omega^2}{c \muz}  \lim_{Z \to -\infty} \! \int\limits_{z = Z} \!\!  d^2 \xperp \, \abs{ \bv{\psi}_u(\xperp, z; k)  }^2  \le  0.
\ee
Because the source-free paraxial equation leads to quantum-like unitary propagation (with respect to longitudinal position) of the field profile in successive transverse planes, the (spectral density) of Poynting flux for any radiation fields at any given frequency must be the same in every transverse plane, such that
\be\label{invariancep1}
\abs{ \bv{\psi}_{\stext{rad}}(\xperp, z_1; k)  }^2  = \abs{ \bv{\psi}_{\stext{rad}}(\xperp, z_2; k)  }^2,
\ee
including in the limits as $z_1 \to -\infty$ and $z_2 \to +\infty$, so 
\be
\begin{split}
\tfrac{1}{\muz}\innerpr{\bv{E}_{\stext{rad}}}{\bv{B}_{\stext{rad}} }_{\Theta} 
&= \tfrac{\omega^2}{c \muz} \! \int  \!  d^2 \xperp \, \abs{ \bv{\psi}_{\stext{rad}}(\xperp, +\infty; k)  }^2 \,- \tfrac{\omega^2}{c \muz} \! \int \!  d^2 \xperp \, \abs{ \bv{\psi}_{\stext{rad}}(\xperp, -\infty; k)  }^2 = 0.
\end{split}
\ee
But far upstream, we have
\bsub
\begin{align}
\bv{\psi}_{{d}}(\xperp, -\infty; k) &=  \lim\limits_{z \to -\infty} \bv{\psi}_{{d}}(\xperp, z; k) =  \bv{0},\\
\bv{\psi}_{\stext{rad}}(\xperp, -\infty; k) &=   \lim\limits_{z \to -\infty} \bv{\psi}_{\stext{rad}}(\xperp, z; k) =    \lim\limits_{z \to -\infty} \bv{\psi}_{u}(\xperp, z; k) = \bv{\psi}_{{u}}(\xperp, -\infty; k),
\end{align}
\esub
while far downstream,  
\bsub
\begin{align}
\bv{\psi}_{{u}}(\xperp, +\infty; k) &=   \lim\limits_{z \to +\infty} \bv{\psi}_{{u}}(\xperp, z; k = \bv{0},\\
\bv{\psi}_{\stext{rad}}(\xperp, +\infty; k) &=  \lim\limits_{z \to +\infty} \bv{\psi}_{\stext{rad}}(\xperp, z; k) = \lim\limits_{z \to +\infty} \bv{\psi}_{d}(\xperp, z; k) = \bv{\psi}_{{d}}(\xperp, +\infty; k),
\end{align}
\esub
so we find
\be
\tfrac{1}{\muz}\innerpr{\bv{E}_{d}}{\bv{B}_{d}}_{\Theta} = -\tfrac{1}{\muz}\innerpr{\bv{E}_{u}}{\bv{B}_{u}}_{\Theta}
\ee
for the Poynting flux of the upstream and downstream components of the radiation fields associated with a given current density, and we may further infer that
\be
\begin{split}
-\realpart \innerpd{ \bv{E}_{\stext{rad}} }{ \bv{J}_{\perp} }_{\Theta}
&= -\realpart \innerpd{ [\bv{E}_d - \bv{E}_u ]}{ \bv{J}_{\perp} }_{\Theta} \\
&= -\realpart \innerpd{ \bv{E}_d }{ \bv{J}_{\perp} }_{\Theta}  + \realpart \innerpd{ \bv{E}_u }{ \bv{J}_{\perp} }_{\Theta} \\
&= + \tfrac{1}{\muz}\innerpr{\bv{E}_{d}}{\bv{B}_{d}}_{\Theta} -  \tfrac{1}{\muz}\innerpr{\bv{E}_{u}}{\bv{B}_{u}}_{\Theta}
=  +2\tfrac{1}{\muz}\innerpr{\bv{E}_{d}}{\bv{B}_{d}}_{\Theta} \ge 0.
\end{split}
\ee
Although we have defined the Poynting flux in terms of a far-field limit, in order to best match up with how we think about energy conservation in the non-paraxial case, one particularly useful feature of the paraxial geometry is that for any radiation fields satisfying the source-free paraxial wave equation and gauge constraint, the (spectral density) of energy flux will be the same in any transverse plane, as asserted in equation \eqref{invariancep1}, and so can be calculated wherever is most convenient, without necessarily effecting any $z \to \infty$ limit.  Often, it will be easier to perform the needed integration in the focal plane of the optical beam, where wavefront curvature vanishes, than in an asymptotic $z \to \infty$ far-field limit, where wavefronts become nearly spherical.

%  if the outgoing (and rightward) radiated power spectral density is calculated to paraxial accuracy over an arbitrarily remote downstream transverse plane, \ie,
% and the mechanical power spectral density exchanged between solenoidal fields and sources is also calculated to the same order in $\Theta,$ \ie,

\subsection{Reciprocity and Surjectivity of the Paraxial Radiation Kernel}

Surjectivity of the radiation kernel is somewhat easier to establish in the paraxial case than in the general three-dimensional case.  Again, this is because paraxial propagation between successive transverse planes itself is unitary, and hence always invertible, such that right-moving, paraxial radiation fields (for any given $\omega$) will be determined uniquely everywhere in space by just specifying $\psi(\xperp,z; k)$ in any one transverse plane, labeled by the longitudinal coordinate $z$.

So if we choose as an effective source $\s(\xperp', z'; k) = \tfrac{i}{2} \, \delta(z'-z_0)\, \bv{\psi}(\xperp',z_0; k)$, then the corresponding paraxial radiation envelope,
\be
\begin{split}
\bv{\psi}_{\stext{rad}}(\xperp,z;k) &= \tfrac{2}{i} \! \int \! dz'   \! \int \! d^2\xperp' \, \D(\xperp,z; \xperp',z'; k) \,\s(\xperp', z'; k)  \\
&=  \int \! d^2\xperp' \, \D(\xperp,z; \xperp',z_0; k) \,\bv{\psi}(\xperp', z_0; k) ,
\end{split}
\ee
satisfies the source-free paraxial equation everywhere, as well as the boundary condition
\be
\begin{split}
\bv{\psi}_{\stext{rad}}(\xperp,z_0;k)  &= \int \! d^2\xperp' \, \D(\xperp,z_0; \xperp',z_0; k) \,\bv{\psi}(\xperp', z_0; k)  \\
&= \int \! d^2\xperp' \, \delta(\xperp - \xperp') \,\bv{\psi}(\xperp', z_0; k)  
= \bv{\psi}(\xperp, z_0; k).
\end{split}
\ee
Moreover, if we demand that
\be
\unitvec{z} \!\cdot \! \bv{\psi}(\xperp,z_0;k) = \epsilon \, \tfrac{i}{k} \grad_{\perp} \!\cdot \bv{\psi}(\xperp,z;k),
\ee
which imposes either the leading-order (for $\epsilon = 0$) or next-order (for $\epsilon = 1$) paraxial gauge condition in one transverse plane, then the same gauge condition will automatically hold in all transverse planes, because
\be
\begin{split}
\unitvec{z} \!\cdot\! \bv{\psi}_{\stext{rad}}(\xperp,z;k)  &= \int \! d^2\xperp' \, \D(\xperp,z; \xperp',z_0; k) \;\unitvec{z} \cdot \bv{\psi}(\xperp', z_0; k)  \\
&= \int \! d^2\xperp' \, \D(\xperp,z; \xperp',z_0; k) \,   \epsilon \, \tfrac{i}{k} \grad'_{\perp} \!\cdot \bv{\psi}(\xperp',z_0;k)  \\
&=   -\epsilon \, \tfrac{i}{k}  \!   \int \! d^2\xperp' \, \bigl[\grad'_{\perp} \D(\xperp,z; \xperp',z_0; k) \bigr] \cdot \bv{\psi}(\xperp,z_0;k)  \\ 
&=  +\epsilon \, \tfrac{i}{k}  \!   \int \! d^2\xperp' \, \bigl[\grad_{\perp} \D(\xperp,z; \xperp',z_0; k) \bigr] \cdot \bv{\psi}(\xperp,z_0;k)  \\
&=  +\epsilon \, \tfrac{i}{k} \grad_{\perp}  \! \cdot \! \int \! d^2\xperp' \,  \D(\xperp,z; \xperp',z_0; k) \, \bv{\psi}(\xperp,z_0;k)
= +\epsilon \, \tfrac{i}{k} \grad_{\perp}  \! \cdot  \bv{\psi}_{\stext{rad}}(\xperp,z;k),
\end{split}
\ee
where we have used some integration by parts, exploiting the facts that both the paraxial radiation kernel and paraxial radiation fields decay exponentially rapidly with respect to transverse distance (so as to maintain normalizability).  So we can see that the gauge condition is also propagated correctly from one transverse plane to the next.

It follows that, in principle, we can generate any source-free paraxial solution by convolution of the radiation kernel with some effective source.

Finally, we may deduce a conjugate-reciprocity property for the paraxial radiation fields and sources, analogous to that established in the full three-dimensional geometry.  Using the fact that the paraxial radiation kernel satisfies $\D(\xperp', z'; \xperp, z; k) = D(\xperp, z; \xperp', z'; k)\cc$, 
we find after a few elementary manipulations that
\be
\innerpd{ \bv{E}_{\stext{rad}} }{ \bv{J}'_{\perp} }_{\Theta}  = \innerpd{\bv{J}_{\perp} }{\bv{E}'_{\stext{rad}} }_{\Theta} = \innerpd{\bv{E}'_{\stext{rad}} }{\bv{J}_{\perp} }_{\Theta}\cc,
\ee 
so that $\realpart \innerpd{ \bv{E}_{\stext{rad}} }{ \bv{J}'_{\perp} }_{\Theta} = \realpart \innerpd{\bv{E}'_{\stext{rad}} }{\bv{J}_{\perp} }_{\Theta}$.

\subsection{Paraxial Version of the Variational Principle}

Starting with these mathematical ingredients, it follows (by arguments similar to those used in the general three-dimensional, free-space case) that the MPVP will also hold exactly within the paraxial framework, right down to the same factor of $\sfrac{1}{2}$ to account for the doubled power in the source-free fields compared to downstream ones.

So to formulate the paraxial MPVP to leading order, we simply replace the electromagnetic fields and work and flux integrals with their paraxial counterparts:
\bsub\label{pvp3}
\begin{align}
\tfrac{1}{\muz} \innerpr{\bv{E}_{\stext{ret}} }{\bv{B}_{\stext{ret}} }_{\Theta}  &\ge \max_{\bv{\alpha}} \bigl[\, \tfrac{1}{\muz} \innerpr{\bv{e}_{\stext{ret}} }{\bv{b}_{\stext{ret}} }_{\Theta} \, \bigr] \\
\text{such that}\!&: \; \tfrac{1}{\muz} \innerpr{\bv{e}_{\stext{ret}} }{\bv{b}_{\stext{ret}} }_{\Theta} = -\tfrac{1}{2} \realpart \innerpd{\bv{e}_{\stext{rad}} }{\bv{J}_{}}_{\Theta} , \\
\text{and}\!&: \; [+i \tfrac{\del}{\del z} + \tfrac{1}{2k} \laplacian_{\perp} ] \, \bv{\psi}(\xperp, z; k; \bv{\alpha}) = \bv{0} , \\
\text{and}\!&: \; \unitvec{z} \!\cdot\! \bv{\psi}(\xperp, z; k; \bv{\alpha}) = 0.
\end{align}
\esub
That is, the actual radiation fields radiate more outgoing power, and therefore extract more energy from the actual sources, than could any trial radiation field, if it were present in the vicinity of the sources.  To gain one additional order in the paraxial expansion, we merely need to add a longitudinal term of the form $\psi_z = \tfrac{i}{k} \grad_{\perp} \! \cdot \bv{\psi}(\xperp, z; k; \bv{\alpha})$ to the variational solution found in the lowest-order optimization.

In fact, the MPVP is most likely to find application in the paraxial regime, because source-free solenoidal trial functions can be more readily characterized and parameterized.  In the general three-dimensional   geometry, except for plane-wave or multipole expansions, few analytic solutions can be found that satisfy the source-free Maxwell equations everywhere, while in the paraxial limit, solutions are uniquely specified just by the carrier frequency and a (complex) square-integrable profile in any one transverse plane, which can be decomposed into a convenient, countable set of expansion modes, as in the familiar Gauss-Hermite or Gauss-Laguerre basis sets.  

%%%%

%%%%

\section{Linear Subspaces and Multipole Expansions}

Sometimes, but not always, the manifold of trial radiation fields may consist of a linear vector subspace, where the variational parameters are identified with the expansion coefficients in some basis spanning this subspace.  In such cases, the optimal variational solution may be seen as an orthogonal projection of the actual radiation fields into the subspace of trial radiation fields, where orthogonality is to be defined with respect to the ``Poynting'' inner-product associated with the far-field outgoing power.

Under such circumstances, the MPVP just reduces to a straightforward consequence of two simple criteria, namely \textit{Bessel's inequality}, which says that the electromagnetic power in any one source-free mode, or any finite superposition of orthogonal source-free modes, cannot exceed the power in all the modes, and an \textit{energy conservation} constraint, which dictates that the power radiated must be attributable to power delivered by the sources, even when self-consistent back-action is ignored.  Accuracy of the variational approximation may be anticipated to increase as the dimensionality of the subspace of trial fields increases.  If the space of trial fields includes the actual radiation fields, then the optimal variational solution becomes exact---but also exactly as difficult to calculate.

In full three-dimensional, free-space geometry, very few basis-sets of exact source-free solutions are known, the most familiar being either transverse plane waves, or multipole ``spherical waves'' which can be expressed in terms of vector spherical harmonics.  As the latter involve a countable orthonormal basis rather than continuous generalized basis, and allow for straightforward separation of ingoing and outgoing components and identification of the asymptotic far fields, and avoid some singularities which otherwise arise in a plane-wave expansion, it may be illuminating to briefly discuss the MPVP in the framework of such multipole expansions.

In principle, specification of the current density (everywhere in space, and for all relevant frequencies) uniquely determines the solenoidal part of the current density, which in turn determines the Coulomb-gauge vector potential assuming outgoing Sommerfeld boundary conditions.  

Decomposing the potentials or associated electromagnetic fields into multipolar contributions may be facilitated by expressing the Green function itself as a sum over spherical-wave contributions, in the form
\be
\gret(\bv{x}, \bv{x}'; \omega) = \sum\limits_{\ell = 0}^{\infty} 
j_{\ell}(kr_{<}) \, h^{+}_{\ell}(kr_{>}) \sum\limits_{m= -\ell}^{+\ell}Y_{\ell m}(\unitvec{r}')\cc \,Y_{\ell m}(\unitvec{r})
\ee
where 
\bsub
\begin{align}
r_{<} &= \min\bigl[\, | \bv{x} |, | \bv{x}' |\, \bigr] \\
r_{>} &= \max\bigl[\, | \bv{x} |, | \bv{x}' | \,\bigr]
\end{align}
\esub
are respectively the smaller and larger radial positions amongst the source and observation points, 
the functions
\be
Y_{\ell m}(\unitvec{r}) = Y_{\ell m}(\theta,\phi) = \sqrt{ \tfrac{2\ell+1}{4\pi} \tfrac{(\ell-m)!}{(\ell+m)!}}\, P^{\ell}_{m}(\cos\theta)\, e^{i m\phi}
\ee
are the usual spherical harmonics,\cite{jackson:75} written in terms of associated Legendre polynomials, 
and
\be
h^{+}_{\ell}(x) = (-x)^{\ell} (\tfrac{1}{x} \tfrac{d}{dx})^{\ell} \tfrac{e^{+ix}}{ix}
\ee
are the spherical Hankel functions of the first kind, representing outgoing waves, which ensure the correct asymptotic boundary conditions, while
\be
j_{\ell}(x) = (-x)^{\ell} (\tfrac{1}{x} \tfrac{d}{dx})^{\ell} \tfrac{\sin x}{x}
\ee
are the spherical Bessel functions, which are regular everywhere, including at the origin, and so can be integrated against the current sources in the interior region.

For real-valued position coordinates and frequencies,  the advanced Green function can just be obtained from this causal Green function by complex conjugation, such that
\be
\gret(\bv{x}, \bv{x}'; \omega) = \sum\limits_{\ell = 0}^{\infty} 
j_{\ell}(kr_{<}) \, h^{-}_{\ell}(kr_{>}) \sum\limits_{m= -\ell}^{+\ell}Y_{\ell m}(\unitvec{r}')\cc \,Y_{\ell m}(\unitvec{r}),
\ee
where
\be
h^{-}_{\ell}(x) = -(-x)^{\ell} (\tfrac{1}{x} \tfrac{d}{dx})^{\ell} \tfrac{e^{-ix}}{ix}
\ee
are the spherical Hankel functions of the second kind, representing ingoing spherical waves (and for any real-valued argument, just equal to the complex conjugate of the outgoing spherical Hankel functions).

The associated radiation kernel is just half the difference between these Green functions,
\be
\dop(\bv{x}, \bv{x}'; \omega) = i \,\sum\limits_{\ell = 0}^{\infty} 
j_{\ell}(kr') \, j_{\ell}(kr) \sum\limits_{m = -\ell}^{+\ell}Y_{\ell m}(\unitvec{r}')\cc \,Y_{\ell m}(\unitvec{r}),
\ee
which is indeed a bounded and otherwise well-behaved solution to the source-free Helmholtz equation everywhere in space, including at the origin.

While we have expressed the scalar Green functions in terms of scalar spherical waves, in order to decompose the associated vector potentials or electromagnetic fields into multipolar contributions, it will be convenient to employ vector spherical harmonics, which elegantly decompose both the spatial and polarization dependence into contributions which transform irreducibly under rotations.  Exterior to the actual sources $\bv{J}(\bv{x}; \omega)$, the causal, Coulomb-gauge vector potential can be written in the form
\be
\bv{A}_{\stext{ret}}(\bv{x}; \omega) = \sum\limits_{\ell = 0}^{\infty} \sum\limits_{m= -\ell}^{+\ell} \Big\{
 a^E_{\ell m}(\omega)\, \tfrac{1}{ik}\grad\!\times\![ h^{+}_{\ell}(kr) \,\bv{X}_{\ell m}(\unitvec{r})]  
+  a^M_{\ell m}(\omega)\, h^{+}_{\ell}(kr) \, \bv{X}_{\ell m}(\unitvec{r})
\Bigr\},
\ee
where the $\bv{X}_{\ell m}(\unitvec{r})$ are solenoidal vector spherical harmonics,\cite{jackson:75,george_gamliel:90,morehead:01} as defined and discussed in Appendix \ref{appendix_vsh}.

Such superpositions satisfy the outgoing Sommerfeld boundary conditions asymptotically as $kr \to \infty$, but are actually solutions to the source-free, frequency-domain Maxwell equations everywhere except right at the origin, where the Hankel functions blow up.  That is to say, the corresponding electromagnetic fields $\bv{E}_{\stext{ret}}(\bv{x}; \omega)$ and $\bv{B}_{\stext{ret}}(\bv{x}; \omega)$ correspond in the asymptotic far field to the outgoing fields actually radiated by the actual sources, but can be evaluated at any non-zero radial position, where they may be interpreted as the actual fields extrapolated backwards  from the far field according to free-space propagation, as if the actual sources were replaced with an effective point source at the origin that would reproduce the same far-field radiation pattern.

An advanced vector potential that reverses the flow of the asymptotic far fields is just
\be
\bv{A}_{\stext{adv}}(\bv{r}; \omega) = \sum\limits_{\ell = 0}^{\infty} \sum\limits_{m= -\ell}^{+\ell} \Big\{
 a^E_{\ell m}(\omega)\, \tfrac{1}{ik}\grad\!\times\![ h^{-}_{\ell}(kr) \,\bv{X}_{\ell m}(\unitvec{r})]  
+  a^M_{\ell m}(\omega)\, h^{-}_{\ell}(kr) \, \bv{X}_{\ell m}(\unitvec{r})
\Bigr\},
\ee
with the same expansion coefficients, but involving incoming spherical waves, while the radiation from the actual source can be determined as the difference between these retarded and advanced spherical wave expansions, such that
\be
\bv{A}_{\stext{rad}}(\bv{r}; \omega) = 2i \sum\limits_{\ell = 0}^{\infty} \sum\limits_{m= -\ell}^{+\ell} \Big\{
 a^E_{\ell m}(\omega)\, \tfrac{1}{ik}\grad\!\times\![ j_{\ell}(kr) \,\bv{X}_{\ell m}(\unitvec{r})]  
+  a^M_{\ell m}(\omega)\, j_{\ell}(kr) \, \bv{X}_{\ell m}(\unitvec{r}),
\Bigr\},
\ee
which is a well-behaved solution to the free-space Maxwell equations everywhere, including at the origin.  In fact, whereas the forms of the causal and advanced ``extrapolants'' depend on where we have situated the origin (because, when evaluated in the near field, in effect they replace the actual sources with an equivalent point source located at the chosen origin), the radiation vector field will be uniquely determined everywhere, independent of the choice of the origin, despite appearances to the contrary.

Using various differential and orthogonality properties of the vector spherical harmonics, as well as the asymptotic form for the spherical Hankel functions, namely
\be
h^{+}_{\ell}(kr) \to (-i)^{\ell+1} \, \tfrac{ e^{+ikr}}{kr} \; \text{ as } kr \to +\infty,
\ee
a straightforward calculation confirms that the ``Poynting'' inner product between outgoing multipolar electric and magnetic fields can be written as
\be
\tfrac{1}{\muz} \innerpr{ \bv{E} }{ \bv{B}' } =  \tfrac{c}{\muz} \sum\limits_{\ell} \sum\limits_{m} \bigl[
 a^E_{\ell m}(\omega)\cc a^E{}'_{\ell m}(\omega) +  a^M_{\ell m}(\omega)\cc a^M{}'_{\ell m}(\omega) \bigr],
\ee
which is just proportional to the familiar $l^2$ inner product involving a sum over products of corresponding multipole expansion coefficients.  In this context, applying the MPVP with these coefficients interpreted as variational parameters would just reduce to calculating overlap integrals between the source $\bv{J}(\bv{x},\omega)$ and the vector spherical harmonics $j_{\ell}(kr) \,\bv{X}_{\ell m}(\unitvec{r})$ or $\grad\!\times\![ j_{\ell}(kr) \,\bv{X}_{\ell m}(\unitvec{r})]$, in order to determine the corresponding expansion coefficients themselves.

Finally, we may note that in order to translate between spherical wave and plane wave representations, one can make use of the well-known Bessel plane-wave expansion formula,
\be
e^{i \bv{k} \cdot \bv{x}} = 4\pi \,\sum\limits_{\ell = 0}^{\infty} i^{\ell} \, j_{\ell}(kr)  \sum\limits_{m = -\ell}^{+\ell}Y_{\ell m}(\unitvec{r})\cc \,Y_{\ell m}(\unitvec{k})
\ee
where $r = | \bv{x} |$, and $k = | \bv{k} |$.

% For propagation in vacuum, the MPVP is somewhat limited in its usefulness by the constraint that trial solutions must be solenoidal source-free solutions to the Helmholtz equation.  In three-dimensional space, convenient closed-form analytic solutions are few, and if the trial radiation fields do not exactly satisfy these source-free Maxwell equations, the results may still be of approximate validity, but the strict lower bound on the radiated power may be lost.

\section{Time Domain}

The MPVP has been established so far in the frequency domain, for radiation fields propagating in free-space (apart from the sources), in either full three-dimensional geometry or paraxial geometry.  While  analogous versions can be proven directly in the time domain, it is perhaps simpler to rely on the frequency-domain results plus some unitary Fourier transforms.  Because the MPVP holds with respect to spectral densities of work and energy flux, separately in each infinitesimal frequency interval, it will also hold \textit{a fortiori} when integrated over any frequency band, and therefore must also hold true in the (integrated) time domain, as a consequence of the Parseval-Plancherel identity governing inner products, and an analog for cross products.

We may start by making various weak technical assumptions about good behavior, such that: all Cartesian components of physical fields are real-valued in the space and time domain;  the integrals needed for various Fourier transforms and functional inner products exist, while the order of iterated integrations, and of any various integrations and differentiations, can be commuted, while integrations by parts can be performed.  Upon integrating over all real frequencies (now including negative frequencies, assuming physical field components are all real), we find
\be
\begin{split}
\int \! d{\omega} \! \int \! d^3\bv{x}\, & \bv{E}_{} (\bv{x}; \omega)\cc \!\cdot\! \bv{J}(\bv{x}; \omega) =
\int \! d^3\bv{x} \! \int \! d{\omega}\, \bv{E}_{} (\bv{x}; \omega)\cc \!\cdot\! \bv{J}(\bv{x}; \omega) 
= \int \! d^3\bv{x} \int \! dt \, \bv{E}_{} (\bv{x}; t)\cc \!\cdot\! \bv{J}(\bv{x}; t) \\
&= \int \! d^3\bv{x}  \int \! dt \, \bv{E}_{} (\bv{x}; t) \!\cdot\! \bv{J}(\bv{x}; t) = \int \! dt  \int \! d^3\bv{x} \, \bv{E}_{} (\bv{x}; t) \!\cdot\! \bv{J}(\bv{x}; t).
\end{split}
\ee
An analogous identity holds for Fourier transforms of cross products, such that
\be
\begin{split}
\int \! d{\omega} \! \int \! d^2\bv{a} & \cdot \bigl[ \bv{E}_{} (\bv{x}; \omega)\cc \!\times\! \bv{J}(\bv{x}; \omega) \bigr] = 
 \int \! d^2\bv{a} \cdot \Bigl[ \int \! d{\omega}\, \bv{E}_{} (\bv{x}; \omega)\cc \!\times\! \bv{J}(\bv{x}; \omega) \Bigr] \\
&= \int \! d^2\bv{a} \cdot \Bigl[ \int \! d{t}\, \bv{E}_{} (\bv{x}; t)\cc \!\times\! \bv{J}(\bv{x}; t) \Bigr] = \int \! d^2\bv{a} \cdot \bigl[ \int \! d{t}\, \bv{E}_{} (\bv{x}; t) \!\times\! \bv{J}(\bv{x}; t) \bigr] \\
 &= \int \! d{t} \int \! d^2\bv{a} \cdot \bigl[ \bv{E}_{} (\bv{x}; t) \!\times\! \bv{J}(\bv{x}; t) \bigr] .
\end{split}
\ee
For time-frequency transform pairs, solenoidal and irrotational characteristics are preserved by Fourier transforms, so
\bsub
\begin{align}
\grad \!\cdot\! \bv{A}(\bv{x},t) &= {0} \text{ for all }  t \in \realsymbol \;\;\textit{ if and only if }\;\; \grad \!\cdot\! \bv{A}(\bv{x},\omega) = {0} \text{ for all } \omega \in \realsymbol \\
\grad\!\times\! \bv{A}(\bv{x},t) &= \bv{0}  \text{ for all }  t \in \realsymbol \;\;\textit{ if and only if }\;\; \grad\!\times\! \bv{A}(\bv{x}, \omega) = \bv{0} \text{ for all } \omega \in \realsymbol.
\end{align}
\esub
Also, a vector field satisfying the homogeneous Helmholtz equation at all real frequencies
will satisfy the homogeneous wave equation (d'Alembert's equation) at all times:
\be
\bigl(\laplacian - \tfrac{\del^2}{\del t^2} \bigr) \bv{A}(\bv{x}, t) = 0 \text{ for all }  t \in \realsymbol \;\;\textit{ if and only if }\;\;
\bigl(\laplacian +  \tfrac{\omega^2}{c^2}\bigr) \bv{A}(\bv{x}, \omega) = 0 \text{ for all } \omega \in \realsymbol.
\ee
We may then infer that, because the variational inequality upon which the MPVP relies holds true locally in the frequency domain, that is, separately at each real frequency $\omega$, it will also hold true in the time domain globally, that is, when integrated overall all real times $t$.

It can sometimes be more convenient to work, or think, in the time domain rather than in the frequency domain, depending on the nature of the Joule work and Poynting flux integrals.  An additional integration over time will be required, while the integrations must still be performable or approximable at different values of the adjustable variational parameters appearing in the trial radiation solutions.  On the other hand, if the current sources are given in the time domain, we can avoid having to calculate their Fourier transforms, as long as we can also express the trial radiation fields in the time domain.

A time-domain picture can also illuminate the differences between radiative and reactive fields, to which we alluded above.  Consider the total energy exchanged between the sources and fields, calculated in terms of the positive or negative work performed on the moving charges by the electric fields (under our assumptions that the charges still follow prescribed trajectories).  Because irrotational and solenoidal vector fields are functionally orthogonal when integrated over all space, we may first decompose the overall time-domain work integral as
\be
\int \!dt \! \int \! d^3\bv{x}\; \bv{J}(\bv{x},t)\! \cdot\! \bv{E}(\bv{x},t) = \int \! dt \! \int \! d^3\bv{x}\; \bv{J}_{\|}(\bv{x},t) \!\cdot\! \bv{E}_{\|}(\bv{x},t) + \int \! dt \! \int \! d^3\bv{x} \; \bv{J}_{\perp}(\bv{x},t) \!\cdot\! \bv{E}_{\perp}(\bv{x},t),
\ee
where
\be
\begin{split}
-\!\int \! dt \! &\int \! d^3\bv{x} \; \bv{J}_{\|}(\bv{x},t) \!\cdot\! \bv{E}_{\|}(\bv{x},t) =  \Bigl. \tfrac{1}{2} \! \int \! d^3\bv{x} \; \rho(\bv{x},t)  \abs{ \phi(\bv{x},t) }^{2} \, \Bigr\rvert_{t = -\infty}^{t = +\infty}
\end{split}
\ee
is just equal to the net change in the instantaneous Coulomb potential energy due to any overall rearrangement in the relative positions of the charges between the remote past and future, and is clearly not associated with any irreversible radiative energy transport, both because these Coulomb fields remain tied to the source charges, and because their rapid $O(\sfrac{1}{r^2})$ fall-off implies that these field components will not contribute to the Poynting flux in the limit of infinitely remote bounding surfaces.  

The work integral involving the solenoidal fields and solenoidal sources can be further decomposed, as
\be
\begin{split}
 \int \! \, d^3\bv{x} \, \bv{J}_{\perp}(\bv{x},t) \!\cdot\! \bv{E}_{\perp}(\bv{x},t)
&= \int dt \! \int \! d^3\bv{x} \, \bv{J}_{\perp}(\bv{x},t) \!\cdot\! \bar{\bv{E}}_{\perp}(\bv{x},t)
+  \tfrac{1}{2} \int \! dt \!\int\!  d^3\bv{x} \, \bv{J}_{\perp}(\bv{x},t) \!\cdot\! \bv{E}_{\stext{rad}}(\bv{x},t) ,\\
\end{split}
\ee
but using the spatiotemporal symmetries of the time-symmetric Green function $\gbar(\bv{x}, t' \bv{x}', t')$, it is straightforward to show that
\be
\lim\limits_{T \to \infty} \int\limits_{-T}^{+T} \!  dt \! \int \! d^3\bv{x} \; \bv{J}_{\perp}(\bv{x},t) \!\cdot\! \bar{\bv{E}}_{\perp}(\bv{x},t) = 0,
\ee
at least in a principal value sense as indicated, verifying that the reactive solenoidal fields really do not participate in any irreversible net exchange of energy between the sources and fields, but rather the integral $\int \! d^3\bv{x} \, [\bv{E} \cdot \bar{\bv{J}}]$ simply represents a reactive power associated with the rate at which the sources reversibly store energy in or recover energy from the non-radiative fields in their vicinity.

%\bsub\label{vp6}
%\begin{align}
%\tfrac{1}{\muz} \innerpr{\bv{E}_{\stext{ret}} }{\bv{B}_{\stext{ret}} }  &\ge \max_{\bv{\alpha}} \bigl[\,
% \tfrac{1}{\muz} \innerpr{\bv{e}_{\stext{ret}} }{\bv{b}_{\stext{ret}} } \, \bigr] \\
%
%\text{such that}\!&: \text{ } \tfrac{1}{\muz} \innerpr{\bv{e}_{\stext{ret}} }{\bv{b}_{\stext{ret}} } = -\tfrac{1}{2} \realpart \innerpd{\bv{e}_{\stext{rad}} }{\bv{J}_{}} , \\
%
%\text{and}\!&: \text{ } \grad \! \times \!  \grad \! \times \! \bv{e}_{\stext{rad}}(\bv{x}; \omega; \bv{\alpha}) = k^2 \, \bv{e}_{\stext{rad}}(\bv{x}; \omega; \bv{\alpha}).
%\end{align}
%\esub

\section{Partial Coherence}

So far, we have implicitly assumed that the source $\bv{J}(\bv{x}; \omega)$ or $\bv{J}(\bv{x}; t)$ is fully prescribed and deterministic, with no residual uncertainty, jitter, or randomness.  At all positions and frequencies, the resulting variational approximation to the radiation will be coherent in the sense of having well-defined phases, or in being representable as a definite linear superposition over a set of modes.  But in many contexts, in characterizing the emitted radiation fields, we would want to account for effects of statistical uncertainty, underspecification, and/or fluctuations in the sources.  In other words, we may be interested in \textit{partially coherent} radiation fields.

Regarding issues of optical coherence, notice that the processes of averaging over any statistical uncertainty in the charged particle trajectories constituting the source $\bv{J}(\bv{x}; t)$, and of performing the variational optimization over adjustable parameters, will not generally commute, if any variational parameters appear nonlinearly in the trial radiation fields.  Furthermore, whether the parameters appear linearly or nonlinearly or both, when applied to a definite, deterministic source, or directly to an averaged source, the MPVP procedure will naturally produce optimized expansion coefficients for fields with definite phase relationships between different modes, rather than any sort of statistical mixture over modes.

So some care will be required if partially coherent radiation is to be modeled, particularly if quantities such as degrees of optical coherence, coherence times and longitudinal or transverse coherence lengths, interference fringe visibilities, or optical emittances are of interest.  If the averaged source $\bigl\langle \bv{J}(\bv{x}; \omega) \bigr\rangle$ is used directly as input to an MPVP optimization, then as a consequence of the linearity of Maxwell's equations, the resulting variational fit would approximate the so-called coherent component of the radiation fields, equal to the expectation value $\bigl\langle \bv{E}_{\stext{rad}}(\bv{x}; \omega) \bigr\rangle$, but higher-order moments may actually be of more interest.  For instance, we may seek to estimate the average Poynting vector $\tfrac{1}{\muz}\bigl\langle \bv{E}(\bv{x}; \omega)\cc \!\times\! \bv{B}(\omega) \big\rangle$, but when applied to $\langle \bv{J}(\bv{x}; \omega) \rangle$, the MPVP naturally generates approximations instead to $\tfrac{1}{\muz} \bigl\langle \bv{E}(\bv{x}; \omega)\cc \bigr\rangle \!\times\! \bigl\langle \bv{B}(\omega) \bigr\rangle$, the Poynting vector associated with the averaged fields.  With any appreciable statistical uncertainty or fluctuations, these will not be equal in general, and indeed, very often $\abs{ \bigl\langle  \bv{E}_{\stext{rad}}(\bv{x}; \omega) \bigr\rangle }^2 \ll \bigl\langle \bigl\lvert  \bv{E}_{\stext{rad}}(\bv{x};\omega) \bigr\rvert^2  \bigr\rangle$ for radiation from relativistic electron beams, because of shot noise or other effects.

Thus, to meaningfully apply the MPVP as formulated to the emission of partially coherent radiation, one should first optimize separately over different possible realizations of the source current density $\bv{J}(\bv{x}; \omega)$ (typically, but not necessarily, using the same family of trial radiation profiles), and only then average over these possibilities in any expressions which are nonlinear in the fields and/or sources.  In this manner, we can approximate, say, the second-order coherence tensor
\be
\Gamma_2(\bv{x}, \omega; \bv{x}', \omega') =  \bigl\langle \bv{E}_{\stext{rad}}(\bv{x}; \omega) \, \bv{E}_{\stext{rad}}(\bv{x}'; \omega')\hc \bigr\rangle,
\ee
(needed to describe standard two-point interference experiments), or higher-order coherence tensors such as $\Gamma_4(\bv{x}, \omega; \bv{x}, \omega; \bv{x}', \omega'; \bv{x}'; \omega')$ (as needed, for instance, to describe Hanbury-Brown-Twiss or other experiments involving intensity correlations).  Again, if instead of optimizing before averaging, we average then optimize, we can only approximate the expectation value of the fields, not higher-order moments or correlations.

When one can specify different realizations of the source in terms of beam parameters that can subsequently be treated as random variables, then the average may be accomplished analytically or semi-numerically.  Otherwise, we may need to resort to some sort of \textit{Monte Carlo} simulation, drawing (pseudo)-random samples from the distribution of possible sources, approximating the resulting fields, and calculating sample averages.

%%%%

\section{Discussion}

\subsection{Summary of the MPVP}
 
Both the electromagnetic fields and energy exchange between these fields and the sources can be analyzed in a Hilbert-space setting, where we use a volumetric or ``Joule'' inner product to assess the work that would be exchanged between current sources and radiation fields satisfying the source-free Maxwell equations, and a surficial or ``Poynting'' inner product to assess the energy radiated in the outgoing far fields.  

The spontaneous emission from prescribed sources is characterized by the requirement that the outgoing far-field energy flux (spectral density) be as large as possible, consistent with that energy having come from virtual work exchanged between the current sources and their radiation fields, which are the unique extrapolant of the actual physical fields which agree with the outgoing far fields but satisfy the source-free Maxwell equations everywhere in space and time.

As a consequence, we are led to a simple variational principle based on this idea.  Given a parameterized family of trial radiation fields satisfying the source-free Maxwell equations everywhere, but depending on some set of adjustable parameters determining the overall amplitude, phase, shape, and polarization of the trial mode,  the values of the parameters may be estimated by constrained maximization of the spectral density of outgoing energy flux, or of the spectral density of mechanical work which would be exchanged between the actual sources and the trial source-free fields, if they were present in the vicinity of the sources while the sources follow the specified trajectories.  Spectral densities for energy exchanged between sources and fields and for energy emitted as radiation are directly related by constraints of energy conservation.  Because a variational inequality holds separately at each frequency, an integrated inquality will hold \textit{a fortiori} when integrated over any frequency band or even over all frequencies, and therefore also if we instead work in the time domain and integrate over all time.

% subject to a constraint enforcing energy conservation, which balances the spectral densities of energy radiated and work exchanged. 
%Power spectral densities for exchange of energy between sources and fields and for emission of radiation are related by constraints of energy conservation

After optimization, the resulting source-free trial solution is the best guess for the actual radiation field, within the manifold of possibilities allowed by the parameterized family.  In particular, its outgoing component is an approximation to the actual outgoing fields in the far field of the sources.  If calculated without further approximations, the optimized Poynting flux (spectral density) provides a true lower bound for the actual Poynting flux (spectral density) emitted by the sources.

Strictly speaking, ``power'' is a bit of a misnomer, since the variational bound involves a constrained maximization of the \textit{ spectral density} of energy radiated or work exchanged, rather than power \textit{per se}.  But that becomes a mouthful, so in a slight abuse of terminology, we just refer to it as a maximum ``power'' variational principle for convenience, which also serves to emphasize that variational functionals start with integrands related to Poynting flux density in the far-field or to power delivered or extracted from sources, not with volumetric electromagnetic energy density, since integrals of the latter can diverge for harmonic sources.

% The optimized trial mode shape $\bv{f}_{\perp}(\bv{x}; \omega; \tilde{\alpha})$ (or more accurately, the outgoing component thereof) is then the best guess of the actual radiation field profile, within the manifold of possibilities allowed by the family of trial solutions parameterized by $\bv{\alpha}$, and its outgoing Poynting flux yields a strict lower bound on the actual power spectral density of the radiation emitted at the frequency under consideration

\subsection{Features and Limitations}

Although the maximum-power variational principle arises in the setting of linear spaces and operators, variational parameters may actually appear either linearly (\eg, as expansion coefficients in some basis-set decomposition, such as Gauss-Hermite modes in the paraxial case) or nonlinearly (\eg, a spot size or waist location in an adjustable Gaussian mode).

Approximations derived from the MPVP will enjoy the usual benefits and suffer the usual drawbacks of other extremal variational principles.  The optimized power (spectral density) can provide a true lower bound for the actual radiated power (spectral density), and the accuracy of the power so estimated, as well as of the corresponding field profile or any other physical observables derived from them, should improve monotonically as additional functionally-independent parameters are included in the variational fit, to allow for more general radiation envelope shapes.

Lower-bound estimates for electromagnetic power (spectral density) are relatively insensitive to errors in the trial-mode profile, being of second order in ``shape'' errors at a local maximum, but conversely, the electromagnetic field values or field profile and polarization are then approximated with comparatively less accuracy than is the emitted power.

In the context of paraxial optics, where the MPVP could be most readily applied, one might wonder whether a variational approach is any simpler than just directly integrating a Fresnel-type diffraction integral.  Some applications have already suggested a useful role for variational approximations.  Keep in mind that at each frequency of interest, the variational principle supplies an approximation to the radiation field everywhere in space, in conveniently parameterized form, but does requires a $3$-dimensional integration for every point in parameter space searched.  Direct convolution of the paraxial Green function would require a $3$-dimensional integration for every observation point.

\subsection{Interpretations of the MPVP}

This maximum-power variational principle can be variously interpreted according to one's inclinations or applications.  From \eqref{vp3}, we see that the best variational approximation maximizes the spectral density of radiated power consistent with the constraint that this power could have arisen from work extracted by the associated radiation fields from the sources.  That is, when source are assumed to follow prescribed trajectories, classical charges must radiate spontaneously ``as much as possible,'' consistent with energy conservation.  As seen in \eqref{vp5},  the variational approximation also minimizes a Hilbert-space distance between the actual fields and the parameterized trial family of solenoidal, homogeneous fields subject to an energy-conservation constraint, where this distance is defined in terms of outgoing spectral Poynting flux in the frequency bands of interest.

The variational solution also maximizes, for each frequency component (or across all frequencies or time), the spatial overlap, correlation, or resemblance, between the actual sources and the radiation fields (extrapolated back from the far field into the region of the actual sources according to source-free propagation).  Because the rate of energy exchange between the electric field and source takes the form of an overlap integral $-\realpart \int \!d^3\bv{x}\, [\bv{E}_{\stext{rad}}(\bv{x};\omega)\cc \!\cdot \bv{J}(\bv{x}; \omega) ]$ between these quantities, an obvious  ``folk theorem'' has suggested itself to many authors, wherein currents should in some sense ``look like'' the fields they produce (apart perhaps from a phase shift).  The MPVP is a precise and quantitative operationalized version of this often vaguely-formulated intuition.

Indeed, this framework vividly confirms the following elementary but interesting fact: the only part of the current density $\bv{J}(\bv{x};\omega)$ that can actually radiate into free space is the transverse, on-shell component, which is to say, the part that can be written as a superposition of transverse harmonic plane waves satisfying the vacuum dispersion relation, and hence looks much like the radiation field it generates.

Equivalently, one can say the optimal radiation field profile is that which, if it were actually incident on the sources, would maximally couple to them and would experience maximal small-signal gain due to energy absorbed from those sources (neglecting saturation or back-action effects), and furthermore, the ``virtual'' gain so delivered would be equal to the estimated power spontaneously radiated. 

\subsection{Comparison to Madey's Theorem}

In FEL amplifier or other situations involving amplification, instabilities, or stimulated emission, we naturally expect to observe, in the presence of gain, primarily that mode which grows the fastest.  But a similar principle is also applicable in the spontaneous-emission regime\cite{beckefi:66}, because arguments along the lines leading to Einstein's derivation of the $A$ and $B$ coefficients\cite{einstein:17}, or its generalization to FEL physics in the form of Madey's theorem,\cite{madey:79} establish definite connections between spontaneous emission, stimulated emission, and stimulated absorption\cite{nikonov_et_al:98}, even when the radiation is completely classical, and even when back-action on the charges can be neglected.

In fact, the primary difference between this MPVP, and Madey's theorem describing the low-gain bandwidth in free electron lasers, is that by assuming completely prescribed sources, we are ignoring any effects of recoil, multiple scattering, dynamical bunching, saturation, or any other feedback, so once emitted, radiation cannot induce recoil of its own source, or be subsequently scattered or absorbed by other parts of the source downstream.  The MPVP can be seen to be maximizing the mode shape for small-signal gain but without any saturation or back-action, with this ``virtual'' gain delivered proportional to the estimated power spontaneously radiated.  As a result, we end up finding a relationship between the spontaneous emission spectrum and that of the``bare'' stimulated emission, not the ``net''  response given by the difference between stimulated emission and absorption as in Madey's theorem, which results in a small-signal gain proportionality to the \textit{derivative} of the spontaneous emission spectrum, rather than to the spontaneous emission spectrum itself.
 
\subsection{Comparison to Other Variational Principles}

We note that the MPVP is reminiscent of, but evidently distinct from, better known variational principles arising in electromagnetic theory,\cite{mikhlin:64,blanchard_bruning:92,hanson_yakovlev:02} including energy principles, such as Thomson, Dirichlet's, and Hadamard's principles, used in electrostatics and circuit theory;  Fermat's principle in optics; maximum entropy and minimum free energy principles in radiation thermodynamics 
% foundations and applications,
Rayleigh-Ritz, Rumsey, Schwinger, and other principles used in antenna, cavity, and waveguide analysis \cite{rumsey:54,cohen:55,bojarsky:83,schwinger:06}; action principles used in Lagrangian and Hamiltonian approaches to electrodynamics and field theory;  variational numerical methods for electromagnetics, such as minimum residual, moment, Ritz-Galerkin, or other finite element or spectral element methods \cite{harrington:68,glowinski:83,mitchell_wait:84,brenner_scott:94,dyakonov:96,wang:91,zhang:91}; and assorted specialized variational principles developed for FEL analysis\cite{xie_deacon:86,amir_greenzweig:86,luchini_motz:88a,yu_et_al:90,hafizi_roberson:92,xie:00,duda_mori:00} or laser propagation\cite{firth:77,anderson_bonnedal:79}.

% MPVP is similar to and reminiscent of, but distinct from, other well known variational principles used extensively in electrostatics and circuit theory (Thomson and Dirchlet?s principle); optics (Fermat?s principle); antenna, cavity, and waveguide theories (Raleigh-Ritz and Rumsey principles); laser/plasma and FEL physics (Hamilton?s principle, least action principle); and general numerical methods for electromagnetics (minimum residual, moment, finite-element, or Raleigh-Ritz-Galerkin methods) (Also similar but not equivalent to the familiar Raleigh-Ritz approximation in quantum

% \textit{action}-based variational principles in Lagrangian formulations of electrodynamics, as well as various \textit{reaction}-based principles commonly used in waveguide and cavity analysis, and also minimum \textit{energy} or  \textit{free-energy} principles that arise in various electrostatic, dynamic, or thermodynamic contexts, as well as Ritz-Galerkin methods commonly used in finite-element or other numerical simulations, the Rayleigh-Ritz variational principle familiar from ordinary quantum mechanics, and assorted specialized variational principles developed for FEL analysis\cite{xie_deacon:86,amir_greenzweig:86,luchini_motz:88a,yu_et_al:90,hafizi_roberson:92,xie:00,duda_mori:00} or laser propagation\cite{firth:77,anderson_bonnedal:79}.

Fundamentally, the MPVP involves finding \textit{extrema} of \textit{power}-related quantities of the form $\mathcal{P}(\omega) = -\realpart \int \! d^{3} \bv{x}\, [\bv{E}\cc \!\cdot\! \bv{J}]$ at frequencies of interest.  In contrast, in reciprocal media (\ie, those with symmetric susceptibility tensors), Rumsey \textit{reaction}-based variational principles\cite{rumsey:54} involve finding merely \textit{stationary} points of quantities of the form $\mathcal{R}(\omega) = \realpart \int\! d^{3}\bv{x}\, [\bv{E}\!\cdot\! \bv{J}]$.  In lossless media (characterized by Hermitian susceptibility tensors), Lagrangian \textit{action}-based variational principles\cite{jackson:75} involve finding \textit{stationary} points of quantities that include terms of the form $\mathcal{A}(\omega) = \impart \int \! d^{3}\bv{x} \, [ \bv{E}\cc \!\cdot\! \bv{J} ]$. Moreover, because of the hyperbolic character of the wave equation, for generically these stationary-action solutions will be saddle-points, rather than maxima or minima, in function space, so no bounds on the radiated power can be thereby obtained---in fact, if one attempts to use a source-free variational basis, the problem becomes degenerate, and no absolute power level can be determined.  (As an aside, also notice that this pattern might suggest a fourth type of variational principle, involving stationary points of the imaginary part of the reaction, $\impart \int\! d^{3}\bv{x}\, [\bv{E}\!\cdot\! \bv{J}]$). 
% We are not aware of any such principle in use).

We might also note some conceptual similarities between the MPVP and a ``Maxwellian'' perspective on particle acceleration suggested by Zolotorev and coworkers \cite{zolotorev:99,huang:04}, which relates work done on a charged particle by external fields to the interference between these accelerating fields and the particle's own radiation fields. 

Mathematically, the MPVP is most closely related to Bessel's inequality, to which it essentially reduces in special cases, and to a family of extremal variational principles described by the \textit{Lax-Milgram} theorem \cite{debnath:90, zeidler:95}, which underpins Ritz-Galerkin and other approximation methods seeking weak solutions to PDEs, and involves bilinear forms which are both bounded (or, equivalently, continuous) and coercive (or synonymously, elliptic), in some norm within a relevant Banach space. Roughly speaking, continuity and coercivity together mean that all elements of the eigenspectrum of the associated linear operators are finite but bounded away from zero.  However, in the case of the MPVP, the relevant sesquilinear form
%namely $-\qamp{ s_{\perp}}{i\omega\dop(\omega)}{\bv{J}_{\perp}}$, 
associated with the radiation kernel is not strictly positive-definite but is only bounded and coercive in a semi-norm, once again because of the existence of non-radiating sources constituting the nullspace of $\dop$.  In order to obtain a well-defined extremal variational principle with unique optima, the trial solutions must be explicitly restricted to divergence-free solutions to the source-free Helmholtz equation, which are uniquely related to the far-field radiation pattern but omit information about the actual non-radiative fields in the vicinity of the sources.

In certain respects, the framework developed here, involving inner products and radiation kernels, is also reminiscent of that of so-called \textit{reproducing kernel} Hilbert spaces\cite{weinhart:82}, which arise in the theory of PDEs as well as in fields such as machine learning.  However, again not all of the technical conditions for such spaces are met.

\subsection{Possible Extensions and Generalizations}

As developed here, the MPVP is applicable to localized sources which emit radiation into otherwise free space.  In principle, it appears that the derivations will generalize to various other cases involving  idealized waveguides or other lossless, linear, possibly structured media, with localized inhomogeneities due to boundaries or variation in dielectric  properties.  Of course, finding source-free solutions in more elaborate geometries may be challenging, and some care will be needed to distinguish macroscopic radiation fields generated by the free sources from the microscopic radiation fields associated with the totality of free and bound sources.

Even for propagation in vacuum, the MPVP will be limited in its usefulness by the constraint that trial solutions must be solenoidal source-free solutions to the Helmholtz equation.  In three-dimensional space, convenient closed-form analytic solutions are few, and if the trial radiation fields do not exactly satisfy these source-free Maxwell equations, the results may still be of approximate validity, but the strict lower bound on radiated energy may be lost.

Source-free solutions are generally easier to find in a paraxial limit.  Since the MPVP is known to hold at both the leading order and next order of an asymptotic paraxial expansion,  and also in the case of full three-dimensional, free-space geometry (in effect, at infinite order in the paraxial expansion), we conjecture, but have not proven, that a self-consistent form for the MPVP may hold exactly at each cumulative order in a generalized paraxial expansion in powers of characteristic diffraction angle $\Theta$.

Regarding issues of optical coherence, we have already noted how, in order to capture statistical properties of partially coherent light using the MPVP as currently formulated, we should optimize first and then average over uncertainty in the sources, rather than
optimizing with respect to the averaged current source.  Some further investigations are warranted into the consequences of any bias introduced by using approximations which themselves are variational lower bounds.  It is also an interesting but open question as to whether the variational principle can be extended to apply \textit{directly} to second-order or higher-order coherence tensors, perhaps based on the {van Cittert-Zernicke} theorem, stating that under certain conditions, a coherence tensor will satisfy a wave equation similar to that governing the underlying fields. 

%  For now, if partially coherent beans are to be modeled, we should optimize first and then average over uncertainty in the sources, rather than optimizing with respect to the averaged current source.

Another natural, if difficult, question concerns prospects for generalizing the variational principle to systems with optical gain or loss.  So far, we have not seen how to extend the MPVP in this direction, except perhaps as part of a perturbative sequence of successive corrections, where we might try to alternate variational approximation of emitted fields from given sources with modification of those sources due to effects of the predicted emission, hoping thereby to converge to some sort of self-consistent estimate for both fields and particles.

Finally, one may wonder about prospects for incorporating possible quantum optical effects.  As currently formulated, the MPVP applies only to sources consisting in effect of charged particles following prescribed classical trajectories, or ordinary statistical mixtures of such trajectories, which coincides exactly with the class of sources leading to completely classical radiation according to the Glauber-Sudarshan criterion (meaning the quantum mechanical density operator for the optical fields will be associated with an everywhere non-negative Glauber-Sudarshan \textit{quasi-distribution} function based on expansion in a Glauber coherent states basis).  However, in most situations involving radiation from relativistic bunches, quantum effects are anticipated to be small and subtle at most, so we hypothesize that modes identified by the classical MPVP procedure may offer good approximate starting points for exploring quantum mechanical corrections via ``wave-packet quantization'' techniques in quantum optics.\cite{chiao:08}

\section{Examples of Undulator Radiation}

However simple or even mundane the MPVP may appear, it is not without practical utility.  Here we will briefly discuss two illustrative applications drawn from beam physics, one mostly analytic (and somewhat over-simplified so as to remain so), and the second, mostly numerical.  The first concerns confirmation of features of the spontaneous radiation from a low-emittance electron beam in an ideal helical undulator.  The second involves efforts towards optimizing designs for low-gain harmonic-cascade free electron lasers (FELs).

% successfully applied to the problem of low-gain FELs potential for wider applicability

\subsection{Spontaneous Emission from a Low-Emittance Electron Beam in a Helical Undulator}

We can approximate analytically various properties of spontaneous emission from a highly relativistic, low-emittance electron bunch traveling through a weak but long undulator magnet.  As these properties are familiar from other textbook derivations, this provides more of a consistency check on the MPVP than a novel application, although some minor differences do emerge.

\subsubsection{Approximate Particle Trajectories}

We will ignore any effects of finite transverse or longitudinal emittance of the electron beam---with some cost in computation effort, more realistic phase-space distributions could be incorporated when needed.  Specifically, we consider a collimated bunch of $N_e$ classical electrons, each of mass $m$ and charge $q = -e$, initially traveling nearly along the $\unitvec{z}$ axis ($x \approx y \approx 0$) with velocity $\bv{v}= +\unitvec{z} \beta c$, corresponding to a per-particle relativistic energy $E = mc^2 \gamma = mc^2 [1 - \beta^2]^{-\frac{1}{2}}$, and incident on an idealized {helical} undulator, whose magnetostatic fields near the $z$ axis are derivable (approximately) from the Coulomb-gauge vector potential
\be
\bv{A}(\bv{x}) \approx \begin{cases}
\unitvec{x} \,A \,\cos(k_uz) \cosh(k_uy) - \unitvec{y}\, A\,\sin(k_u z) \cosh(k_u x) &\text{ if }  0 < z < L_u \\
0 & \text{ otherwise}
\end{cases},
\ee
where $A$ is a constant proportional to the peak field strength, $k_u = \tfrac{2\pi}{\lambda_u} > 0$ is a positive wavenumber fixing the spatial periodicity $\lambda_u$ of the magnetic field, $N_u$ is an integer specifying the number of wiggler periods in the magnet, and hence $L_u = N_u\lambda_u$ is the overall length of the undulator.  The vector potential $\bv{A}(\bv{x})$ is divergence-free and satisfies the free-space, static Maxwell equations everywhere except right at the entrance and exit of the magnet, (precisely at $z = 0$ or $z = L_u$), where in reality the fields can taper off, but not infinitely abruptly in longitudinal position $z$ as modeled here, so we have introduced a small error.  Imperfections in the magnets will lead to other corrections that will be neglected here.  The corresponding ideal undulator magnetostatic fields are 
\be
\begin{split}
\bv{B}(\bv{x}) = \grad\!\times\! \bv{A} =  &+A k_u \bigl[\unitvec{x}\,\cos(k_u z) \cosh(k_u x) - \unitvec{y}\,\sin(k_u z) \cosh(k_u y) \bigr]\\
&- A k_u  \unitvec{z}\, \bigl[ \sin(k_u z) \sinh(k_u x) +   \cos(k_u z) \sinh(k_u y) \bigr] 
\end{split}
\ee
for $0 < z < L_u$, and $\bv{B}(\bv{x}) = \bv{0}$ otherwise.

We assume that the incident particles remain mono-energetic, collimated, highly relativistic (in the sense that $\gamma \gg 1$), and that the undulator extends over many periods, (\ie, $N_u \gg 1$).  We neglect effects of space-charge forces and radiation reaction, and assume that electrons follow prescribed spatial trajectories determined by initial conditions and the prescribed undulator fields.  As magnetic fields can perform no mechanical work on point charges (neglecting intrinsic spin), these fields will not change the speed of the incident electrons, so the kinematic Lorentz factor inside the undulator remains equal to its initial value $\gamma$ upstream.

Assuming the spatial extent of transverse excursions remains small, in the sense that $\abs{k_u x} \ll 1$ and $\abs{k_u y} \ll 1$, we can set $\cosh(k_u x) \approx \cosh(k_u y) \approx 1$ and $\sinh(k_u x) \approx \sinh(k_u y) \approx 0$, so the vector potential experienced by any electron inside the wiggler is approximately $\bv{A}(\bv{x}) \approx \unitvec{x} A \,\cos(k_u z) - \unitvec{y} A\,\sin(k_u z)$, which is independent of both $x$ and $y$.   This in turn implies that the components of transverse \textit{canonical} momentum $P_x = p_x - e A_x(z)$ and $P_y = p_y - e A_y(z)$ will be conserved.  For an initially on-axis electron (for which $p_x = p_y = 0$, and $A_x(z) = A_y(z) = 0$ just before entering the undulator), the components of transverse momentum will then start off and hence remain equal to zero.  Keeping in mind that $\gamma$ also remains constant (and presumed large), it follows under our various approximations that, for any one electron in the region $0 \le z \le L_u$, the transverse velocity inside the undulator satisfies
\bsub
\begin{align}
mc \gamma \beta_x(z) = p_x(z) \approx e A_x(z) \approx +e A \cos(k_u z) \\
mc \gamma \beta_y(z) = p_y(z) \approx e A_y(z) \approx -e A \sin(k_u z),
\end{align}
\esub
or
\bsub\label{xybetas}
\begin{align}
\beta_x(z)  = +\tfrac{a_u}{\gamma} \cos(k_u z) \\
\beta_y(z) = -\tfrac{a_u}{\gamma} \sin(k_u z),
\end{align}
\esub
expressed in terms of a so-called dimensionless \textit{undulator parameter} $a_u = \tfrac{eA}{mc}$.
From the fact that
\be
(m c^2 \gamma)^2 = (m c^2)^2 + c^2 (p_x^2 + p_y^2 + p_z^2) = m^2 c^4 + c^2 p_z^2 + c^2 e^2 A^2
\ee
remains constant, we may infer that for any electron inside the undulator, its normalized longitudinal velocity
\be
\beta_z = +\sqrt{ 1 -  \tfrac{1}{\gamma^2} [ 1 + a_u^2 ] }
\ee
also remains constant, but is slightly lower than the normalized longitudinal velocity $\beta$ outside the undulator, as some kinetic energy now resides in the transverse ``quiver'' motion.  It can then be convenient to decompose the overall Lorentz factor as $\gamma = \gamma_{\perp} \gamma_{\|}$, where the contribution $\gamma_{\perp} = \sqrt{1 + a_u^2}$ incorporates the effects of the transverse quiver in the undulator fields, and $\gamma_{\|} = \tfrac{1}{\sqrt{1 - \beta_z^2}}$ includes only the effects of longitudinal motion.  Such an approximate trajectory can remain consistent with special relativistic kinematics only when $\beta_x^2 + \beta_y^2 = \tfrac{a_u^2}{\gamma^2} < 1 - \frac{1}{\gamma^2} = \beta^2 < 1$.

Consider the $j$th electron, that enters the undulator at time $t = t_j$, and leaves the undulator at a later time $t = t_j + T_u$, where $T_u = \tfrac{L_u}{c \beta_z}$ is the duration of time spent by inside the undulator.  As a function of time, the longitudinal position of the particle will then be
\be
z_j(t) = \begin{cases}
c \beta (t - t_j)  &\text{ if } t < t_j \\
c\beta_z(t - t_j) &\text{ if } t_j \le  t \le t_j + T_u \\
c\beta(t - t_j ) + L_u - c\beta T_u &\text{ if } t >  t_j + T_u 
\end{cases},
\ee
while the transverse position of the undulating electron can be given explicitly as a function of longitudinal position by
\bsub
\begin{align}
x_j(z) &= \tfrac{a(z)}{\gamma} \tfrac{1}{k_u \beta_z} \sin\bigl( k_u z \bigr) \\
y_j(z) &= \tfrac{a(z)}{\gamma} \tfrac{1}{k_u \beta_z} \cos\bigl( k_u z \bigr),
\end{align}
\esub
in which we have introduced
\be
a(z) = 
\begin{cases}
  a_u &\text{ if } 0 \le z \le L_u  \\
  0 &\text{ otherwise } 
\end{cases},
\ee
as an effective $z$-dependent wiggler parameter.  The corresponding velocity components are
\be
\dot{z}_j(t) = 
\begin{cases}
  c\beta_z&\text{ if } t_j \le  t \le t_j + T_u \\
  c\beta&\text{ otherwise }
\end{cases},
\ee
in the longitudinal direction, and, after just re-scaling equations \eqref{xybetas},
\bsub
\begin{align}
\dot{x}_j(z) &= +c \, \tfrac{a(z)}{\gamma} \, \sin\bigl( k_u z \bigr) \\
\dot{y}_j(z) &= -c \, \tfrac{a(z)}{\gamma} \, \cos\bigl( k_u z \bigr),
\end{align}
\esub
in the transverse plane.  Further demanding that transverse excursions remain small, in the sense that $k_u \lvert x(t) \rvert  \ll 1$ and $k_u \lvert x(t) \rvert  \ll 1$, we will require that 
\be
| \tfrac{a_u}{\gamma} | \ll 1,
\ee
which will be satisfied supposing $0 < | a_u | \lesssim 1$ but $\gamma \gg 1$.

\subsubsection{Estimating Optical Properties}

The characteristic angle for synchrotron emission from relativistic charged particles is about $\sfrac{1}{\gamma}$, and in order for the angular deflection of an electron's trajectory to remain no larger than this angle, we must ensure that 
\be
\min\bigl[ \tfrac{\lvert \beta_{x}(z) \rvert}{| \beta_z |} , \tfrac{\lvert \beta_{y}(z) \rvert}{| \beta_z |}] \le \tfrac{1}{\gamma},
\ee
which is equivalent to $\lvert a_u \rvert  \le 1$, provided that $\gamma \gg 1$. 

These approximate classical electron trajectories should be reasonably accurate to the extent that longitudinal electron motion is highly relativistic, quantum effects are negligible, initial and final deflections at the ends of the undulator are small, and the transverse motion of any electron remains small compared to the characteristic range of transverse variation in the vector potential, which will be of order $k_u\inv$.  

To estimate the peak wavelength $\lambda_1$ for the undulator radiation emitted near the forward direction, we can make use of a simple resonance argument: constructive interference between radiation emitted by the electrons at different points along the wiggler would be maximized if the radiation (traveling at the vacuum speed of light $c$) slips ahead of the electrons (traveling at a speed a bit less than $c$) by one optical wavelength $\lambda$ in the time it takes the electrons to advance by one undulator period $\lambda_u$ in the undulator, so that electrons will continue to oscillate in phase with the radiation.  (The distance of travel over which the emitted radiation fields slip ahead of the emitting charged particle by one radiation wavelength, after which the radiation can be truly considered to have separated from the source, is referred to as the \textit{formation} length.  The lab-frame formation length is macroscopic for undulators, on the order of $\lambda_u$).  That is, in a time $\tfrac{\lambda_u}{c \beta _z}$, an electron will move forward by a distance of one undulator period $\lambda_u$.  At resonance, in the very same time interval, the radiation from that electron should slip ahead by one radiation wavelength relative to the electron, or a lab-frame distance of $\lambda_u + \lambda$ in all. So  $\tfrac{\lambda_u}{c \beta _z} c = \lambda_u + \lambda$, or 
\be
\lambda_1 = (\tfrac{1}{\beta_z} -1) \lambda_u \approx  \tfrac{1 + a_u^2}{2\gamma^2} \, \lambda_u.
\ee
The subscript on $\lambda_1$ is intended to indicate this is the fundamental resonant wavelength, in contrast to higher harmonics (which are can arise in planar wigglers or in the presence of nonlinear bunching effects).  Two powers of $\gamma$ enter because of a redoubled relativistic effect:  in their average rest frame, the electrons see a Lorentz-contracted spatial periodicity in the magnetic field causing them to wiggle, but the resulting  radiation emitted is then Doppler shifted upon observation back in the lab frame.
 
Let the corresponding central frequency of this emitted light be $\omega_1 = c k_1 =  \tfrac{2\pi}{\lambda_1}$.  During its time in the undulator, each electron will emit, predominately in the nearly forward direction, a wave-packet consisting of $N_u$ wavelengths of radiation in all, for a total optical pulse duration of $\delta t \approx N_u \lambda_1/c$ on-axis. The RMS (power-weighted) temporal duration will be somewhat shorter, around $\Delta t \approx \tfrac{1}{2\sqrt{3}} \delta t$, supposing that $N_u \gg 1$.  Assuming the (RMS) bandwidth of the optical pulse is governed by the Fourier-Heisenberg uncertainty principle, $\Delta \omega \, \Delta t \gtrsim \tfrac{1}{2}$, we find $\tfrac{\Delta \omega}{\omega} \approx \tfrac{\sqrt{3}}{2\pi} \tfrac{1}{N_u}$.  For $N_u \gg 1$, this relative bandwidth will be small, so by Taylor expansion we can infer 
\be
\tfrac{\Delta \lambda}{\lambda} \approx \tfrac{\Delta k}{k}  = \tfrac{\Delta \omega}{\omega} \approx \tfrac{\sqrt{3}}{2\pi} \tfrac{1}{N_u} 
\ee
as well.
 
For spontaneous undulator radiation emitted within this so-called \textit{coherent mode}, the spatial profile of the radiation should be approximately that of a diffraction-limited light beam, with an RMS emission half-angle $\Delta \theta$ centered on $+\unitvec{z}$, and a minimum spot size, or waist, of some RMS radius $\Delta r$, achieved at some longitudinal location $z_0$ corresponding to the effective focal plane.  Far downstream, this undulator radiation from the electron beam will appear approximately as if it has been diffracting from an illuminated aperture of radius about $\Delta r$ centered at $x = y = 0$ and $z = z_0$.

If we imagine geometrically tracing the light rays emitted within an RMS half-angle $\Delta \theta$, we notice that, according to an observer downstream, all rays could be observed to have emanated from a transverse aperture located at the midpoint $z = z_0 = \tfrac{1}{2} L_u = \tfrac{1}{2} N_u \lambda_u$ of the undulator, and with transverse radius 
\be
\Delta r \approx \tfrac{1}{2} N_u \lambda_u  \tan \Delta \theta \approx \tfrac{1}{2} N_u \lambda_u \, \Delta \theta,
\ee
where the last approximation is justified because we know the angles of emission for highly relativistic electrons should be small due to ``headlighting'' effects, such that $\tan \Delta \theta \approx \Delta \theta$.  Supposing the radiation in the coherent mode is diffraction-limited, the Fourier-Heisenberg uncertainty principle dictates that $\Delta r\, \Delta \theta \approx \tfrac{\lambda_1}{4\pi}$.  Combining this with our preceding results, we find
$\tfrac{1}{2} N_u \lambda_u (\Delta \theta)^2 \approx \tfrac{\lambda_1}{4\pi}$, or 
\be
\Delta \theta \approx \tfrac{1}{2\sqrt{\pi}} \tfrac{\sqrt{1+ a_u^2}}{\sqrt{N_u} \gamma},
\ee
which is indeed small under our assumptions, and, notably, even smaller by a factor of order $\sfrac{1}{\sqrt{N_u}}$ than the characteristic angle for synchrotron radiation from a relativistic particle in a simple bend magnet, which is of order $\sfrac{1}{\gamma}$.  Because deflection angles for the undulating electrons remain small by assumption, radiation from successive bends in a weak undulator can coherently superpose, leading to an almost diffraction-limited beam for radiation near the resonant frequency, as was just assumed.

The corresponding focused spot size for this mode will be about
\be
\Delta r \approx  \tfrac{1}{ 2\sqrt{\pi}}  \tfrac{\sqrt{N_u} \gamma \lambda_1}{ \sqrt{1 + a_u^2} }  \approx \tfrac{\sqrt{N_u} \sqrt{1+a_u^2}}{ 4\sqrt{\pi} \gamma} \lambda_u.
\ee
Since we expect that $\sqrt{N_u} > 1$ but $a_u < 1$, this spot size is typically somewhat larger than the transverse excursion of any one beam electron, which is about $\delta r \approx \tfrac{a_u}{\gamma} \tfrac{1}{k_u} = \tfrac{a_u \lambda_u}{2 \pi \gamma}$.

The effective \textit{Rayleigh range}, or \textit{Fraunhofer distance}, characterizing the longitudinal distance required for the coherent mode to diffract appreciably, is approximately
\be
z_R \approx \tfrac{ \pi \,\Delta r^2}{\lambda_1} \approx  \tfrac{1}{8}  N_u \lambda_u,
\ee
which is somewhat less than the overall length of the undulator itself. This means that by the time the radiation has traveled much past the end of the undulator, it can be assumed to be in the diffractive far field. 

From the Larmor-Li{\'e}nard formula, recognizing that an electrons's acceleration due to the undulator fields remains orthogonal to its velocity, the total electromagnetic energy radiated by any one electron inside the undulator may be estimated to be about
\be
\mathcal{E} \approx \mathcal{P} \tfrac{L_u}{\beta_z c} = \tfrac{\mu_0}{6\pi c}  e^2 \gamma^6 
( \tfrac{a_u k_u \beta_z c^2}{\gamma})^2 (1 - \beta^2)  \tfrac{N_u \lambda_u}{\beta_z c}
\approx
\tfrac{1}{3} \tfrac{e^2}{\epsz} a_u^2 \gamma^2 k_u  N_u 
\approx
\tfrac{2\pi}{3} N_u\, \hbar \omega_1\, \alpha\, a_u^2(1+ a_u^2),
\ee
% \be
% \mathcal{E} \approx \mathcal{P} \tfrac{L_u}{c} = \tfrac{\mu_0}{6\pi c}  e^2 \gamma^4  c^2 c^2 k_u^2 \tfrac{a_u^2}{\gamma^2} \tfrac{N_u \lambda_u}{c} 
% =  \tfrac{1}{3} N_u \gamma^2 k_u  \tfrac{e^2}{\epsilon_0 }  a_u^2 = 
%\tfrac{2\pi}{3} N_u\, \hbar \omega\, \alpha\, a_u^2,
%\ee
% \be
%\mathcal{E} \approx \mathcal{P} \tfrac{L_u}{c} \approx \tfrac{\mu_0}{6\pi c}  e^2 \gamma^6  c^4 k_u^2 % \beta_z^2 \tfrac{a_u^2}{\gamma^2} \tfrac{1}{\gamma^2(1+a_u^2)}  \tfrac{N_u \lambda_u}{c} 
%\ee
where $\alpha  = \tfrac{e^2}{4\pi \epsilon_0 \hbar c}$ is the fine structure constant.  For our approximations to be self-consistent, this energy should be much smaller than the relativistic energy $\gamma mc^2$ per electron, or else the motion of an electron would eventually become substantially affected by its own electromagnetic emission.

To make a more meaningful local comparison of power balance, we can look at the radiation reaction force in relation to the Lorentz force.  We could use the Abraham-Dirac-Lorentz expression for the reaction force, but for oscillatory motion, we know that this force is just engineered so that $\mathcal{P}  = \bv{F} \!\cdot\! \bv{v}$, where $\bv{F}$ is the applied force.  On average, the reaction force should act in the $-\unitvec{z}$ direction, and be of typical magnitude $F \approx  \tfrac{1}{3} k_u \hbar \omega\, \alpha\, a_u^2$.  The average Lorentz force must be zero, but the RMS force is approximately of magnitude $F_B \sim ec |A| k_u =  mc^2 k_u  |a_u|$.  Therefore the reaction force will be much smaller than the typical Lorentz force inside the undulator provided that $\alpha |a_u|  \tfrac{\hbar \omega}{mc^2}  \ll 1$, but we know $\alpha \approx \tfrac{1}{137} \ll 1$, that $|a_u| \le 1$ in the weak undulator regime, and that $\hbar \omega \ll mc^2$ at least until emitted photon energies climb unrealistically into the gamma-ray regime.

However, only a fraction of this energy is actually radiated into the coherent mode. This fraction can be roughly estimated by thinking about the photons emitted in the average rest frame of the electron, where the emission will be be approximately dipolar and monochromatic, but due to Lorentz contraction of the undulator period, the emission frequency will be about $\omega' \approx \tfrac{2\pi c \beta_z}{(\lambda_u/\gamma)} \approx \gamma c k_u$, while the emission for any one electron will occur over a time interval of about $\Delta t' = \tfrac{N_u (\lambda_u/\gamma)}{\beta_z c} \approx \tfrac{N_u \lambda_u}{c \gamma}$.  From the Larmor formula, we can estimate the total number of photons emitted per electron to be around
\be
\mathcal{N}_1 \approx 
 \tfrac{2}{3} \alpha \,\hbar c \,\tfrac{ \omega'{}^4 (x^2 + y^2)}{c^3} \, \Delta t' \tfrac{1}{\hbar \omega'}
\approx
 \tfrac{2}{3} \alpha \tfrac{\omega'{}^3}{c^2}  \tfrac{a_u^2}{\gamma^2 k_u^2 \beta_z^2} \tfrac{N_u \lambda_u}{c \gamma} \approx \tfrac{4\pi}{3} \alpha \,a_u^2 \,N_u,
\ee
which must be a Lorentz invariant.  But after accounting for the relativistic headlighting effect, the coherent-mode radiation, confined within a narrow, forward cone of half-angle $\Delta \theta$ in the lab-frame, will correspond to photons emitted within a half-angle $\Delta \theta' \approx 2\gamma \,\Delta \theta \approx \tfrac{1}{\sqrt{\pi}} \tfrac{\sqrt{1+ a_u^2}}{\sqrt{N_u}}  $ in the average rest frame.  Assuming $N_u$ is large, some simple integrations over a dipolar radiation angular pattern reveal that this amounts to a fraction $f_c \approx \tfrac{3}{4 \pi} \tfrac{1+a_u^2}{{N_u}}$ of the total photons emitted.  Hence the expected number of coherent-mode photons emitted per electron is about 
\be
\mathcal{N}_c = f_c \, \mathcal{N}_{1}  \approx \alpha \, a_u^2(1+a_u^2).
\ee
Notice that this is independent of the number $N_u$ of undulator periods, which makes sense because the coherent bandwidth shrinks in frequency in inverse proportion to $N_u$ even as the total energy radiated increases linearly with $N_u$.

%%%%

\def\e{\epsilon}

\subsubsection{A Gaussian-Mode Variational Approximation to the Radiation Fields}

We may wonder how these back-of-the-envelope estimates compare to approximations generated from the MPVP.  Using the latter also has the advantage of allowing some assumptions to be relaxed.  For example, we might continue to assume that $\tfrac{a_u}{\gamma} \ll 1$, but not necessarily that $a_u \lesssim 1$.

To find a simple variational approximation, we will first need the current density, which may be determined from the spatial positions of the electrons and their velocities, worked out above.  In the time domain, the current density for the $j$th electron can be written as
\bsub
\begin{align}
J_{x_j}(x,y,z,t) &=  -e \,\dot{x}_j(z) \, \delta\bigl( x - x_j(z)  \bigr) \, \delta\bigl( y - y_j(z) \bigr) \, \delta\bigl( z - z_j(t) \bigr) \\
J_{y_j}(x,y,z,t) &=  -e \,\dot{y}_j(z) \, \delta\bigl( x - x_j(z)  \bigr) \, \delta\bigl( y - y_j(z) \bigr) \, \delta\bigl( z - z_j(t) \bigr) \\
J_{z_j}(x,y,z,t) &=  -e \,\dot{z}_j(z) \, \delta\bigl( x - x_j(z)  \bigr) \, \delta\bigl( y - y_j(z) \bigr) \, \delta\bigl( z - z_j(t) \bigr). 
\end{align}
\esub
Fourier transforming in time, and using that fact that we can re-write the last Dirac delta function as
\be
\delta\bigl( z - z_j(t) \bigr) = \tfrac{1}{c\beta_z(z)} \, \delta\bigl( \tfrac{z}{c\beta_z(z)} +  \tau_j - t \bigr),
\ee
in which we have introduced a $z$-dependent longitudinal velocity
\be
\beta_z(z) = \begin{cases}
\beta_z  &\text{ if } 0 \le z \le L_u \\
\beta &\text{ otherwise }
\end{cases},
\ee
and $z$-dependent temporal offset
\be
\tau_j(z) = 
\begin{cases}
t_j &\text{ if } z \le L_u \\
t_j + T_u - \tfrac{L_u}{c\beta} &\text{ if } z > L_u
\end{cases},
\ee
the frequency-domain current density components for the $j$th electron become
\bsub
\begin{align}
J_{x_j}(\bv{x}; \omega) &=  -\tfrac{1}{\sqrt{2\pi}} \tfrac{e}{c\beta_z(z)} \,\dot{x}_j(z) \, \delta\bigl( x - x_j(z)  \bigr) \, \delta\bigl( y - y_j(z) \bigr) \, e^{+i \tfrac{\omega}{c \beta_z(z)}z} e^{+i \omega \tau_j(z)}  \\
J_{y_j}(\bv{x}; \omega) &=  -\tfrac{1}{\sqrt{2\pi}} \tfrac{e}{c\beta_z(z)} \,\dot{y}_j(z) \, \delta\bigl( x - x_j(z)  \bigr) \, \delta\bigl( y - y_j(z) \bigr) \, e^{+i \tfrac{\omega}{c \beta_z(z)}z} e^{+i \omega \tau_j(z)}   \\
J_{z_j}(\bv{x}; \omega) &=  -\tfrac{1}{\sqrt{2\pi}} \tfrac{e}{c\beta_z(z)} \,\dot{z}_j(z) \, \delta\bigl( x - x_j(z)  \bigr) \, \delta\bigl( y - y_j(z) \bigr) \,  e^{+i \tfrac{\omega}{c \beta_z(z)}z} e^{+i \omega \tau_j(z)} .
\end{align}
\esub
It will be convenient to re-express the current density as
\be
\bv{J}_j(\bv{x}; \omega) =  J_{+_j}(\bv{x}; \omega)\, \unitvec{\e}_{+} + J_{-_j}(\bv{x}; \omega) \,\unitvec{\e}_{-} + J_{z_j}(\bv{x}; \omega)\, \unitvec{z},
\ee
where
\be
\unitvec{\e}_{\pm} = \unitvec{\e}_{\mp}\cc= \tfrac{ \unitvec{x} \pm i \unitvec{y}}{\sqrt{2}}
\ee
are right/left circular polarization basis vectors in the transverse plane, and
\be
{J}_{\pm_j}(\bv{x}; \omega) = (\mp i) \tfrac{e}{2\sqrt{\pi}}  \tfrac{a(z)}{\gamma \beta_z(z)} e^{\mp i k_u z} \, \delta\bigl( x - \tfrac{a(z)}{k_u \gamma \beta_z(z)} \sin( k_u z )  \bigr) \, \delta\bigl( y -  \tfrac{a(z)}{k_u \gamma \beta_z(z)} \cos( k_u z )  \bigr) \, e^{+i \omega ( \frac{z}{c\beta_z(z)} + \tau_j )}
\ee
are the corresponding right and left circularly polarized components of $\bv{J}(\bv{x}; \omega)$.

As a simple sanity check, we can first verify that in the absence of the undulator, electrons traveling at constant velocity are not predicted to radiate at all according to the MPVP, in agreement with the full Maxwell theory.  Setting $a(z) = 0$, the frequency-domain current density reduces to
\bsub
\begin{align}
J_{x_j}(\bv{x}; \omega) &= \phantom{+}0  \\
J_{y_j}(\bv{x}; \omega) &= \phantom{+}0   \\
J_{z_j}(\bv{x}; \omega) &= -\tfrac{e}{\sqrt{2\pi}}  \, \delta\bigl( x  \bigr) \, \delta\bigl( y \bigr) \, e^{i \frac{\omega}{c \beta}} e^{+i \omega t_j} .
\end{align}
\esub
Any solenoidal radiation electric field can in principle be expressed in terms of some linear superposition over transverse, harmonic plane waves of the form
\be
\bv{E}(\bv{x}; \bv{k}; \omega) \propto \unitvec{\e}(\bv{k}) \, e^{i \bv{k} \cdot \bv{x}},
\ee
where $\bv{k} \!\cdot\! \unitvec{\e}(\bv{k}) = 0$ and $\bv{k} \!\cdot\! \bv{k} = \omega^2$.  For any such component of any such trial field, the Joule work integral is
\be
\int \!d^3 \bv{x} \; \bv{E}(\bv{x}; \bv{k}; \omega)\cc \!\cdot\! \bv{J}_j(\bv{x}; \omega) =
-\tfrac{e}{\sqrt{2\pi}} \,\e_z(\bv{k}) \, e^{+i \omega t_j} \!\!\! \int\limits_{-\infty}^{+\infty} \!\! dz \, e^{-i k_z z} \, e^{i \frac{\omega}{c\beta} z} = -e \,\sqrt{2\pi} \, \e_z(\bv{k}) \, \delta\bigl(k_z - \tfrac{\omega}{c \beta} \bigr).
\ee
However, because $0 \le \beta < 1$, and $0 \le | k_x | \le \abs{\bv{k}} = \tfrac{\omega}{c}$, it follows that $c \beta | k_x | < \omega$, so the argument of the Dirac delta function can never vanish, implying $\delta\bigl(k_z - \tfrac{\omega}{c \beta} \bigr) = 0$, and hence $\int \!d^3 \bv{x} \; \bv{E}(\bv{x}; \bv{k}; \omega)\cc \cdot \bv{J}_j(\bv{x}; \omega) = 0$.  As a consequence of linearity, this remains true for any trial field satisfying the source-free Maxwell equations everywhere in space, and so any variational approximation to the radiated power will indeed vanish, as would be anticipated from the Li{\'e}nard-Larmor formula.

In the presence of the undulator fields, we can estimate the radiation profile and power of the coherent mode within a paraxial approximation.  The lowest-order Gaussian paraxial mode propagating in the $+\unitvec{z}$ direction can be written as 
\be
\bv{E}(\xperp,z;k) =  [ a_{+}e^{i\phi_+}  \unitvec{\e}_{+}  + a_{-} e^{i\phi_-} \unitvec{\e}_{-} ] \, i\omega\,\psi(\xperp,z - z_0;k)\, e^{+ikz},
\ee
where the wavenumber $k = k(\omega)$ satisfies the one-dimensional vacuum dispersion relation $\omega = c k$,
% note sign conventions!
\be
q(z; k) = z - i z_R = z - i \tfrac{\pi \sigma^2}{\lambda} = z - i \tfrac{1}{2} k \sigma^2
\ee
is the so-called \textit{complex beam parameter} expressed in terms of the focused spot size $\sigma$, or associated Rayleigh range $z_R = z_R(\sigma,k) = \tfrac{1}{2} k \sigma^2$, and
\be
\psi(\xperp,z; k) = \sqrt{\tfrac{z_R}{\pi k}}  \, \tfrac{1}{q(z;k)} \,e^{+ik \frac{\lvert\xperp\rvert^2}{2 q(z;k)}}
\ee
is the complex spatial mode shape, including effects of diffraction, wavefront curvature, and a Gouy phase shift through the focus.

% The
In the field profile $\psi(\xperp,z - z_0;k)$, the pre-factor has been chosen to provide a convenient normalization.  For any such paraxial mode, the spectral density of Poynting flux can be most easily calculated in the focal plane $z = z_0$, in which
\be
\begin{split}
\tfrac{1}{\muz}\innerpr{\bv{E}_d}{\bv{B}_d} &=
(\abs{a_+(\omega)}^2 +\abs{a_-(\omega)}^2) \tfrac{\omega^2}{c \muz} \! \int  \!  d^2 \xperp \,\psi(\xperp, 0; k)\cc \, \psi(\xperp, 0; k) \\
&= (\abs{a_+(\omega)}^2 +\abs{a_-(\omega)}^2) \tfrac{\omega^2}{c \muz} \,  \tfrac{z_R}{\pi k}  \! \int\limits_{0}^{\infty} \! 2\pi r \, dr\, \tfrac{e^{-\frac{k r^2}{z_R}}}{z_R^2} \\
&= (\abs{a_+(\omega)}^2 +\abs{a_-(\omega)}^2) \,\tfrac{\omega^2}{c \muz}  \tfrac{ z_R}{\pi k}  \tfrac{\pi}{k z_R}
= (\abs{a_+(\omega)}^2 +\abs{a_-(\omega)}^2) \,\,\tfrac{c}{\muz}.
\end{split}
\ee

Now, the MPVP tells us to look for a constrained maximum of the absolute value of the ``work'' integral
\be
\begin{split}
P_j(\omega)\cc &= -\! \int \! d^3\bv{x} \,  \bv{E}(\bv{x}; \omega) \!\cdot  \bv{J}(\bv{x}; \omega)\cc \\
&= +a_{+}(\omega) \,e^{i\phi_+ -i\omega t_j}  \tfrac{ec}{2{\pi}}  \tfrac{a_u}{\gamma \beta_z} \sqrt{k z_R}
\!\int\limits_{0}^{L_u }\!\!  \tfrac{dz}{q(z-z_0)} \, e^{\frac{+i k a_u^2}{k_u^2 \beta_z^2 \gamma^2 \,2 q(z-z_0)}}  \, e^{+i(k + k_u) z - i \frac{k}{\beta_z}z} \\
&= -a_{-}(\omega)\, e^{i\phi_- -i\omega t_j} \tfrac{ec}{2{\pi}} \tfrac{a_u}{\gamma \beta_z} \sqrt{k z_R}
\!\int\limits_{0}^{L_u } \!\!  \tfrac{dz}{q(z-z_0)} \,e^{\frac{+i k a_u^2}{k_u^2 \beta_z^2 \gamma^2 \,2 q(z-z_0)}}  \, e^{+ i (k-k_u) z - i \frac{k}{\beta_z} z},
\end{split}
\ee
with respect to the independent adjustable shape and polarization parameters, under the assumptions that $k z_R \gg \sqrt{k z_R} \gg 1$, consistent with the paraxial approximation.  Subsequently, the overall phases  can be adjusted to make the integral real and positive, and the amplitude can be chosen in accord with the additional energy conservation constraint.

Unfortunately, these overlap integrals cannot be expressed in any simple closed form.  For any particular set of physical beam and beamline parameters, we could proceed numerically, to find the best approximation within the family of Gaussian paraxial modes.  For truly optimal approximations, such numerical integration and parameter searches would seem to be unavoidable.

But to keep the discussion more general, let us see how far we can progress analytically, by making some reasonable additional approximations based on stationary-phase type considerations and treatment of $\tfrac{a_u}{\gamma}$ and/or $\frac{1}{k\sigma}$ as small parameters.  Notice that at leading order, the work integral will already be proportional to $\tfrac{a_u}{\gamma}$, so if we think of expanding the exponential
\be
e^{\frac{+i k a_u^2}{k_u^2 \beta_z^2 \gamma^2 \,2 q(z-z_0)}} \approx 1 + \tfrac{i k a_u^2}{k_u^2 \beta_z^2 \gamma^2 \,2 q(z-z_0)} + \dotsb
\ee
in powers of $\tfrac{a_u}{\gamma}$, simply truncating after just the zeroth-order term will only introduce a small relative error of order $O\bigl( \tfrac{a_u^2}{\gamma^2} \bigr)$ in the integrals.  If we proceed by using such a Taylor expansion, approximating
\be
\begin{split}
P_j(\omega)\cc \approx 
&+a_{+}(\omega)\, e^{i\phi_+ - i\omega t_j}  \tfrac{ec}{2{\pi}}  \tfrac{a_u }{\gamma \beta_z} \sqrt{k z_R}
\!\int\limits_{0}^{L_u }\! dz\, \tfrac{e^{+i(k + k_u - \frac{k}{\beta_z}) z} }{z-z_0 - iz_R} \\
&-a_{-}(\omega)\, e^{i\phi_- - i\omega t_j} \tfrac{ec}{2{\pi}} \tfrac{a_u}{\gamma \beta_z} \sqrt{k z_R}
\!\int\limits_{0}^{L_u }\! dz\, \tfrac{e^{+ i (k-k_u - \frac{k}{\beta_z}) z} }{z-z_0 - iz_R} ,
\end{split}
\ee
then the rapidly-varying phase in the first overlap integral will become stationary when
\be
(1 - \tfrac{1}{\beta_z} ) k + k_u = 0,
\ee
or equivalently when $\lambda = (\tfrac{1}{\beta_z} - 1) \lambda_u = \lambda_1$, which is identical to the resonance condition inferred above.  In contrast, the stationary phase condition in the overlap integral for the other polarization component would correspond to
\be
(1 - \tfrac{1}{\beta_z} )k - k_u = 0,
\ee
which cannot be satisfied for any positive values of $k$ and $k_u$, so this second integral will tend to oscillate away via destructive interference, particularly in relation to the former integral.  We are led to  the conclusion that the polarization is predominately circular, in the same sense as the gyration of the electrons in the transverse plane as they pass through the undulator, as would be anticipated intuitively.  Accordingly, we will just set $a_{-}(\omega) \approx 0$ in comparison to the presumably non-zero variational approximation to $a_{+}(\omega)$.

Our variational principle also clarifies what might be regarded as loopholes in our earlier reasoning deducing the peak emission frequency and the angular spread of the undulator radiation.  In order to deduce the frequency, we had invoked a resonance argument, wherein an electron, in advancing one undulator period, should slip behind the on-axis radiation by exactly one optical wavelength.  But if there is only one electron, there is no actual radiation behind the electron that catches up to it, and ostensibly no opportunity for coherent constructive interference. But the MPVP justifies thinking in terms of the virtual work that \textit{would} be exchanged between an electron and a harmonic radiation field, if it were present.  Similarly, to argue why the emission angle can be of order $\sfrac{1}{\sqrt{N_u} \gamma}$ for an undulator, rather than $\sfrac{1}{\gamma}$ is in a bending magnet or wiggler, we were thinking in terms of interference of light rays that could overlap over successive periods of the undulator when electrons are quivering sufficiently gently.  But again, with a single electron, radiation emitted at a given (small angle) 

% The
In order to make a rough estimate of the bandwidth for the emitted radiation, we can ask under what shift in $k$, from $k_1$ to $k_1 + \delta k$, will the rapidly varying phase in the integrand first result in completely destructive interference, if still ignoring the more slowly varying $1/q(z;k)$ term, as well as the $k$ dependence in the pre-factors.  Using the usual trick of looking for pairwise cancelations, we predict
$(1 - \tfrac{1}{\beta_z}) \, \delta k \, \tfrac{L_u}{2} \approx \pm \pi$, or
\be
k_u \tfrac{\delta k}{k}  L_u \approx \pm \tfrac{\pi}{2},
\ee 
 which in turn implies
 \be
 \tfrac{\delta k}{k}  \approx \pm  \tfrac{1}{N_u} ,
 \ee
 which is at least in the same ballpark as our earlier estimate using the uncertainty principle.  Again, if needed, somewhat better approximations to the bandwidth could be obtained numerically.
 
In order to estimate the shape parameters $z_0$ and $z_R$ for emission near the resonant frequency $\omega_1$, we can  maximize
\be\label{op2}
\frac{\abs{\innerpd{\bv{E}_{\stext{rad}}}{\bv{J}} }^2 }{ \innerpr{\bv{E}_d}{\bv{B}_d} } \propto g(z_0,z_R;k_1) 
\approx   k_1 z_R \, \Bigl\lvert \, \int\limits_{0}^{L_u }\!\! \tfrac{dz}{z-z_0 - iz_R}  \, \Bigr\rvert^2  =  {k_1 z_R} \abs{ \, \ln\bigl[ \tfrac{ z_0 - L_u + i z_R  }{ z_0 + i z_R } \bigr]  \, }^2,
\ee
where we have, once again, expanded the Gaussian exponential in the integrand to leading order in the small parameter $\sfrac{a_u}{\gamma}$.

This function is obviously nonnegative and continuous, and it is not difficult to verify that it is symmetric in $z_0$ about $z_0 = \tfrac{L_u}{2}$, so this point will correspond to a local maximum or minimum.   Somewhat less obviously, near the optimal values of $z_R$, \eqref{op2} will also be monotonically increasing for $0 < z_0 < \tfrac{L_u}{2}$, and it follows that the optimal value of $z_0$ will always occur at $z_0 = \tfrac{L_u}{2}$ exactly, just as we might anticipate intuitively.  Alas, actually finding the optimal value of $z_R$ is somewhat more involved, but after some algebra, it ends up corresponding to the positive root of the transcendental equation
\be
\tfrac{2 z_R}{L_u}\, \tan\bigl( \tfrac{4 z_R L_u}{4 z_R^2+L_u^2}  \bigr) = 1,
% \tfrac{2 z_R}{L_u} \, \tan\Bigl[ \tfrac{2 \tfrac{2 z_R}{L_u} }{ \bigl(\frac{2 z_R}{L_u} \bigr)^2 +1} \Bigr] = 1,
\ee
which is about  $\tfrac{z_R}{L_u} \approx 0.35926116\dotsc$, suggesting a somewhat longer Rayleigh range (and hence larger focused spot size) than was previously estimated, which was about $\tfrac{z_R}{L_u} \approx 0.125$.

Finally, the amplitude at the fundamental frequency $\omega = \omega_1$ can be estimated from the constraint that
\be
1 = \frac{ \muz \abs{\innerpd{\bv{E}_{\stext{rad}}}{\bv{J}} } }{2 \innerpr{\bv{E}_d}{\bv{B}_d} } 
\approx \frac{|a_{+}(\omega_1)|\, \tfrac{ec}{2{\pi}}  \tfrac{a_u }{\gamma \beta_z} \sqrt{k_1 z_R}
\abs{ \, \ln\bigl[ \tfrac{- L_u + 2 i z_R  }{ +L_u + 2i z_R } \bigr] } }{2\, |a_+(\omega)|^2 \! \frac{c}{\muz}},
\ee
or
\be
| a_+(\omega_1) | \approx { \tfrac{e\muz}{4{\pi}}  \tfrac{a_u }{\gamma \beta_z} \sqrt{k_1 z_R}
\abs{ \, \ln\bigl[ \tfrac{- L_u + 2 i z_R  }{ +L_u + 2i z_R } \bigr] } } ,
\ee
where, at the optimized value of $z_R$, we find
\be
\bigl\lvert  \, \ln\bigl[ \tfrac{- L_u + 2 i z_R  }{ +L_u + 2i z_R } \bigr] \bigr\rvert 
=  \bigl\lvert \arctan \big[ \tfrac{4z_R L_u}{4 z_r^2 -L_u^2} \bigr] \bigr\rvert = \tfrac{  8 z_R L_u }{ 4 z_R^2 + L_u^2   } \approx 1.89549\dotsc.
\ee
The associated variational bound on the peak spectral Poynting flux density, radiated by one electron, is then
\be
\begin{split}
\tfrac{1}{\muz}\innerpr{\bv{E}_d}{\bv{B}_d} &\ge \abs{a_+(\omega_1)}^2 \,\tfrac{c}{\muz}
\approx  \tfrac{e^2 c \muz}{16 \pi^2}  \tfrac{a_u^2 }{\gamma^2 \beta_z^2} {k_1 z_R}
\bigl[ \, \tfrac{8 z_R L_u}{4 z_R^2 + L_u^2} \,\bigr]^2 .
\end{split}
\ee

Strictly speaking, in order to trace out the variational approximation to the power spectrum as a function of frequency, we should re-optimize the trial mode parameters at every frequency $\omega = \omega_1 + \delta \omega$, but away from resonance the integrals will become more cumbersome.  Here we shall be content with making an additional approximation based on the intuition that the far-field coherent signal will look more-or-less like $N_u$ repetitions of a sinusoidal oscillation at the central frequency $\omega_1$, so that the power spectrum should look ``sinc''-like.  

In this case, in order to convert (approximately) from the peak spectral density of radiated energy (at central frequency $\omega_1$) to integrated emitted energy over one entire pulse from one electron, given an assumed sinc-like spectrum, we can simply multiply by the factor
\be
\tfrac{ \frac{1}{2} N_u \frac{2\pi}{\omega_1} }{ \frac{1}{2\pi}  \frac{N_u^2 \pi^2}{\omega_1^2}  } = \tfrac{2 \omega_1}{N_u},
\ee
which is the ratio of the peak of a sinc-shaped power spectrum to its integral over all frequencies, and represents an effective total bandwidth.  Per electron, the energy radiated into the coherent mode is then estimated to be
\be
\mathcal{E}_c \approx   \tfrac{e^2 c \muz}{16 \pi^2}  \tfrac{a_u^2 }{\gamma^2 \beta_z^2} {k_1 z_R}
\bigl[ \, \tfrac{8 z_R L_u}{4 z_R^2 + L_u^2} \,\bigr]^2 \tfrac{2 \omega}{N_u}
=  \tfrac{e^2}{4 \pi \epsz} k_1  \tfrac{a_u^2 }{\gamma^2 \beta_z^2} \tfrac{k_1}{k_u} (\tfrac{z_R}{L_u})^3
\Bigl[ \, \tfrac{8 }{4 \frac{z_R^2}{L_u^2} + 1} \, \Bigr]^2, 
\ee
or
\be
\mathcal{E}_c \approx  \tfrac{e^2}{4 \pi \epsz} k_1  \tfrac{a_u^2 }{\gamma^2 \beta_z^2} \tfrac{2 \gamma^2}{(1+a_u^2)} (\tfrac{z_R}{L_u})^3
\Bigl[ \, \tfrac{8 }{4 \frac{z_R^2}{L_u^2} + 1} \, \Bigr]^2
\approx \alpha \, \hbar \omega  \tfrac{2 a_u^2}{1+a_u^2} \tfrac{1}{\beta_z^2} (\tfrac{z_R}{L_u})^3
\Bigl[ \, \tfrac{8 }{4 \frac{z_R^2}{L_u^2} + 1} \, \Bigr]^2,
\ee
where at the variational optimum,
\be
(\tfrac{z_R}{L_u})^3
\Bigl[ \, \tfrac{8 }{4 \frac{z_R^2}{L_u^2} + 1} \, \Bigr]^2 \approx  1.29079\dotsc,
\ee
verifying our earlier rule-of-thumb that each electron radiates about $O(\alpha)$ photons into the coherent mode, over a complete pass through the undulator.

\subsection{Multiple Electrons}

Neglecting energy and angular spread as we have done, each electron in the beam will radiate into the same mode, just offset in time.  If we know the locations of the electrons, we can then just approximate the net radiation electric field as 
\be
\bv{E}_{\stext{rad}}(\xperp,z;\omega) \approx \bv{E}_1(\xperp,z;k) \, \sum\limits_{j=1}^{N_e} e^{i \omega t_j},
\ee
where $\bv{E}_1(\xperp,z;k)$ is the variational trial mode optimized using the current of one electron that enters the undulator ($z = 0$) at $t = 0$, and, as above, the $t_j$ specify the actual arrival times for the electrons $j = 1, \dotsc, N_e$ in the bunch.  If, more realistically, we have only limited information about the number and positions of the electrons, then we can perform statistical averages.  Usually, we are justified in assuming that the electron bunch can be modeled as an inhomogeneous Poisson process, where the \textit{expected} number of electrons that cross the entrance to the undulator at $z = 0$ in any specified time interval $t_a < t \le t_b$ is given by 
\be
\bar{N}(t_a, t_b) = \int\limits_{t_a}^{t_b} \! dt\, \bar{n}(t),
\ee 
for some average bunch profile $\bar{n}(t)$ which is real and nonnegative.
Then in the frequency domain, the expectation value of the radiation electric field is
\be
\langle \bv{E}_{\stext{rad}}(\xperp,z;\omega)  \rangle \approx 
\bv{E}_1(\xperp,z;k) \;  \Bigl\langle \sum\limits_{j=1}^{N_e} e^{i \omega t_j} \Bigr\rangle
= \bv{E}_1(\xperp,z;k) \, \sqrt{2\pi} \, \bar{n}(\omega),
\ee
where
\be
\bar{n}(\omega) = \tfrac{1}{\sqrt{2\pi}} \! \int \! dt \, \bar{n}(t) \, e^{i \omega t} 
\ee
is a longitudinal ``structure function'' for the bunch, given by the Fourier transform of $\bar{n}(t)$, and satisfying, for all  $\omega \in \realsymbol$, the following properties:
\bsub
\begin{align}
\bar{n}(\omega) &\text{ is a uniformly continuous function of frequency } \omega,\\
\bar{n}(\omega)\cc &= \bar{n}(-\omega), \\
\abs{\bar{n}(\omega)} &\le \bar{n}(0) = \tfrac{1}{\sqrt{2\pi}} \bar{N}(-\infty, +\infty), \\
\bar{n}(\omega_j - \omega_k) \; &\text{defines a positive-semidefinite matrix, for any finite sequence of real frequencies}.
\end{align}
\esub
These properties follow because $\bar{n}(\omega)$ is proportional to the \textit{characteristic function} of the probability density function for the (real-valued) arrival time of an arbitrarily chosen electron in the bunch.

The  (two-point) spectral correlation function or coherence function (i.e., matrix of second-order, frequency-domain moments) can be written as
\be
\begin{split}
\langle \bv{E}_{\stext{rad}}(\xperp,z;\omega) \, \bv{E}_{\stext{rad}}(\xperp',z';\omega')\hc \rangle &\approx 
\bv{E}_1(\xperp,z;k) \, \bv{E}_1(\xperp',z';k')\hc  \; \Bigl\langle \sum\limits_{j=1}^{N_e}  \sum\limits_{k=1}^{N_e} e^{i \omega (t_j - t_k)} \Bigr\rangle \\
&=\bv{E}_1(\xperp,z;k) \, \bv{E}_1(\xperp',z';k')\hc \bigl[ \sqrt{2\pi} \, \bar{n}(\omega -\omega') + 2\pi \,\bar{n}(\omega) \, \bar{n}(\omega')\cc \bigr],
\end{split}
\ee
indicating that the spontaneous undulator radiation emitted by different electrons will tend to add incoherently (i.e., in intensity rather than amplitude) unless there is appreciable micro-bunching, yielding significant Fourier content in the structure function $\bar{n}(\omega)$ at the same frequencies where $\abs{ \bv{E}_1(\xperp,z;\omega/c) }^2$ is important.

\subsection{A Numerical Example}

With a bit of numerical quadrature and optimization, we can improve our approximations a bit by working with the full overlap integral.  We will choose parameters similar to those of typical IOTA experiments, considering an electron bunch with $N_e \approx 1$, and nominal energy $E = 100$ MeV, corresponding  to $\gamma = 195.7$.  Interpreted in terms of uncertainties rather than variability (since we may only have one electron in total), the  normalized transverse emittances are about $\epsilon_x = \epsilon_y = 8.6$ nm, while the RMS relative momentum spread is about $\sigma_p = 1.29 \cdot\! 10^{-4}$.  However, we will continue to ignore emittance effects for now, and consider an electron of nominal energy on an on-axis trajectory.  Because the particle emittance is much smaller than the intrinsic RMS emittance for a coherent radiation mode, which is about $\tfrac{\lambda_1}{4\pi}$, and the relative spread in particle energy is much smaller than the intrinsic bandwidth of the spontaneous radiation, this neglect is expected to be a reasonable approximation.  (On the other hand, the entropic emittance of a single coherent mode is zero, so any finite emittance in the electron bunch will translate into \textit{some} partial coherence in the radiation).

We assume the bunches pass through a magnetostatic undulator with $a_u = 0.8$, $\lambda_u = 12.9$ cm, and $N_u = 6$.  While at present the actual undulators used in IOTA are planar, for the sake of simplicity, we will focus on the case of a helical undulator---the planar case can also be handled numerically, but is a bit more complicated because of the ``figure-of-eight'' motion in the average electron rest frame, and concomitant harmonics in the emission spectrum.

For an electron arriving at $t = 0$, upon re-scaling some variables, we have for the work integral
\be
\begin{split}
P_1(h)\cc &= a_{+}(h)\, e^{i\phi_+ +2\pi i N_u(1-h)\zeta_0}
\tfrac{ec}{\sqrt{{\pi}}}  \tfrac{\sqrt{N_u} a_u }{\sqrt{1+a_u^2} \beta_z} \!\!\! \int\limits_{-\zeta_0}^{1-\zeta_0} \!\! \!d\zeta\,\tfrac{ \sqrt{\zeta_R}}{\zeta - i \zeta_R} \,
e^{\frac{+ih a_u^2}{(1+a_u^2) 2\pi N_u \beta_z^2 (\zeta-i \zeta_R)}}
e^{i 2\pi N_u(1 - h) \zeta}
\end{split}
\ee
for the co-rotating polarization component, in which $\zeta_0 = \tfrac{z_0}{L_u}$ and $\zeta_R = \tfrac{z_R}{L_u}$ are the scaled waist location and Rayleigh range, respectively, and the effective ``harmonic number'' $h = \tfrac{\!k}{k_1} = \tfrac{\!\omega}{\omega_1}$ is a scaled frequency (not necessarily an integer).

First, we need to maximize $\tfrac{| P_1(h) |}{|a_+(h)| }$ with respect to the adjustable parameters $\zeta_0$ and $\zeta_R$, for all scaled frequencies $h$ of interest.  Then, we can choose $| a_+(h) |$ so as to satisfy the energy conservation constraint, which requires $| a_+(h) | = \max \tfrac{| P_1(h) |}{|a_+(h)| }$.  The corresponding variational bound on spectral density of Poynting flux is then $\tfrac{c}{\muz} | a_+(h) |^2$.

One can foresee two sorts of tradeoffs that influence the nature of the maximization.  A longer Rayleigh range would keep the on-axis light intensity from diffracting away as much over the region of interaction inside the undulator, but must also increase the focused spot size, and hence decrease the peak value of the on-axis light intensity.  Second, in comparison to the phase of the quivering electrons, the paraxial wavefronts experience a gradual extra Gouy phase shift (eventually accumulating to $\pi$ in all) as the light passes across the focal plane, so shifting the location of the waist far upstream or downstream would decrease the net de-phasing, but as a consequence of diffraction, also decrease the average on-axis light intensity as experienced by the electrons.

Once again, because the magnitude of this integral will be invariant under the reflection $\zeta_0 \to (1-\zeta_0)$, either the optimum will occur exactly at $\zeta_0 = \tfrac{1}{2}$, or there will be a degenerate maximum for some pair $\zeta_0 < \tfrac{1}{2}$ and $(1-\zeta_0) > \tfrac{1}{2}$.  Intuitively, it seems plausible that the transverse profile of the radiation mode should be symmetric about the midpoint of the source, suggesting that when the variational optimum splits, we should switch the trial mode shape, and  employ a superposition of several modes, or even just a shifted pair of Gaussian modes, in order to keep the field profile reflection symmetric.  This in itself could be an interesting prediction of the MPVP---not the fact that two Gaussian can in principle do better than one (that is a trivial consequence of the variational nature of the approximation), but that the variational approach starting with one Gaussian mode actually points to its own limitations, and suggests to us when we should consider a different trial solution.

\begin{figure}
	\includegraphics[height=0.16\textheight]{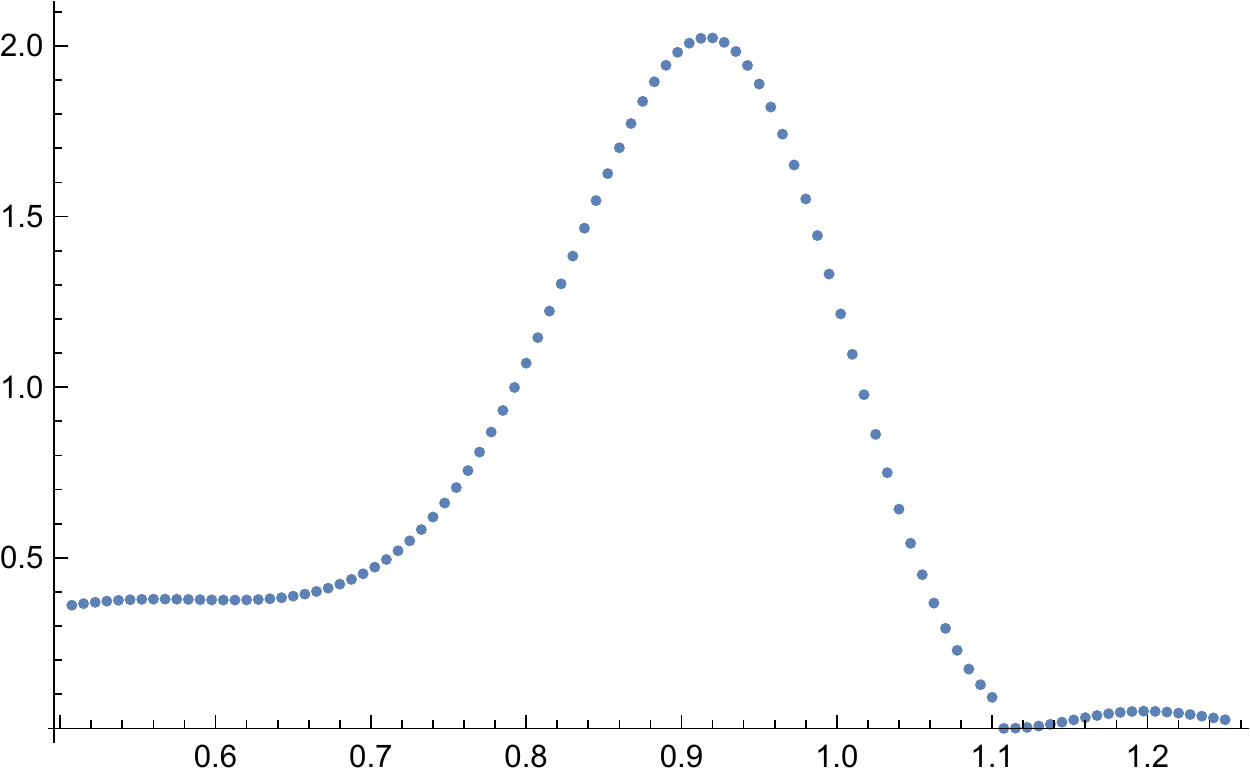}
	\includegraphics[height=0.16\textheight]{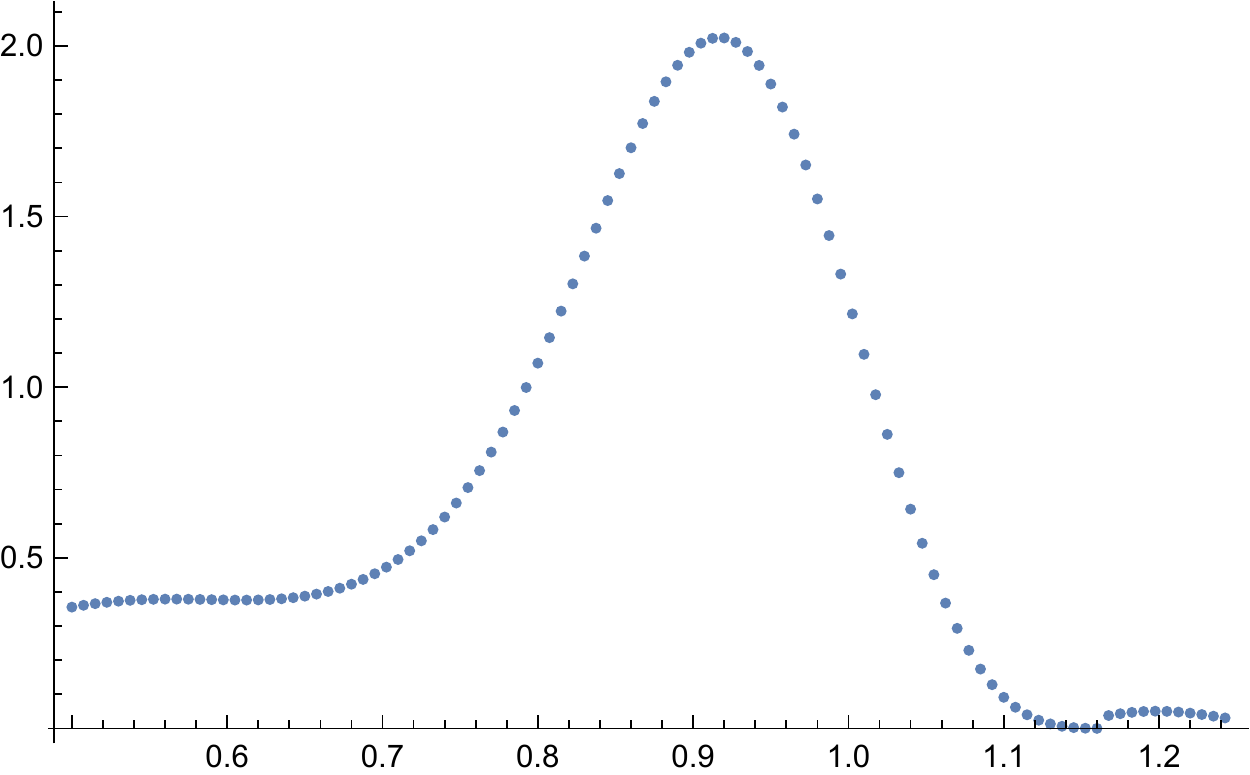}
 	\caption{Variational approximation to the power spectrum for spontaneous radiation in a weak helical undulator, in arbitrary units (vertical axis) versus the normalized frequency $\sfrac{\omega}{\omega_1}$.  Two different but typical optimizations are shown, to help assess numerical convergence to the global maximum.}\label{varA}
\end{figure}

\begin{figure}
	\includegraphics[height=0.16\textheight]{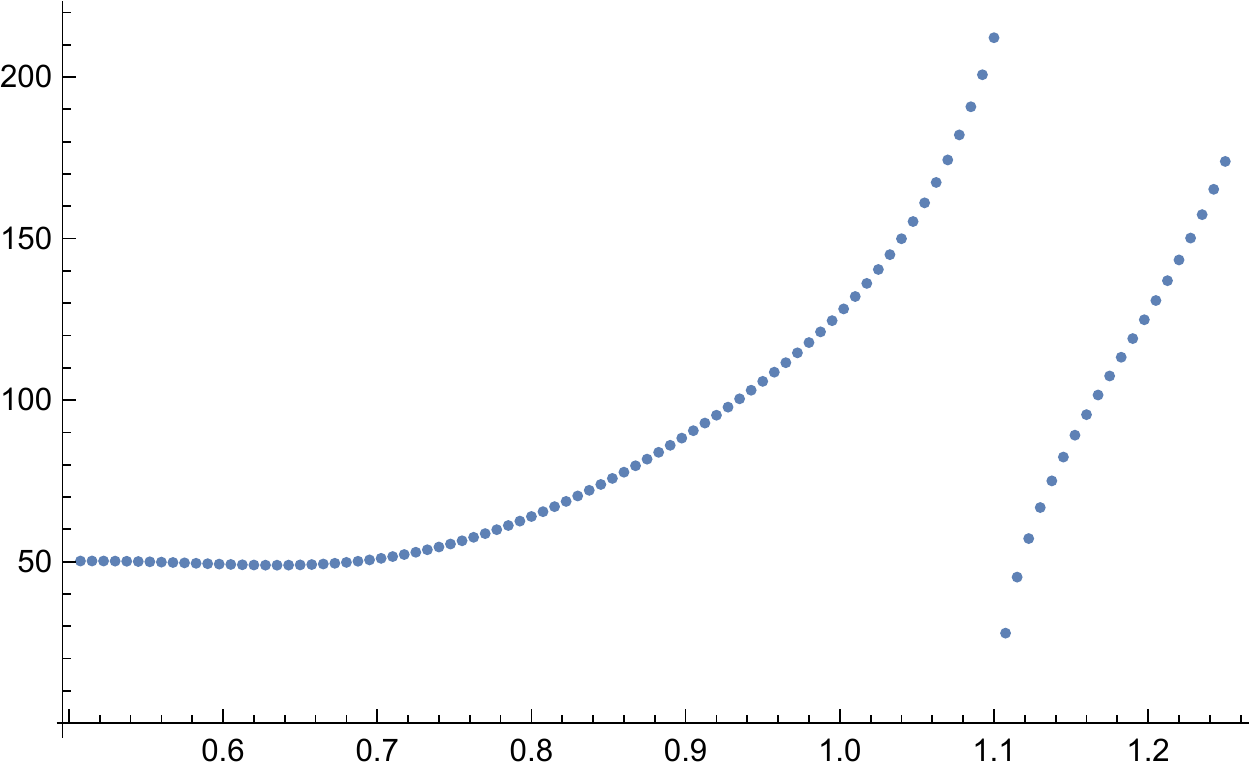}
	\includegraphics[height=0.16\textheight]{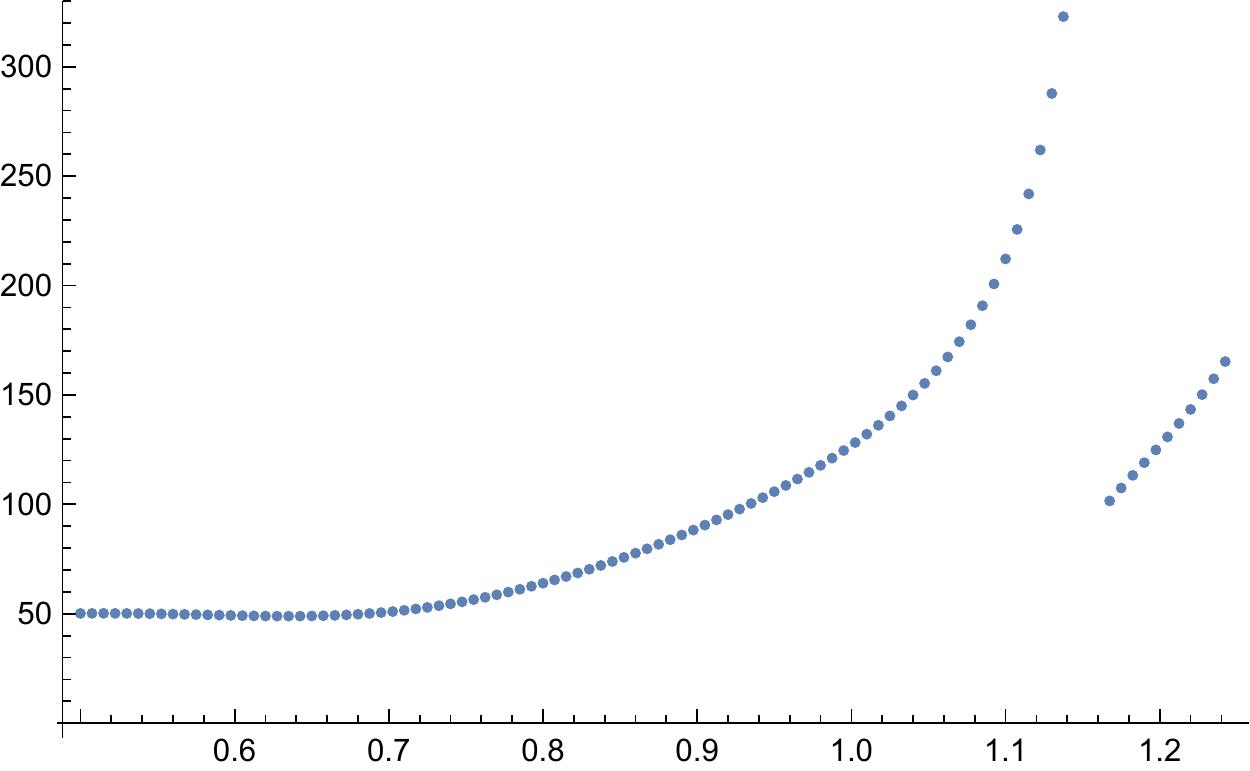} \caption{Optimized normalized spot size $\sfrac{\sigma}{\lambda_1}$ versus the normalized frequency $\sfrac{\omega}{\omega_1}$.  Two different but typical optimizations are shown, to help assess numerical convergence to the global maximum.}\label{varB}
\end{figure}

Such aspects will be explored elsewhere.  For now, we stick with the a single Gaussian mode, but force it to be symmetric by fixing $\zeta_0 = \tfrac{z_0}{L_u} = \tfrac{1}{2}$.  Results of numerical integration and optimization are shown in Figure \ref{varA} and Figure \ref{varB}, where we have optimized as a function of the one remaining free parameter corresponding to the Rayleigh range or the focused spot size, when other beam and beamline parameters are fixed at the values specified above.

%\begin{comment}
%\end{comment}

%\begin{comment}
%\end{comment}

The approximate power spectrum exhibits some features which are not obviously explained (by us).  Overall, even in the absence of any gain, the peak emission frequency is downshifted relative to the textbook resonance by about $8\%$, suspiciously close to $\tfrac{1}{2N_u}$.  The spectrum is also asymmetric, with a gently rolling plateau on the low-frequency side rather than pronounced sinc-like wiggles. (We may be seeing similar features in some GENESIS computer simulations in our next example, albeit in somewhat different context).

Over most of the central bandwidth, the paraxial parameter remains small, such that the spot size spans about $50$ wavelengths or more, except very near the local minimum in the emitted power, where the optimization algorithm struggled a it, presumably because the objective function becomes both small in value and flat.  In order to capture this emission minimum, the optimized spot size ends up jumping rapidly in value, but the precise behavior is hard to capture---different realizations of the function optimization with different initial guesses end up being fairly robust almost everywhere except close to the bottom of the dip, where apparently different local optima exist.

In fact, if we allow $\zeta_0 = \tfrac{z_0}{L_u}$ to vary, then the numerically optimized value also bifurcates around this same dip and secondary bump seen in the power spectrum (around $\tfrac{\omega}{\omega_1} \ge 1.1$), as well as in the low-frequency region (around $\tfrac{\omega}{\omega_1} \le 0.75$) where the variational power spectrum may also be trying to exhibit a secondary bump or at least an inflection point as it transitions from the central peak to a more rolling plateau.

%\begin{comment}
\begin{figure}
\includegraphics[height=0.125\textheight]{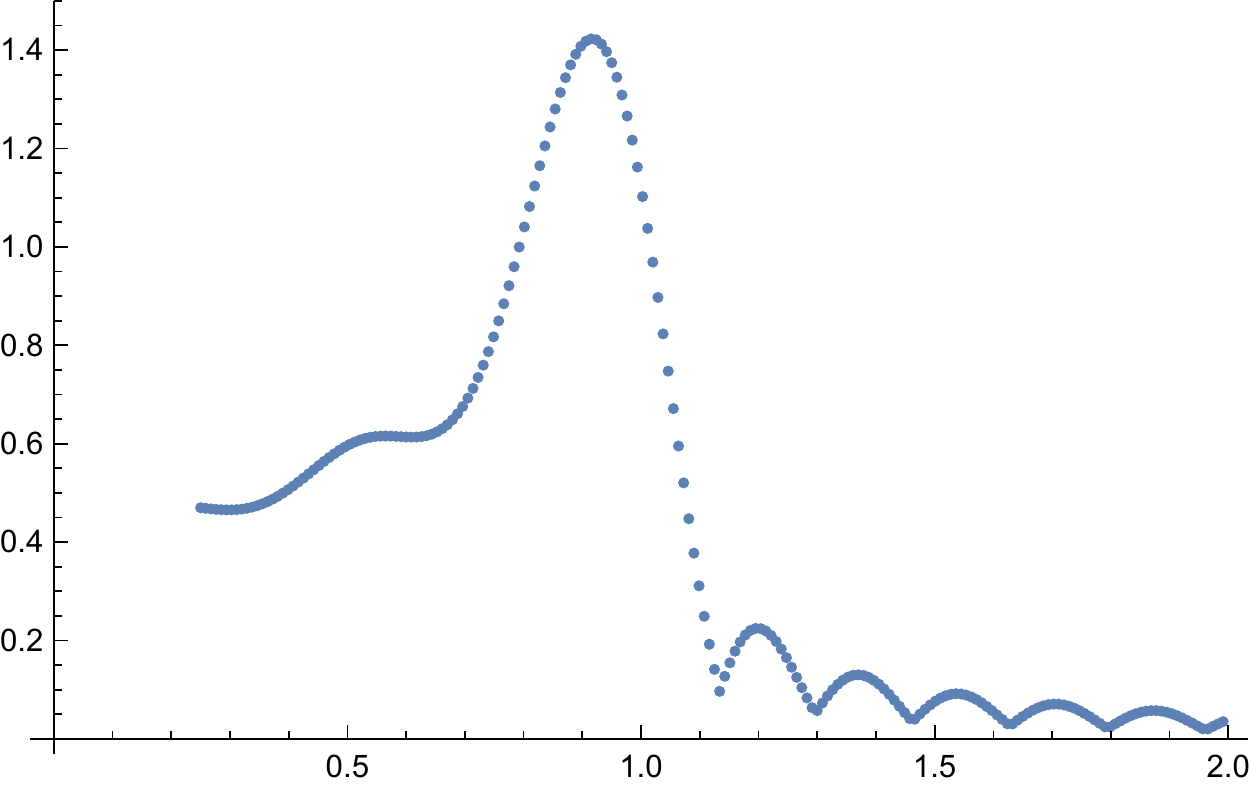}
\includegraphics[height=0.125\textheight]{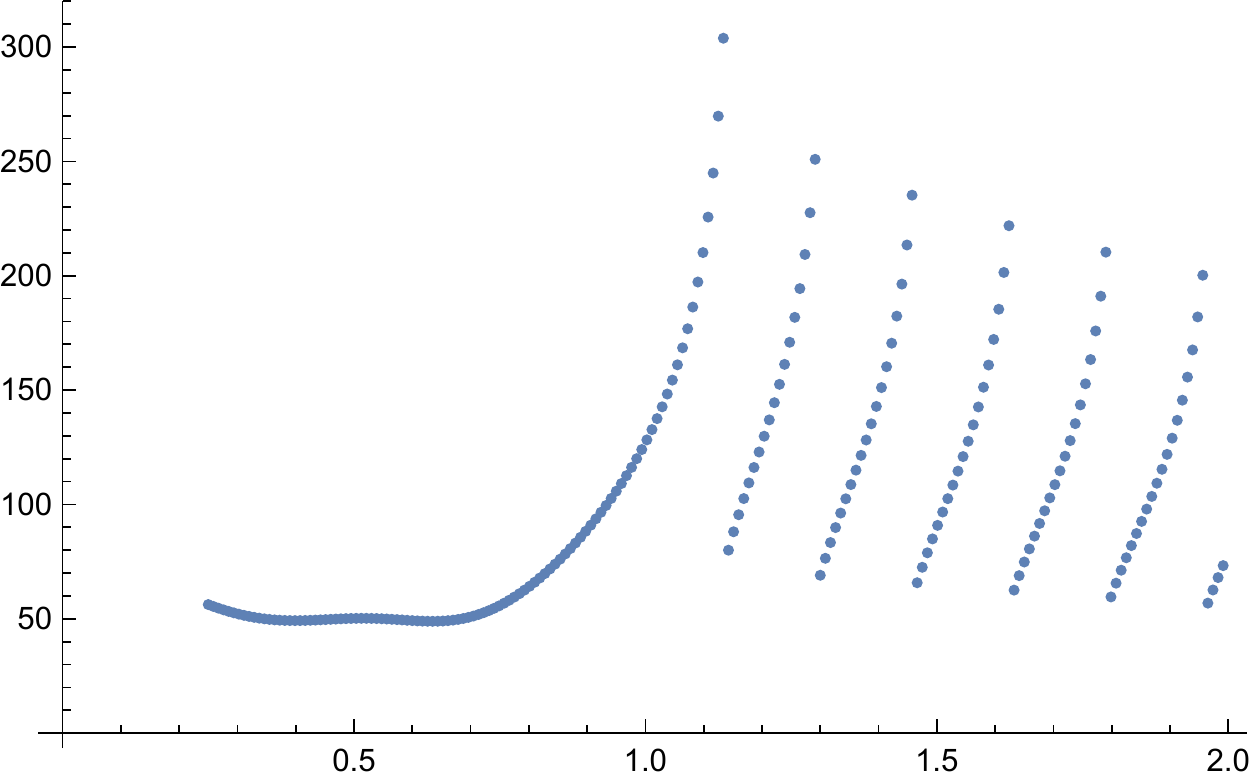}
\includegraphics[height=0.125\textheight]{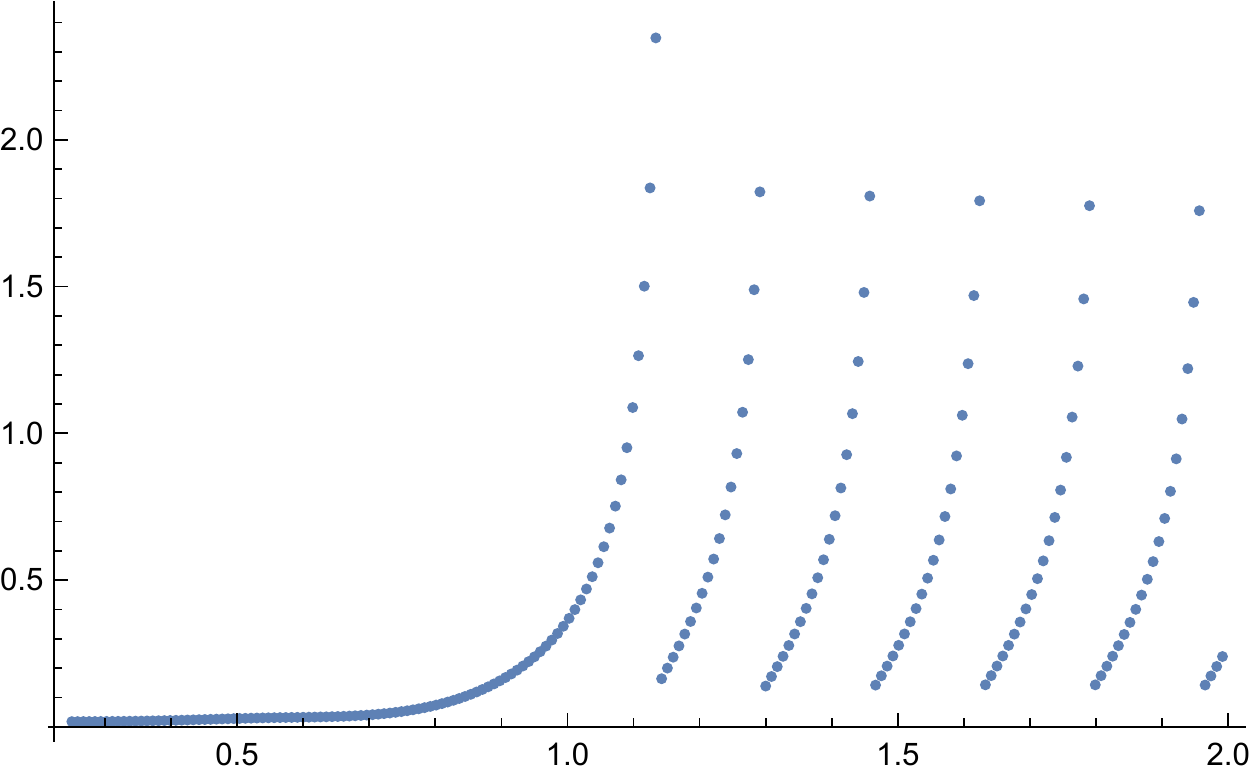}
 \caption{Variational approximation to the power spectrum (left) in arbitrary units,  relative focused spot size $\sfrac{\sigma}{\lambda_1}$ (middle), and normalized Rayleigh range $\sfrac{z_R}{L_u}$, as a function the normalized frequency $\sfrac{\omega}{\omega_1}$.}\label{varE}
\end{figure}
%\end{comment}

This is evident in another realization over a somewhat expanded frequency range and with somewhat higher numerical accuracy, shown in Figure \ref{varE}.  We see a few attempts at a sequence of successively smaller lobes in the power spectrum at frequencies above the peak, accompanied by rapid shifts in the fitted spot size in apparent attempts to fit the dips.  At frequencies below the peak, there is a slight ripple on top of a higher plateau.  We conjecture that the flatter plateau may correspond to a broad-spectrum synchrotron emission background underneath what we consider the coherent mode proper.  Very roughly, by matching each undulation to a circular path, the critical frequency for bending magnet emission can be estimated to be about $\omega_ c \sim \tfrac{6}{\pi^2} a_u\, \omega_1$.

\subsection{Applications to Harmonic Cascade FEL Radiation}

We actually stumbled upon the MPVP during an analysis of high-brightness x-ray generation via harmonic cascade in relativistic electron beams traveling through a sequence of undulators.\cite{penn:04a,penn:04b}
 
To start the harmonic emission process, energy modulations are induced in a relativistic electron beam passing through a modulator-undulator while overlapping a seed laser, and these energy modulations are converted downstream into spatial modulations (micro-bunching) via a specialized, highly dispersive beamline (chicane).  Micro-bunching will occur at the fundamental as well as higher harmonics, due to nonlinearities, and subsequently the electron beam is induced to radiate at a chosen harmonic in a suitably tuned, radiator-undulator.  Such modulation, dispersion, pre-bunching, and harmonic emission can be cascaded, where output radiation at a chosen harmonic from the previous stage can be used as the seed in the next stage, overlapping with a fresh part of the beam in a suitably-tuned downstream modulator-undulator, in order to induce energy modulation at the shorter wavelength, so as to produce still higher harmonics.
 
If the actual gain is sufficiently low in each radiator-undulator, so that prior micro-bunching from the upstream modulator/chicane dominates over dynamic self-bunching due to the FEL instability, then the MPVP may be used to estimate the profile and power of the output radiation in each stage.  Instead of resorting to detailed but time-consuming numerical simulations, the electromagnetic mode structure of the radiation can be approximated in terms of a paraxial trial field described by certain adjustable parameters such as spot size, waist location, amplitude, and carrier wavelength and phase. Some of these parameters are subsequently constrained by dynamical considerations, but some remain free within this model, and may be approximated at the end of the calculation by maximizing the resulting power.
 
 The MPVP provided mathematical justification for this intuitively plausible and appealing procedure, and the variational trial-function approach offered a reasonably accurate yet highly efficient ``analytic'' (or semi-numerical) approximation tool for estimating radiation power and optimizing beam-line design, much simpler and faster than either intricate FEL computer simulations (using GENESIS or other codes), or else single-particle algorithms based on summations over Li{\'e}nard-Wiechart fields, thus allowing for more economical parameter search during preliminary design optimization. 

\begin{figure}
	\includegraphics[height=0.22\textheight]{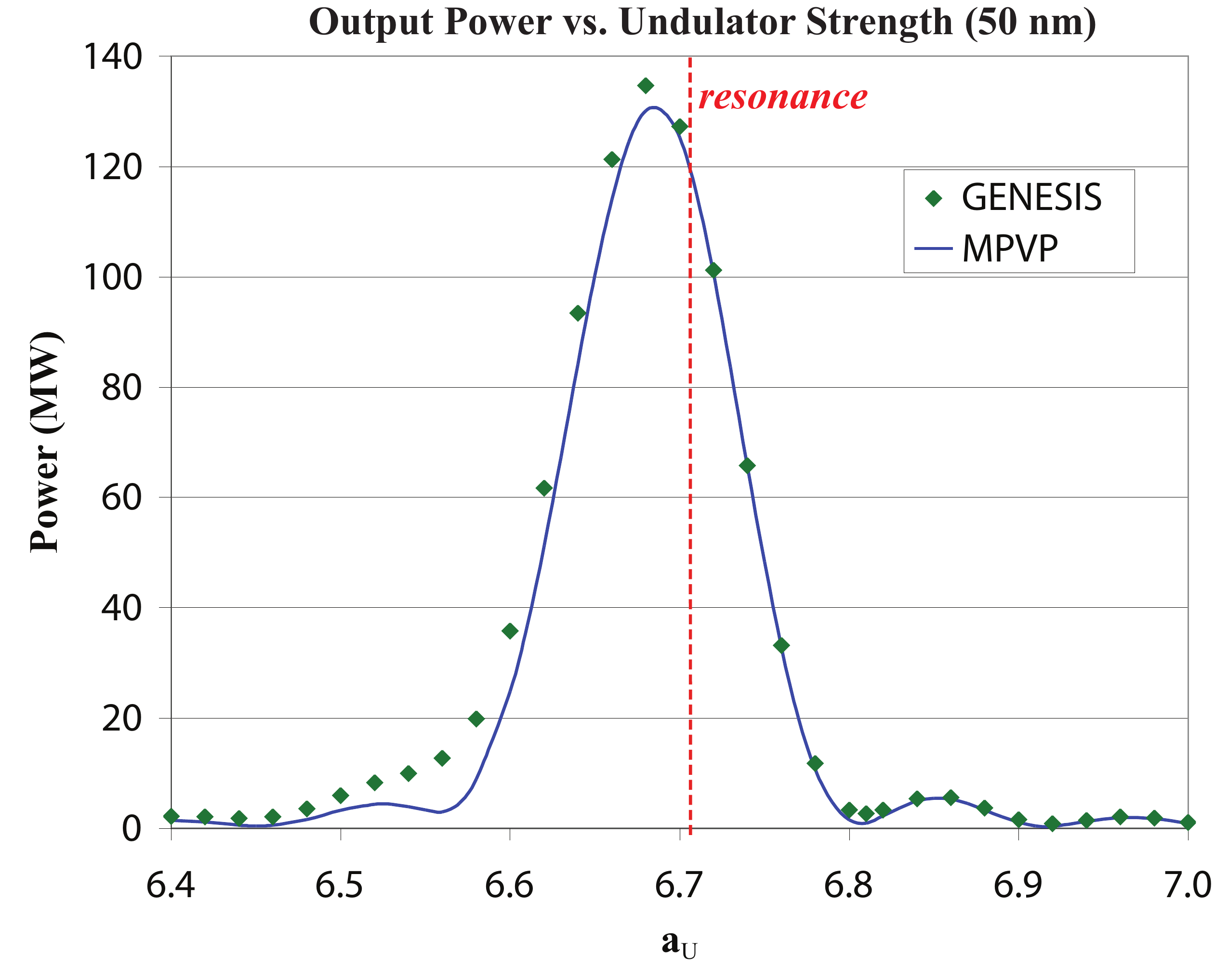}
	\includegraphics[height=0.22\textheight]{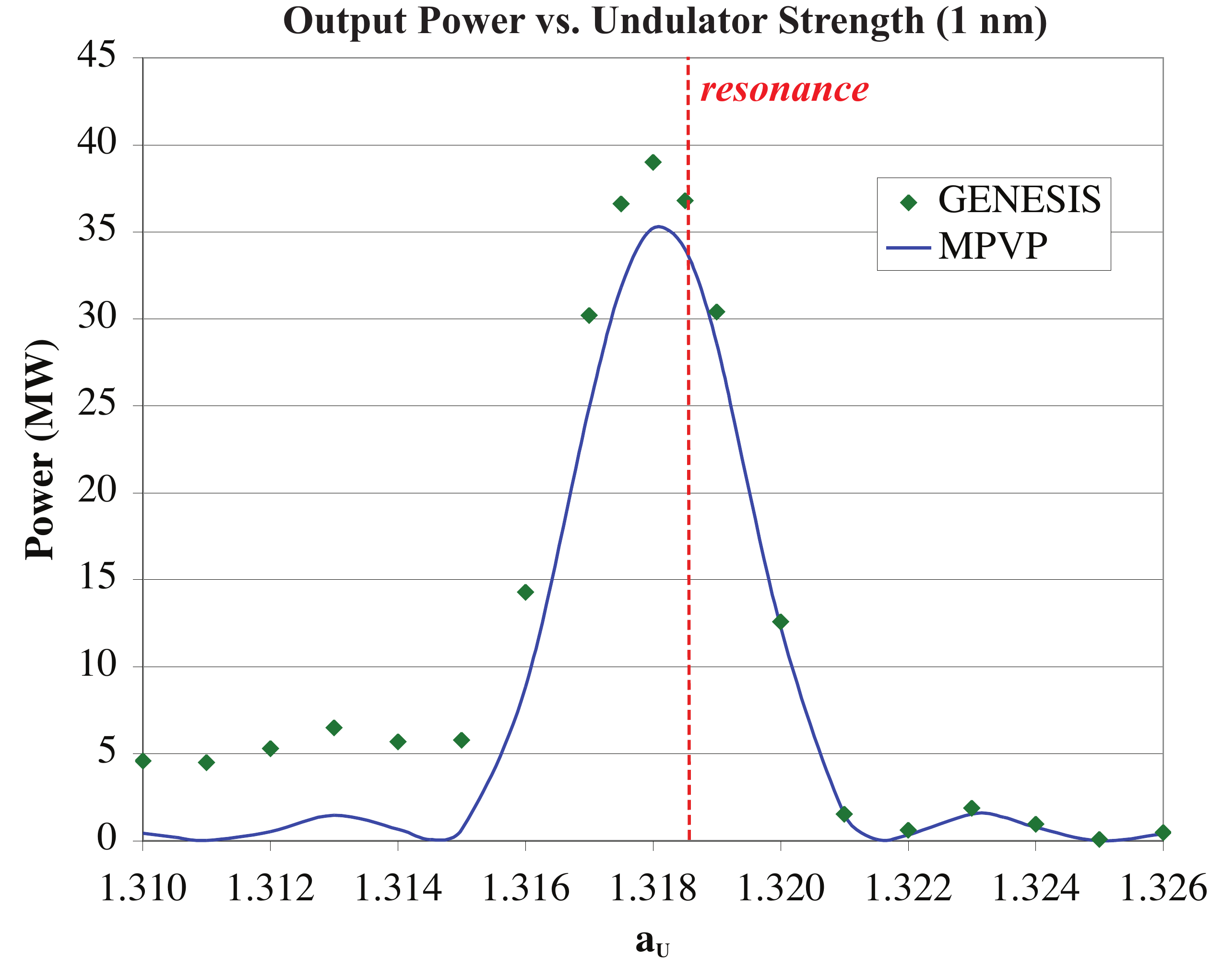}
 	\caption{Comparison of single-stage GENESIS FEL simulation with variational approximation based on a Gaussian trial mode, showing predicted output power with respect to normalized undulator strength $a_\stext{u}$, for configurations leading to emission at the $4$th harmonic, at a wavelength of $\lambda = 50$ nm, or emission at the a $3$rd harmonic, at a wavelength of $\lambda = 1$ nm.}\label{fig:comparison1}
\end{figure}

\begin{figure}
	\includegraphics[height=0.24\textheight]{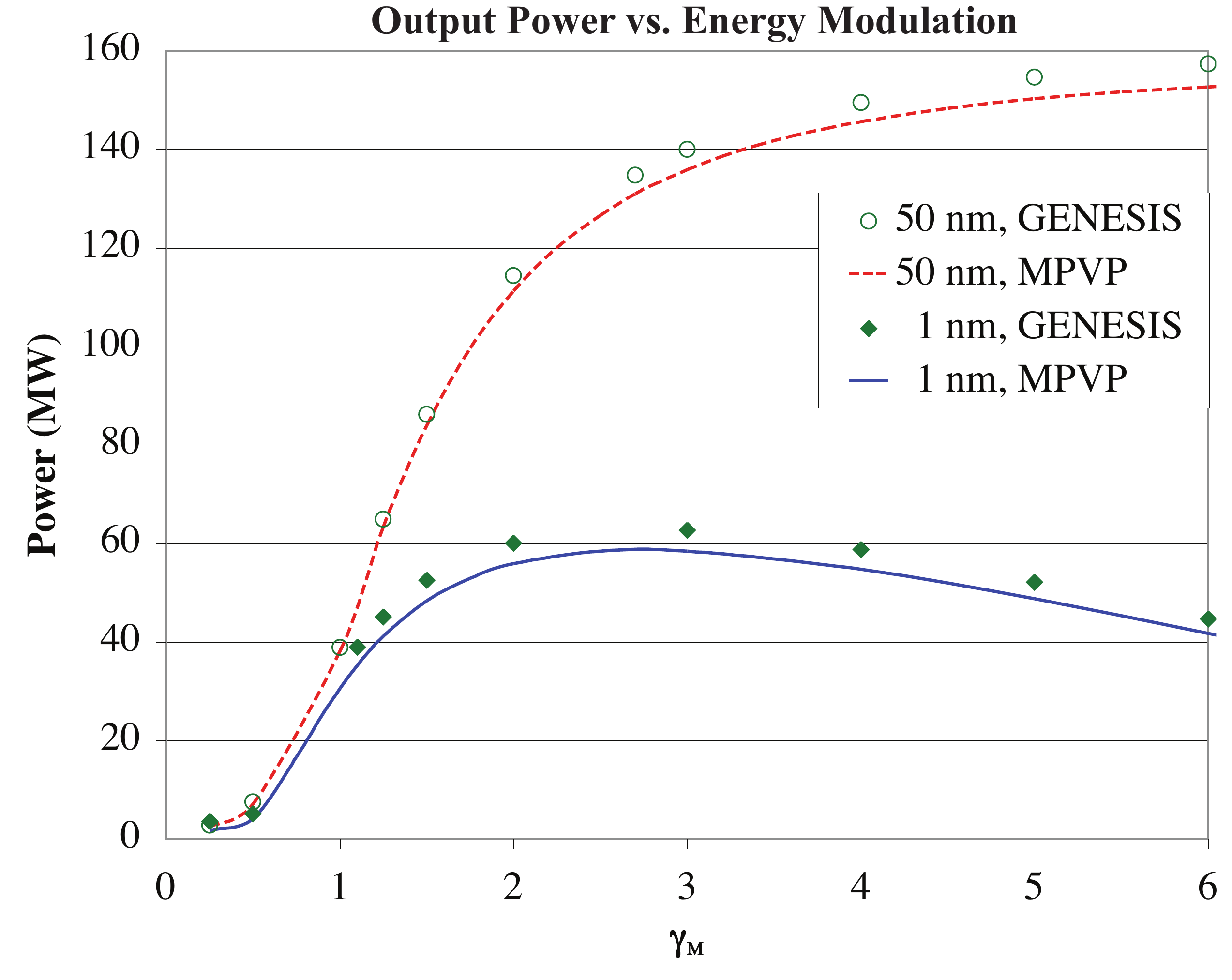}
 \caption{Comparison of single-stage GENESIS FEL simulation with variational approximation based on a Gaussian trial mode, showing predicted output power with respect to energy modulation parameter $\gamma_{\stext{M}}$, for the same two cases as described above.}\label{fig:comparison2}
\end{figure}

Some typical direct comparisons between the GENESIS simulations and simple variational approximations are shown in Figure \ref{fig:comparison1} and Figure \ref{fig:comparison2}, for the cases of an electron beam of energy $\gamma m_e c^2 = 3.1\mbox{ GeV}$ and normalized transverse emittance $\varepsilon_{\stext{R}} = 2 \mbox{ }\mu\mbox{m}$ in single stages of two different configurations, producing either $\lambda = 50\mbox{ nm}$ radiation (at the $4$th harmonic) or $\lambda = 1 \mbox{ nm}$ radiation (at the $3$rd harmonic).  Note that magnetic insertion devices were not necessarily assumed to be in the weak undulator regime.

%\begin{comment}
%\end{comment}

%\begin{comment}
%\end{comment}

In these studies, the trial solution was chosen to be a single fundamental Gaussian paraxial mode, with adjustable spot size and waist location, much like the previous example.  In efforts to improve designs through iterative parameter search, the MPVP proved far simpler and orders-of-magnitude faster than the detailed computer simulation using the GENESIS code, while still providing reasonable accuracy.  Because of the extremal nature of the MPVP and the high accuracy anticipated from the GENESIS simulations, as expected the variational estimates systematically underestimate the simulated numerical solutions, by an average of about $3\%$ for the $50\mbox{ nm}$ case and about $10\%$ for the $1\mbox{ nm}$ case.  Accuracy could be further improved by including additional Gauss-Hermite modes in superposition, or by adding other adjustable parameters to allow for ellipticity, annularity, skew or misalignment, etc., in the radiation profile, but this simple Gaussian trial mode proved adequate for preliminary design purposes.

In addition, the MPVP is particularly well suited to this sort of task, because the power-maximization over adjustable parameters in a trial radiation mode can be incorporated naturally into the iterative optimization cycle searching for other design parameters, such as energy modulation, undulator strength, chicane slippage factors, etc.
 
%%%%

\section{Conclusions}

Although similar to other well-known variational principles widely used in electromagnetic theory, the MPVP appears to be an independent result, and thus adds to the large family of variational techniques and tools available for electromagnetic problems in general, and undulator radiation in particular.  Mathematical details aside, at its most essential, the MPVP is really just a formalization of the ideas that current sources should ``look'' as much as possible like the radiation fields that they generate, or that classical charges radiate ``as much as possible'' consistent with maintaining consistency with energy conservation and the electromagnetic dispersion relation.  However simple, or even trivial, these relationships are not without practical content, and find useful application to problems involving emission by relativistic electrons in magnetic undulators, and possibly other sorts of classical radiation.

%%%%

\section{Acknowledgements}

% This research was supported by the US Department of Energy Office of Science, High Energy Physics Grant DE-SC0020062
% This work was supported by the US Department of Energy, Office of Science, Office of High Energy Physics,  under grant DE-SC0020062 and contract number DE-AC02-05CH11231

This research was supported by the US Department of Energy, Office of Science, Office of High Energy Physics, under grant DE-SC0020062, and under contract number DE-AC02-05CH11231.

We dedicate this work to the memory of our friend, colleague, and teacher, Max S.\ Zolotorev, 1941--2020.

%%% appendices
\appendix

\section{Fourier Transform Conventions}\label{fourier}

We take all Fourier transforms to be \textit{unitary}, in the sense of the Parseval-Plancherel identity.  With a limited alphabet, in physics we conventionally use the same symbol to denote various representations of the same physical vector field but connected via Fourier transforms; which domain is being assumed will be indicated explicitly by the arguments, or should otherwise be clear from the context.

That is, starting with a (possibly complex) vector field $\bv{f}(\bv{x}; t)$ which is a function of spatial position $\bv{x}$ and time $t$, we can transform between the \textit{time domain} and the so-called angular \textit{frequency domain} or \textit{spectral domain}, via the Fourier transform/ inverse transform pair:
\bsub
\begin{align}
\bv{f}(\bv{x}; \omega) &= \tfrac{1}{\sqrt{2\pi}} \! \int\! d t\, e^{+i \omega t} \bv{f}(\bv{x}; t),\\
\bv{f}(\bv{x}; t) &= \tfrac{1}{\sqrt{2\pi}} \!\int \!d \omega \, e^{-i \omega t}\bv{f}(\bv{x}; \omega),
\end{align}
\esub
We can also transform between the \textit{position-space}, or \textit{real-space} representation, and 
\textit{wavenumber-space}, or \textit{reciprocal space} representation, via the pair
\bsub
\begin{align}
\bv{f}(\bv{k}; t) &= \tfrac{1}{(2\pi)^{3/2}} \! \int \! d^3 \bv{x} \, e^{-i \bv{k} \cdot \bv{x} } \bv{f}(\bv{x}; t),\\
\bv{f}(\bv{x}; t) &= \tfrac{1}{(2\pi)^{3/2}} \! \int \! d^3  \bv{k} \, e^{+i \bv{k} \cdot \bv{x}}  \bv{f}(\bv{k}; t).
\end{align}
\esub
Opposite sign conventions are employed for time and space transforms because the generic complex plane wave is taken to be $e^{i\bv{k} \cdot \bv{x} - i \omega t}$ in most physics textbooks, particular in relativistic electrodynamics.

These transforms can also be composed to transform between the space-time representation $\bv{f}(\bv{x};t)$ and the full four-dimensional Fourier representation $\bv{f}(\bv{k}; \omega)$.  We can also restrict the spatial transform in an obvious manner to either just the transverse coordinates, $\bv{x}_{\perp}$, leading to a representation with respect to the transverse wavevector $\bv{k}_{\perp}$, or else just with respect to the longitudinal position $z$, obtaining a Fourier representation in the longitudinal wavenumber $k_z$.

Two complex vector fields will then satisfy the  temporal Parseval-Plancherel relations:
\bsub
\begin{align}
\int \! dt \, \bv{f}(\bv{x}; t)\cc \!\cdot\! \bv{g}(\bv{x}; t) &= \int d\omega\,  \bv{f}(\bv{x}; \omega)\cc \!\cdot\! \bv{g}(\bv{x}; \omega),\\
\int \! dt \, \bv{f}(\bv{k}; t)\cc \!\cdot\! \bv{g}(\bv{k}; t) &= \int d\omega\,  \bv{f}(\bv{k}; \omega)\cc \!\cdot\! \bv{g}(\bv{k}; \omega),
\end{align}
\esub
at all positions or wavevectors where the fields are square-integrable in time or frequency.  Similarly, the vector fields will satisfy the spatial Parseval-Plancherel relations:
\bsub
\begin{align}
\int \! d^3 \bv{x}\,  \bv{f}(\bv{x}; t)\cc \!\cdot\! \bv{g}(\bv{x}; t) &= \int d^3 \bv{k} \, \bv{f}(\bv{k}; t)\cc \!\cdot\! \bv{g}(\bv{k}; t),\\
\int \! d^3 \bv{x}\,  \bv{f}(\bv{x}; \omega)\cc \!\cdot\! \bv{g}(\bv{x}; \omega) &= \int d^3 \bv{k} \, \bv{f}(\bv{x}; \omega)\cc \!\cdot\! \bv{g}(\bv{x}; \omega),
\end{align}
\esub
at all times or frequencies where the fields are square-integrable in real or reciprocal space.

For vector fields which are both smooth and square-integrable over all space, we can relate functional transversality or irrotationality to geometric properties but in $\bv{k}$-space.  Such a vector field is solenoidal, or functionally transverse, if and only if it is in effect geometrically transverse to the wavevector everywhere in Fourier space:
\be
\grad\!\cdot\! \bv{f}(\bv{x}; t) = 0  \; \text{ for all } \bv{x} \in \realsymbol^3 \;\; \iff\;\; \bv{k}\!\cdot\! \bv{f}(\bv{k}; t) = 0  \;  \text{ for all } \bv{k} \in \realsymbol^3 ,
\ee
Similarly, a smooth, square-integrable vector field is irrotational, or functionally longitudinal, if and only if it is geometrically longitudinal in Fourier space, i.e., always parallel to the wavevector:
\be
\grad\!\times\! \bv{f}(\bv{x}; t) = \bv{0} \; \text{ for all } \bv{x} \in \realsymbol^3 \;\;\iff\;\; \bv{k}\!\times\! \bv{f}(\bv{k}; t) = \bv{0} \;  \text{ for all } \bv{k} \in \realsymbol^3.
\ee
Inside domains of more complicated topology, decompositions into of solenoidal, irrotational, and harmonic vector fields satisfying differential constraints and appropriate boundary conditions  are somewhat more subtle.

%%%%

\section{Vector Spherical Harmonics}\label{appendix_vsh}

Just as the usual spherical harmonics form a basis for scalar functions on the unit sphere which are eigenstates of ``orbital'' angular momentum, and hence transform simply under rotations, we may also utilize a basis for vector-fields on the unit sphere which are eigenstates of ``total'' angular momentum and which therefore will also transform simply under rotations.

\subsection{Basic Definitions and Properties}

Regarding vector spherical harmonics, various definitions, conventions, and normalizations abound.  For integer $\ell$ and $m$, we will define the primary vector spherical harmonics as
\be
\bv{X}_{\ell \, m}(\unitvec{r}) =
\begin{cases}
 \tfrac{1}{i \sqrt{\ell(\ell+1)}}\,  \bv{x} \!\times\! \grad Y_{\ell \, m}(\unitvec{r}) & \text{ if } \ell >0,\, \abs{m} \le \ell\\
 0 & \text{ if } \ell = m = 0,
 \end{cases}
\ee
in terms of the ordinary (scalar) spherical harmonics $Y_{\ell \, m}(\unitvec{r}) = Y_{\ell m}(\theta,\phi)$.  In addition, we define two other related sets of vector fields on the unit sphere, 
\be
\bv{Z}_{\ell \, m}(\unitvec{r}) = \unitvec{r} \times\! \bv{X}_{\ell \, m}(\unitvec{r}),
\ee
for which a useful alternate construction for  $\bv{Z}_{\ell \, m}$ is
\be
\bv{Z}_{\ell \, m}(\unitvec{r}) =
\begin{cases}
 \tfrac{i}{\sqrt{\ell(\ell+1)}} \, r \, \grad Y_{\ell \, m}(\theta, \varphi) & \text{ if } \ell >0,\, \abs{m} \le \ell\\
 0 & \text{ if } \ell = m = 0
 \end{cases},
\ee
and
\be
\bv{R}_{\ell \, m}(\unitvec{r}) = \unitvec{r} \, Y_{\ell \, m}(\unitvec{r}),
\ee
which are purely radial.

That $\bv{X}_{0 \, 0}(\unitvec{r}) $ and $\bv{Z}_{0 \, 0}(\unitvec{r})$ both vanish identically may be viewed as a consequence of Brouwer's ``hair-combing'' theorem, which establishes the impossibility of constructing a continuous non-singular vector field everywhere tangent to the surface of a sphere.  This is closely related to the fact that spherically-symmetric solutions to the free-space Maxwell equations can exist only in the static ($\omega = 0$) and hence non-radiative limit.

For any allowed $\ell$ and $m,$ these vector spherical harmonics are geometrically orthogonal as vectors at every point $\unitvec{r}$ on the unit sphere:
\be
\bv{X}_{\ell \, m}^{*}(\unitvec{r}) \cdot \bv{Z}_{\ell \, m}(\unitvec{r})= \bv{X}_{\ell \, m}^{*}(\unitvec{r}) \cdot \bv{R}_{\ell \, m}(\unitvec{r}) = \bv{Z}_{\ell \, m}^{*}(\unitvec{r}) \cdot \bv{R}_{\ell \, m}(\unitvec{r}) = 0,
\ee
 As vector fields, these fields are also functionally orthonormal when integrated over solid angles:
\be
\int d^2\Omega(\unitvec{r}) \, \bv{W}_{\ell\,m}^{*}(\theta, \varphi) \!\cdot\!  \bv{W}'_{\ell'\,m'}(\theta, \varphi) = \delta_{\bv{W}\,\bv{W}'}\delta_{\ell\, \ell'}\delta_{m \,m'}\left(1 - \delta_{\bv{W}\bv{X}}\delta_{\ell\, 0}\right)\left(1 - \delta_{\bv{W}\bv{Z}}\delta_{\ell\, 0}\right),
\ee
where $\bv{W}, \bv{W}' \in \{\bv{X}, \bv{Z}, \bv{R} \}$.  In addition, they also collectively constitute a complete set for vector fields defined on the unit sphere:
\be
\sum_{\bv{W} = \bv{X}, \bv{Z}, \bv{R}} \sum_{\ell = 0}^{\infty} \sum_{m = -\ell}^{\ell}  \bv{W}_{\ell\,m}(\theta, \varphi) \,  \bv{W}_{\ell\,m}(\theta', \varphi')\hc =   \delta(\varphi - \varphi')\, \delta(\cos \theta - \cos \theta') \, \mathcal{I}_{3},
\ee
where the dagger denotes the Hermitian transpose operation, and $\mathcal{I}_{3}$ is the $3 \times 3$ identity matrix on the ``polarization'' degrees of freedom.

Using various vector identities and the properties of the scalar spherical harmonics, these spherical vector harmonic basis fields also can be shown to satisfy a number other useful of geometric and differential properties.  Letting $f(r)$ be an arbitrary (but differentiable, where needed) function of the radius $r$, we find:
\begin{subequations}
\begin{align}
\unitvec{r} \cdot \!\bv{X}_{\ell \, m} &= 0,\\
\unitvec{r} \cdot \bv{Z}_{\ell \, m} &= 0,\\
\unitvec{r} \cdot \bv{R}_{\ell \, m} &= Y_{\ell\, m};
\end{align}
\end{subequations}
and
\begin{subequations}
\begin{align}
\unitvec{r} \times \!\bv{X}_{\ell \, m} &= \phantom{-}\bv{Z}_{\ell \, m},\\
\unitvec{r} \times \bv{Z}_{\ell \, m} &=-\bv{X}_{\ell \, m},\\
\unitvec{r} \times \bv{R}_{\ell \, m} &= \phantom{-}\bv{0};
\end{align}
\end{subequations}
as well as
\begin{subequations}
\begin{align}
\grad \!\cdot\! \left[ f(r) \bv{X}_{\ell \, m}\right] &= \phantom{+} 0,\\
\grad \!\cdot\! \left[ f(r) \bv{Z}_{\ell \, m}\right] &= -i \sqrt{\ell(\ell+1)} \frac{f(r)}{r} Y_{\ell\, m},\\
\grad \!\cdot\! \left[ f(r) \bv{R}_{\ell \, m}\right] &= \phantom{+}\left[  \frac{d f(r)}{dr} +  \frac{2 f(r)}{r} \right]Y_{\ell\, m};
\end{align}
\end{subequations}
together with
\begin{subequations}
\begin{align}
\grad\! \times \!\left[ f(r) \bv{X}_{\ell \, m}\right] &=  \phantom{-1} i\sqrt{\ell(\ell+1)} \frac{f(r)}{r} \bv{R}_{\ell \, m}  + \left[  \frac{d f(r)}{dr}   + \frac{f(r)}{r}  \right]  \bv{Z}_{\ell \, m},\\
\grad\!  \times\! \left[ f(r) \bv{Z}_{\ell \, m}\right]  &= -\left[ \frac{f(r)}{r} + \frac{d f(r)}{d r} \right]  \bv{X}_{\ell \, m},\\
\grad\!  \times\! \left[ f(r) \bv{R}_{\ell \, m}\right] &= -i \sqrt{\ell(\ell+1)} \frac{f(r)}{r}  \bv{X}_{\ell \, m}.
\end{align}
\end{subequations}

\subsection{Expansions of Vector Fields}

A vector field $\bv{f}(\bv{r})$  can be decomposed as a linear combination of the vector spherical harmonics with expansion coefficients that are functions only of the radial position, namely
\be
\bv{f}(\bv{r}) = \sum\limits_{\ell = 0}^{\infty} \sum\limits_{m = -\ell}^{\ell}  
f^{X}_{\ell\, m}(r) \, \bv{X}_{\ell \, m}(\unitvec{r}) + f^{Z}_{\ell\, m}(r) \, \bv{Z}_{\ell \, m}(\unitvec{r}) +f^{R}_{\ell\, m}(r)  \, \bv{R}_{\ell \, m}(\unitvec{r}),
\ee
where 
\be
f^{R}_{\ell\, m}(r) = \int d^2\Omega(\unitvec{r})\,  \bv{R}^{*}_{\ell \, m}(\unitvec{r}) \!\cdot \! \bv{f}(\bv{r}) 
=\int d^2\Omega(\unitvec{r})\,  Y^{*}_{\ell \, m}(\unitvec{r})\,  \unitvec{r} \!\cdot \! \bv{f}(\bv{r}),
\ee
for all allowed $\ell$ and $m$,
and
\be
f^{X}_{\ell\, m}(r) = \int d^2\Omega(\unitvec{r})\,  \bv{X}^{*}_{\ell \, m}(\unitvec{r}) \!\cdot \! \bv{f}(\bv{r}) 
= \tfrac{1}{\sqrt{\ell(\ell+1)}} \! \int d^2\Omega(\unitvec{r})\,  Y^{*}_{\ell \, m}(\unitvec{r})\, \tfrac{1}{i} \bv{x} \!\times\! \grad \!\cdot \! \bv{f}(\bv{r}),
\ee
and
\be
\begin{split}
f^{Z}_{\ell\, m}(r) &= \int d^2\Omega(\unitvec{r})\,  \bv{Z}^{*}_{\ell \, m}(\unitvec{r}) \!\cdot \! \bv{f}(\bv{r}) 
= -\tfrac{i}{\sqrt{\ell(\ell+1)}} \! \int d^2\Omega(\unitvec{r})\,  r \grad Y^{*}_{\ell \, m}(\unitvec{r}) \cdot  \bv{f}(\bv{r}) \\
&=  
\tfrac{i}{\sqrt{\ell(\ell+1)}} \! \int d^2\Omega(\unitvec{r})\,  Y^{*}_{\ell \, m}(\unitvec{r}) \left[  r \grad \!\cdot\! \bv{f} - \tfrac{\del}{\del r} \left( \bv{r}\!\cdot\! \bv{f} \right) -\unitvec{r} \!\cdot\! \bv{f} \right] \\
&= \tfrac{i}{\sqrt{\ell(\ell+1)}} \! \int d^2\Omega(\unitvec{r})\,  Y^{*}_{\ell \, m}(\unitvec{r})\, r \, \grad  \cdot \left[   (1 - \unitvec{r}\unitvec{r}\trans)  \bv{f} \right], \\
\end{split}
\ee
for $\ell > 0$ and $| m | \le \ell$.

These basis fields may also be useful in performing Helmholtz-Hodge type decompositions.  The gradient of a scalar function $\Phi(\bv{x})$ can be expressed as 
\be
\grad\Phi(\bv{x}) = \sum\limits_{\ell = 0}^{\infty} \sum\limits_{m = -\ell}^{\ell}  \bigl[  -i \sqrt{ \ell(\ell+1) } \, \tfrac{\Phi_{\ell \, m}(r)}{r}  \, \bv{Z}_{\ell \, m}(\unitvec{r})  +  \tfrac{d \Phi_{\ell \, m}(r)}{dr}  \,  \bv{R}_{\ell \, m} (\unitvec{r})
  \bigr].
\ee
It is straightforward to verify explicitly that this combination of vector spherical harmonics has vanishing curl, and conversely, that any vector field that lies in the span of the vector spherical harmonics, and is curl-free, can be expressed in this form.  That is,  if $\grad\!\times\! \bv{g}(\bv{x}) = \bv{0}$ everywhere in space, then there exists some scalar field $\Psi(\bv{r})$ such that
\be
\bv{g}(\bv{x}) = \grad \Psi(\bv{x}) = \sum\limits_{\ell = 0}^{\infty} \sum\limits_{m = -\ell}^{\ell}  \bigl[  -i \sqrt{ \ell(\ell+1) }  \tfrac{\Psi_{\ell \, m}(r)}{r}  \, \bv{Z}_{\ell \, m}(\unitvec{r})  +  \tfrac{d \Psi_{\ell \, m}(r)}{dr} \, \bv{R}_{\ell \, m} (\unitvec{r}),
\ee
where the $\Psi_{\ell\, m}(r)$ can be determined most easily from the condition $\frac{d}{dr} \Psi_{\ell \, m}(r) = g^{R}_{\ell\, m}(r)$, together with
\be
\int d^2\Omega(\unitvec{r})\, Y^{*}_{\ell \, m}(\unitvec{r})\, \grad\!\cdot\!\bv{g}(\bv{x}) = - \ell(\ell+1) \tfrac{\Psi_{\ell \, m}(r)}{r^2}  + \tfrac{d^2 \Psi_{\ell \, m}(r)}{dr^2}  + \tfrac{1}{r} \tfrac{d \Psi_{\ell \, m}(r)}{dr}.
\ee

Supposing instead we have a solenoidal vector field $\bv{b}(\bv{x})$ such that  $\grad \cdot \bv{b}(\bv{x}) = 0$ everywhere in space, then $\bv{b}(\bv{x})$ can be written as the curl of some vector potential $\bv{a}(\bv{x})$ (which can itself be taken as divergence-free), in the form:
\be
\bv{b}(\bv{x}) = \grad\!\times\! \bv{a} (\bv{x}) =  \sum\limits_{\ell = 0}^{\infty} \sum\limits_{m = -\ell}^{\ell}  
b^{X}_{\ell\, m}(r)  \bv{X}_{\ell \, m}(\unitvec{r}) +  \grad\!\times\! \left[ a^{X}_{\ell\, m}(r) \bv{X}_{\ell \, m}(\unitvec{r}) \right]  
\ee

Knowing just $\bv{b}(\bv{x})$, the $a^{X}_{\ell\, m}(r)$ can be determined most simply from
\be
\int d^2\Omega(\unitvec{r})\,  Y^{*}_{\ell \, m}(\unitvec{r})\, \unitvec{r}\!\cdot\!\bv{b}(\bv{r}) = b^{R}_{\ell\, m}(r) = i \sqrt{\ell(\ell+1)} \,  \frac{1}{r} a^{X}_{\ell\, m}(r).
\ee
The $b^{Z}_{\ell\, m}(r)$ coefficients do not need to be evaluated independently, since 
\be
b^{Z}_{\ell\, m} = \tfrac{d  a^{X}_{\ell\, m}}{dr}  + \tfrac{a^{X}_{\ell\, m}}{r}.
\ee
The other expansion coefficients of the vector potential may be calculated by solving  the differential equations
\be
i \sqrt{\ell(\ell+1)} \,\tfrac{a^{Z}_{\ell\, m}(r)}{r} = \tfrac{d a^{R}_{\ell\, m}(r)}{r} + \tfrac{2a^{R}_{\ell\, m}(r)}{r},
\ee
ensuring that the vector potential is also solenoidal, and
\be
\tfrac{a^{Z}_{\ell\, m}(r)}{r}  + \tfrac{d a^{Z}_{\ell\, m}(r)}{dr}  + i \sqrt{\ell(\ell+1)} \tfrac{a^{R}_{\ell\, m}(r)}{r}   + b^{X}_{\ell\, m}(r) = 0,
\ee
ensuring that its curl correctly reproduces the original vector field.  Alternatively, we may determine the remaining components in two stages. First, we can replace the $\grad\!\cdot\!\bv{a}(\bv{x}) = 0$ gauge condition with a $a^{Z}_{\ell\, m}(r) = 0$ gauge condition, and solve algebraically for the $a^{R}_{\ell\, m}(r)$ using 
\be
 i \sqrt{\ell(\ell+1)}\, \tfrac{a^{R}_{\ell\, m}(r)}{r} = - b^{X}_{\ell\, m}(r).
\ee
Then we can shift to a new gauge via a generator $\chi(\bv{x})$ whose expansion coefficients can be chosen (via solution of a differential equation) so that $\bv{a}(\bv{r}) + \grad \chi(\bv{r})$ is divergence-free.

\subsection{Hansen Multipoles}

When representing solutions of the Helmholtz equation, it is convenient to define a related family of vector fields, known as ``Hansen multipoles,'' from which we can construct general solutions to the homogeneous Helmholtz equation with various boundary conditions:
\bsub
\begin{align}
\bv{M}^{\sigma}_{\ell\, m}(\bv{x}; k) &= +z^{\sigma}_{\ell}(kr) \, \bv{X}_{\ell \, m}(\unitvec{r}),\\
 \bv{N}^{\sigma}_{\ell\, m}(\bv{x}; k) &=+\tfrac{i}{k}\, \grad\!\times\! \bv{M}^{\sigma}_{\ell\, m}(\bv{x}; k),\\
\bv{L}^{\sigma}_{\ell\, m}(\bv{x}; k) &= -\tfrac{i}{k} \,\grad \left[ z^{\sigma}_{\ell}(kr) Y_{\ell\, m}(\unitvec{r})  \right],
\end{align}
\esub
where $\sigma \in \left\{ -, 0, + \right\}$, for which  $z^{0}_{\ell}(kr)  = j_{\ell}(kr)$ and $z^{\pm}_{\ell}(kr) = h_{\ell}^{\pm}(kr)$ are spherical Bessel and spherical Hankel functions.

The Hansen multipoles satisfy the divergence relations
\bsub
\begin{align}
\grad\!\cdot\! \bv{M}^{\sigma}_{\ell\, m}(\bv{x}; k) &=  0,\\
\grad\!\cdot\! \bv{N}^{\sigma}_{\ell\, m}(\bv{x}; k) &= 0,\\
\grad\!\cdot\! \bv{L}^{\sigma}_{\ell\, m}(\bv{x}; k) &=  ik\,z^{\sigma}_{\ell}(kr) \, Y_{\ell\, m}(\unitvec{r}),
\end{align}
\esub
and the curl relations
\bsub
\begin{align}
\grad\!\times\! \bv{M}^{\sigma}_{\ell\, m}(\bv{r}; k) &=  -ik \, \bv{N}^{\sigma}_{\ell\, m}(\bv{x}; k),\\
\grad\!\times\! \bv{N}^{\sigma}_{\ell\, m}(\bv{r}; k) &=  +ik \, \bv{N}^{\sigma}_{\ell\, m}(\bv{x}; k),\\
\grad\!\times\! \bv{L}^{\sigma}_{\ell\, m}(\bv{x}; k) &=  \phantom{+}\bv{0}.
\end{align}
\esub
Therefore, for $\ell > 0$, the $\bv{M}^{\sigma}_{\ell\, m}(\bv{r}; k)$ and $\bv{N}^{\sigma'}_{\ell'\, m'}(\bv{r}; k)$ vector fields are all linearly-independent solenoidal solutions of the homogeneous vector Helmholtz equation for all $r > 0$ (and even for all $r \ge 0$ in the $\sigma = 0$ case), and are also geometrically orthogonal in the sense that
\be
\bv{M}^{s}_{\ell\, m}(\bv{x}; k) \cdot  \bv{N}^{s}_{\ell\, m}(\bv{x}; k) = 0,
\ee
while, for all $\ell > 0$, the vector fields $\bv{L}^{\sigma}_{\ell\, m}(\bv{x}; k)$ are irrotational solutions, whose divergences $\grad\!\cdot\! \bv{L}^{\sigma}_{\ell\, m}(\bv{r}; k)$ also satisfy the free-space scalar Helmholtz equation for $r > 0$ (and for all $r \ge 0$ if $\sigma = 0$).  Because of their completeness properties of the vector harmonics, all solutions to the source-free Helmholtz equation can be expressed as linear combinations of these basic irreducible families.

\section*{References}

\bibliography{maxpower_2020}

%%%%%%%%%%%%%

\end{document}